\documentstyle[psfig]{article}
\def\begeq{\begin{equation}}
\def\endeq{\end{equation}}
\def\zbar{\bar{z}}
\def\begeqar{\begin{eqnarray}}
\def\endeqar{\end{eqnarray}}
\def\partialbar{\bar{\partial}}
\def\wbar{\bar{w}}
\def\phibar{\bar{\phi}}
\def\xhat{\hat{x}}
\def\phat{\hat{p}}
\def\abar{\bar{a}}
\def\e{\epsilon}
\def\t{\theta}
\def\a{\alpha}

\begin{document}
\begin{titlepage}
\title{Lectures on Non Perturbative Field Theory\\
 and Quantum Impurity Problems.}
\author{H. Saleur\thanks{
Department of Physics, University of Southern California, 
Los-Angeles, CA 90089-0484. email: saleur@physics.usc.edu}}
\date{\today}
\maketitle

\begin{abstract}
These are lectures presented at the Les Houches Summer School ``Topology and Geometry in Physics'',
July 1998. They provide a simple introduction to non perturbative
methods of field theory in $1+1$ dimensions, and their application
to the study of strongly correlated condensed matter 
problems -  in particular quantum impurity
problems. The level is moderately advanced, and takes the student all the way 
to the most recent progress in the field: 
many exercises and additional references are provided. 

In the first part, I give a  sketchy introduction to {\sl conformal field theory}. I then
explain how boundary conformal invariance 
can be used to classify and study low energy, strong coupling fixed points in quantum impurity
 problems.
In the second part, I discuss {\sl quantum integrability} from the point of view of 
perturbed conformal field theory, with a special emphasis on the recent ideas of 
massless scattering. 
I then explain how these ideas allow the 
computation of (experimentally measurable) transport properties  in cross-over regimes. 
The case of edge states
tunneling in the fractional quantum Hall effect is used throughout the lectures as an example 
of application.

\end{abstract}

\end{titlepage}

\huge \noindent{\bf Introduction}
\normalsize
\bigskip
\bigskip

Quantum impurity problems have been for many years, and increasingly 
so recently, a favorite subject of investigations,  for theorists and 
experimentalists alike. There are many good reasons for that. 

First, these
problems often represent the
simplest setting in which some qualitatively fascinating physical properties can 
be observed and probed. For instance, the Kondo model (for a comprehensive review\footnote{These
 being only  lecture notes, I have tried to refer to papers that were pedagogically inclined,
if at all possible, rather than to original works.} 
 , consult 
\cite{hewson}) provides a clear cut example
of asymptotic freedom. The basic experimental fact is that 
normal metals with dilute impurities exhibit an  unusual minimum in the temperature 
dependence  of the electrical resistivity. The ultimate explanation is that the 
interaction of the electrons with
the impurity spins produces an {\sl increase} of the resistivity as $T$ is lowered,
counteracting the usual {\sl decrease} in resistivity arising from interactions with lattice 
phonons. Low temperature means low energy, or large distance: we thus have a problem 
where interactions {\sl increase} at large distance, a characteristic of asymptotic 
freedom
\footnote{The analogy with QCD 
can be made more complete, including the logarithmic dependences
encountered in both problems, and the ``dimensional transmutation''.}. As another example
I would like to mention  the recent experiments \cite{Glattli,Weizmann} about  point contact tunneling
in the fractional quantum Hall effect (for a review on this active topic, see \cite{Sarma}) at filling fraction $\nu={1\over 3}$. Measurments
 of the shot noise reveal a behaviour 
$\left<I^2\right>\propto {e\over 3} I_B$ in the limit of weak backscattering,
where few quasiparticles tunnel, and do so independently. 
If one compares
this formula with the standard Schottky formula for Fermi liquids, one sees that 
this noise has to be  due to the  tunneling of {\sl fractional charges} $e^*={e\over 3}$:
although the existence of these (Laughlin quasiparticles) had been conjectured
for a long time, the noise is the first direct evidence of their existence \footnote{The conductance
itself is not a measure of the charge of the carriers.}. As a final example, 
let me recall that the basic archetype of dissipative quantum
mechanics (for a review, look at \cite{sudiprmp}, \cite{uli}), the two state problem coupled to a bath of oscillators with Ohmic dissipation,
is described by another quantum impurity problem: the anisotropic Kondo model. 
Crucial
fundamental issues are at stake here, as well as a large array of applications in chemistry
and biology.  
 
The second reason of our fascination for quantum impurity problems is that they are,
to a large extent, manageable by analytic methods. This has led to incredibly fruitful progress 
in the past. For instance, 
 the renormalization group was, to a large extent, borne out of the 
efforts of Kondo, Anderson and Wilson to understand the low temperature behaviour of the 
Kondo model (see \cite{hewson}). Also, the works of Andrei \cite{andrei}, Wiegmann \cite{wiegmann}
and others
showed that the Bethe ansatz could be used to analyze situations of
experimental relevance: this  spurred a new interest in quantum integrable
models, an area which, together with its various off-springs like quantum 
groups, knot theory and others,  has become one of most lively in 
mathematical physics.

Finally, it is fair to say that quantum impurity problems are not only of fundamental
interest: they are at the center of the most challenging problems of today's condensed 
matter, like Kondo lattices, heavy fermions, and, maybe, high $T_c$ superconductors.  

In these lectures, I will concentrate on what is usually considered the most important
about quantum impurity problems: their properties as strongly 
interacting systems. There is no doubt that strongly correlated electrons 
are of the highest interest. On the practical side, besides high $T_c$ superconductivity, 
let me mention the 
remarkable recent developments  in manufacturing and understanding  small systems
like quantum wires, carbon nanotubes and the like,
 where, because of the reduced dimensionality, Fermi liquid theory
is not applicable, and the interactions have to be taken into account non perturbatively.
On a more fundamental level, strongly interacting systems exhibit rather counter intuitive 
properties, the most spectacular being probably spin charge separation. It is a challenge
for the theorist to understand these properties, and quantum impurity problems no doubt 
provide  the best theoretical and experimental laboratory to do so.

Maybe it is time now to define what I mean by a quantum impurity problem. The general class of 
systems 
I have in mind have the following features: $(i)$ There are extended gapless (critical) quantum 
mechanical degrees of freedom,
which live in an infinite spatial volume, the ``bulk'' $(ii)$ These  interact with an impurity,
localized at one point in position space. This impurity may carry quantum mechanical 
degrees of freedom.

To have an example in mind, consider the Kondo problem:
(i) The extended degrees of freedom are those of the bulk
metal. The presence of a Fermi surface  means
that the metal sits at a RG fixed point (see e.g. 
\cite{Shankarrmp}). Physically this is easily
understood  since the system of electrons has
(``particle-hole'') excitations of arbitrarily
low energy about its Fermi-sea ground state,
providing the critical degrees of freedom in the bulk of the metal
(ii) The impurity spin, located at one point in space
(say the origin), is a dynamical quantum mechanical
degree of freedom (the dynamical process is the spin-flip).

In the Kondo model we also see another feature
of quantum impurity problems: they are generically
one-dimensional. The problem of an impurity in a Luttinger liquid to be
discussed
below is inherently one-dimensional, but the Kondo model needs to be
reduced to one dimension. Since the impurity spin sits
only at one point in space, it is only the
s-wave wavefunctions of the metal electrons
that can interact with the spin. Second-quantizing
this s-wave theory, we get a (non-interacting) quantum
field theory of one-dimensional Fermions,
defined on  half-infinite (radial)
position space (a half-infinite line), which interacts with the
quantum spin at the end of the line.
Considering a path-integral representation of
the 1D theory of Fermions,
we have a $(1+1)$-dimensional Lagrangian field
theory, one dimension from the (half-infinite)
radial space coordinate, and another dimension
from the (say, euclidean) time coordinate.
All interactions take place at one point in space,
the end of the line, where the impurity
is located. In the $(1+1)$ dimensional
space-time picture, the impurity sits
at the ``boundary'' of space-time, which
can be viewed as the upper-complex plane,
the ``boundary'' being the real axis (these points of view are schematically
illustrated in figure 1 and 2).
This will be the general picture that we use
in the study of all quantum impurity problems.

\begin{figure}[tbh]
\centerline{\psfig{figure=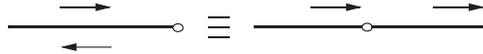,height=.25in,width=2.5in}}
\caption{We will think of quantum impurity problems in various ways. The figure
on the  left represents a $1+1$ quantum point of view where the bulk right and left
degrees of freedom are confined to a half line, with the impurity at the origin. 
Altrnatively, because the theory is massless in the bulk, one can unfold this picture
to get only right degrees of freedom on the full line, as indicated on the  right.}
\end{figure}
\begin{figure}[tbh]
\centerline{\psfig{figure=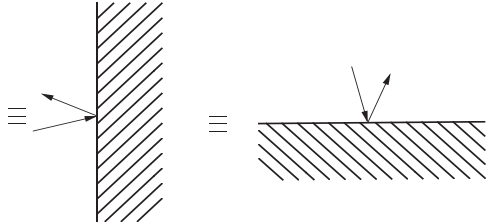,height=1.15in,width=2.5in}}
\caption{Alternatively again, one can go to imaginary time and obtain a $2$D statistical mechanics point
of view, with a theory defined in a half plane, that often we will chose to be 
the upper complex plane. In this figure, the arrows are supposed to ``represent'' 
the bulk degrees of freedom; later, we will see that they can be associated
with integrable quasiparticles. }
\end{figure}

In the Kondo problem we see one  way  how a quantum
impurity problem can be realized experimentally:
a bulk system (here the 3D metal) contains a finite
but small  concentration $x$ of quantum impurities.
In the limit of very dilute impurities ($x < 1$)
the impurities do not interact with each other
(to lowest order in $x$), and the single-impurity
theory  may be used to describe the physics
of the bulk  material in the presence of
dilute impurities. Actually, experiments  performed at very
low concentrations  are known to  be in good
agreement with the single-impurity theory for the
ordinary (one-channel) Kondo model.

A quite different realization  of quantum impurity models
occurs in the context of point contacts. These
are basically electronic devices: two leads (capable of
transporting  electrical current) are attached to
a single  quantum impurity. Each one of the leads is connected
to a battery, so that electrical current is
driven through the quantum impurity. One can
then measure experimentally the electrical
current $I$, flowing through the quantum impurity,
as a  function of the applied driving
voltage $V$ (from the battery) \cite{KF}.  The $I(V)$
curve, the differential (non-linear) conductance
$G_{diff}= \partial I(V)/\partial V$,
or the temperature dependence
of the linear response conductance
$G_{lr}=\lim_{V\to 0}
I(V)/V$ are examples of experimental
probes characteristic of the quantum impurity.
Notice that since the $I(V)$ curve as well as the differential
conductance are non-equilibrium properties,
the point-contact realizations
of quantum impurity problems
are theoretically more challenging
than other realizations, in that 
more than equilibrium statistical mechanics
is involved to achieve a theoretical
understanding of these quantities.

Ideally, the most interesting point contact situation would
involve one-dimensional leads, where electrons are
described by the Luttinger model,  the
simplest non-fermi-liquid metals \cite{Hald}. It consists
of left- and right-moving gapless excitations
at the two fermi points in an interacting 1-dimensional
electron gas.  In the past, this model had been difficult to
realize experimentally however. This is simply because
in a one-dimensional conductor (such
as  a quasi-one-dimensional quantum
wire so thin that the transverse modes
are frozen out at low temperature), random impurities
occur in the fabrication. These impurities lead to
localization due to backscattering processes
between the excitations at the two fermi points.
In other words, the random impurities generate a mass
gap for the fermions.

Fortunately, there is another possiblity:  the edge excitations
at the boundary of samples prepared in  a fractional
quantum Hall state should be  extremely  clean realizations of
the Luttinger non-fermi liquids, as was observed by Wen  \cite{Wen}. In
contrast to quantum wires, these are stable systems because
 for $1/\nu$ an odd integer, the excitations only move in one
direction on a given edge.  Since the right and left edges are
far apart from each
other, backscattering processes
due to random impurities in the bulk cannot localize  those
extended edge states. Moreover, the Luttinger interaction
parameter is universally related  to the filling fraction
$\nu $ of the quantum Hall state in the bulk sample
by a topological argument based on the underlying Chern-Simons
theory, and does therefore not renormalize.
The edge states should thus provide an extremely
clean  experimental realization of the Luttinger model.

We now describe an experimental set-up in more details \cite{exper,Moon} (see figure 3).

\begin{figure}[tbh]
\centerline{\psfig{figure=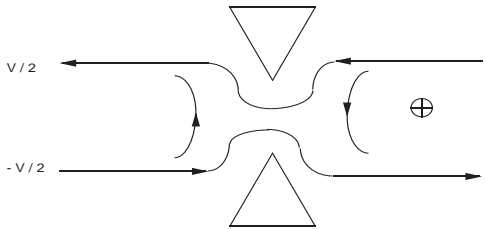,height=1.15in,width=2.5in}}
\caption{A schematic experimental set up to study point contact tunneling 
in the fractional quantum Hall effect (the magnetic field points towards 
the reader). Details are provided in the text.}
\end{figure}

A fractional quantum Hall state  with filling fraction $\nu=1/3$
is prepared in the
bulk of a quantum Hall bar which is long in the $x$-direction and short in the $y$-direction.
This means that the bulk quantum Hall state
is prepared  in a Hall insulator state
(longitudinal conductivity $\sigma_{xx}=0$),
and that the (bulk) Hall
resistivity  is on the $\nu=1/3$ plateau
where $\sigma_{xy}=(1/3)e^2/h $.  This is achieved
by adjusting the applied magnetic field, perpendicular to
the plane of the bar. Since the plateau is broad, the applied
magnetic field can be varied over a significant range without
affecting the filling of $\nu=1/3$.
Next, a gate voltage $V_g$ is applied perpendicular to
the long side of the bar, i.e. in the $y$ direction 
at $x=0$. This has the effect of bringing the right and left
moving edges close to each other near $x=0$, forming
a point contact. Away from the contact there is no backscattering
(i.e. no tunneling of charge carriers)
because the edges are widely separated, but now
charge carriers can hop from one edge to
the other at the point contact.

The left-moving (upper)  edge of the Hall bar
 can now be connected to battery on the right
such that the charge carriers are injected into
the left-moving  lead of the
Hall bar with an equilibrium thermal
distribution at chemical potential $\mu_L$. Similarly,
the right-moving carriers  (propagating in the lower edge)
are injected from the left, with a thermal distribution
at chemical potential $\mu_R$. The difference
of chemical potentials of the injected
charge carriers is the driving voltage $V= \mu_R-\mu_L$.
If $V>0$, there are more carriers injected from
the left than from the right, and a ``source-drain''
current flows from the  left to the right,
 along the $x$-direction of the Hall bar.
In the absence of the point contact,
the driving voltage places the right and left
edges at different potentials (in the $y$-direction,
perpendicular to the current flow),  implying that
the ratio of source-drain
current to the driving voltage $V$ is the Hall conductance
 $G=\nu e^2/h$ (both in linear response and at finite driving voltage
$V$).
When the point-contact interaction is included,
at finite driving voltage, more of the right moving carriers injected from
the left are backscattered than those injected
from the right, resulting in a loss of
charge carriers from the source-drain current.
In this case we write the total source-drain current
as $I(V) = I_0(V) + I_B(V)$, where $I_B(V)$ is
the (negative) backscattering current, quantifying
the loss of current due to backscattering at the point
contact.
It is this backscattering current that I ultimately want to 
show how to compute.

Let me write up some formulas as a preamble. I will not have time in these
lectures to discuss bosonization or edge states in the fractional quantum Hall effect:
I will thus simply claim that, in its bosonized form, the problem is described by the hamiltonian
\begin{equation}
H={1\over 2}\int_{-\infty}^\infty dx \left[\Pi^2+
(\partial_x \phi)^2\right]+
\lambda \cos{\beta\over \sqrt{2}}(\phi_L-\phi_R)(0),
\end{equation}
where ${\beta^2\over 8\pi}=\nu$. Here, the free boson part describes the massless edge states 
\cite{Wen}, and the 
cosine term describes the effect of the gate voltage, with $\lambda\propto V_g$. In
general of course, the backscattering term induced by this gate voltage should be represented
by a complicated interaction; but we keep only the most relevant term (the only one for 
$\nu={1\over 3}$),
which is all that matters in the scaling regime (see below) 
we will be interested in.  

The interaction is a relevant term, that is, in a renormalization group transformation,
one has, $b$ being the rescaling factor \footnote{In particle physics
language, $ {d\lambda\over db}=-\beta(\lambda)$, so our relevant operator
corresponds to a negative beta-function, ie an asymptotically free theory.}  
\begeq
{d\lambda\over db}=(1-\nu)\lambda +O(\lambda^3)
\endeq
This means that at large gate voltage, or, equivalently, at small temperature
(since then, elementary excitations have low energies, so the barrier appears big to them),
the point contact will essentially split the system in half, and no current will flow through
\footnote{This feature is actually remarkable. What it means is that, for one dimensional electrons
with short distance repulsive interactions, an arbitrarily small impurity leads to no transmittance 
at $T=0$\cite{KF}: compare with the effects of barriers on non-interacting electrons you studied
in first year quantum mechanics.}. The questions the theorist wants to answer are:
how do we study the vicinity of the weak-backscattering limit? How do we find out 
more precisely what the strong back-scattering limit looks like? How about its vicinity?
Finally, can we be more ambitious and compute say the current at any temperature, voltage and gate voltage?

For this latter question, let me stress that we are interested in the universal, or scaling, regime,
which is the only case where things will not depend in an complicated way on the microscopic details
of the gate and other experimental parameters. In practice,
what the experimentalist will do is 
first  sweep through values of the gate voltage, the conductance
signal showing a number of resonance peaks, which
sharpen as the temperature is lowered.
These resonance peaks occur for particular values
$V_g = V_g^*$ of the gate voltage, due to tunneling through
localized states in the vicinity of the point contact.
Ideally, on resonance, the source-drain conductance
is equal to the Hall conductance without point contact,
i.e. $G_{resonance} = \nu e^2/h$. This
value is independent of temperature, on resonance. Now, measuring for instance
the linear response conductance
as a function of the gate voltage near the
resonance, i.e. as a function of
$ \delta V_g \equiv V_g-V_g^* $, at a number of different
temperatures $T$,
 one gets resonance curves, one for each
temperature.
These peak at $\delta V_g =0$. Finally, these
conductance  curves should collapse, in the limit of very small
$T$ and $\delta V_g$, 
onto a single universal curve when plotted
as a function of $\delta V_g/T^{1-\nu}$. This is what the field 
theorist wants to compute. 

To proceed, it is useful to formulate the problem as a boundary problem.  
For this, a few manipulations are needed . We 
decompose $\phi=\phi_L+\phi_R$ and set \footnote{More details are given in Part I.
 A common objection 
to the following manipulations is that they  are good only  for free fields, but not
when there is a boundary interaction. This, in fact, depends 
on what one means by ``fields'' - the safest attitude is 
to imagine one does perturbation theory in $\lambda$. Then, all the quantities
are evaluated within the free theory, on which one can legitimately do all 
the foldings, left right decompositions, etc.} :
\begin{eqnarray}
\varphi^e(x+t)={1\over\sqrt{2}}\left[\phi_L(x,t)+\phi_R(-x,t)\right]
\nonumber\\
\varphi^o(x+t)={1\over\sqrt{2}}\left[\phi_L(x,t)-\phi_R(-x,t)\right]
\end{eqnarray}
Observe that these two fields are left movers. We now fold the system
by setting, for $x<0$~:
\begin{eqnarray}
&\phi^e_L=\varphi^e(x+t),~~~~~~~
\phi^e_R=\varphi^e(-x+t),
\nonumber\\
&\phi^o_L=\varphi^o(x+t),~~~~~~~
\phi^o_R=-\varphi^o(-x+t), 
\end{eqnarray}
and introduce new fields $\phi^{e,o}=\phi^{e,o}_L+\phi^{e,o}_R$, both 
defined on the half infinite line $x<0$.
The odd field $\phi^e$ simply obeys  Dirichlet boundary conditions
at the origin $\phi^o(0)=0$, and decouples from the problem.
The field $\phi^e$, which we call rather $\Phi$ in the following, 
has a non trivial dynamics
\begin{equation}
\label{hamil}
H\equiv H^e=\frac{1}{2}\int_{-\infty}^0 dx \
\left[\Pi^{2}+(\partial_x\Phi)^2\right]+
\lambda \cos\frac{\beta\Phi(0)}{2}.
\end{equation}

The aspects we have to understand are, by increasing order
of complexity: the fixed points, their vicinity, and what is in between. This is the order I will
follow in these lectures.

\vfill
\eject
\part{Conformal field theory and fixed points}

The first difficulty one encounters in that field is how
to describe the low energy fixed points. This may sound rather simple in the 
tunneling problem, but in other cases, for instance in a tunneling problem for electrons 
with spin, the matter is quite involved. The reason for this is, that fixed points
are not necessarily described in terms of nice linear boundary conditions for
the bulk degrees of freedom. It does seem to be true however, 
that even if the  quantum
impurity has internal degrees of freedom,
interaction and renormalization effects
do turn the  dynamical quantum impurity
into a boundary condition
on the extended bulk degrees of freedom,
at large distances, low energy or  low temperature. At low
temperatures the system may be in the strong
coupling regime (for instance, this is where Kondo's result diverges).
The boundary condition is thus a way to think
about the strongly interacting system.
Nozi\`eres' physical picture of screening \cite{Nozieres}
illustrates how this works for the simplest case, the one-channel Kondo
model: the antiferromagnetic interaction of the impurity spin
with the spin of the conduction electrons, which has renormalized
to large values at low temperatures, causes complete
screening of the impurity spin. A modified boundary condition on
the electrons that are not involved in the screening,
is left. This mechanism, however, appears to be much more general,
and seems to apply to all quantum impurity problems.

The boundary conditions generated in this process may be highly
non-trivial (see e.g.\ \cite{ALnpbii,ALgreens}).
 However, since the bulk is massless (critical),
the induced boundary condition is scale invariant
asymptotically at large distances
and low temperatures. Actually, it is, in most cases, conformally
invariant.

Quantum impurity problems are thus intimately related
with scale-invariant boundary conditions: these are 
 RG fixed points, and, like in bulk $1+1$ quantum field theories,
(recall that  the bulk is always critical in
the type of systems that we are considering here), conformal symmetry 
is the best way to describe them.

Now, conformal invariance is a long story. All I can do is
provide, in the next  sections, what I believe is 
the minimal set of ideas necessary to understand what is going on, and tackle without 
fear the literature on the subject. 
 In several instances, 
I will have  to  discard entire  discussions of key issues, substituting 
them with some intuitive comments, and only providing the
final result. Additional bits and pieces are then provided in the text
in small characters, together
with specific  references, to help the  reader bridge the gaps. Good reviews
on this subject are the Les Houches Lectures of 1988 \cite{houches88}, the article
by J. Cardy \cite{Cardyreview}, the lectures by J. Polchinski \cite{polchinski}, and  the
 textbook \cite{philippe}. The relevant chapters in \cite{tsvelikbook} can also
be quite useful. 

In the following, I will intimately mix path integral and hamiltonian points of view. 
The two are of course equivalent, but each has its own advantages.

\section{Some notions of conformal field theory}

\subsection{The free boson via path integrals} 

We consider the free bosonic theory, with action
\begeq
S={1\over 2}\int dx_1dx_2\left[\left(\partial_1\Phi\right)^2+
\left(\partial_2\Phi\right)^2\right]
\endeq
To start, let us discuss briefly the issue of correlators and regularization. To keep
in the spirit of condensed matter, we initially 
 define the Gaussian model on a discrete periodic square lattice
of constant $a$ by setting
\begeq
S_{latt}={1\over 2}\sum_{\left<jk\right>}\left[\Phi(r_j)-\Phi(r_k)\right]^2,
\endeq
where the sum is taken over all pairs of nearest neighbours. Introduce
the lattice Green function
\begeq
G_{latt}(r)={1\over L^2}\sum'_{n_1,n_2}{e^{ik.r}\over 4-2\cos{2\pi n_1a\over L}
-2\cos{2\pi n_2a\over L}}
\endeq
where the sum is restricted to the first Brillouin zone
$|n_i|\leq {L\over 2a}$, and the prime means the zero mode is excluded 
\footnote{The zero mode divergence simply occurs because the 
action is invariant under the symmetry $\Phi\to\Phi+\hbox{ cst}$.}.
One has then (where the points $r,r'$ belong to the lattice)
\begeq
\left< \Phi(r)\Phi(r')\right>_{latt}=G_{latt}(r-r')
\endeq
while $G_{latt}$ satisfies the discrete equation (where $\Delta_{latt}$ is the discrete
Laplacian)
\begeq
\Delta_{latt}G_{latt}=-\delta_{latt}(r)
\endeq
The important points here are the behaviours
 $G_{latt}(0)\approx -{1\over 2\pi}\ln{a\over L},L>>a$, and $G_{latt}(r)\approx -{1\over 2\pi}
\ln{r\over L}$
for $a<<r<<L$.

We recall now  Wick's theorem, according to which the average
of any quantity can be obtained as a sum of all pairwise 
contractions. It follows that
\begeqar
\left< e^{i\beta_1\Phi(r)}e^{i\beta_2\Phi(r')}\right>_{latt}=&\nonumber
\exp\left[-{1\over 2}(\beta_1+\beta_2)^2\left<\Phi^2(r)\right>_{latt}\right]\\
&\times\exp\left[ -\beta_1\beta_2\left(\left<\Phi(r)\Phi(r')\right>_{latt}-\left<
\Phi^2(r)\right>_{latt}\right)\right]
\endeqar
To define a continuum limit for this 
model, we look at  distances large compared to the lattice spacing but 
small compared to $L$, where the right hand side of the previous 
expression simplifies into
\begeq
\left< e^{i\beta_1\Phi(r)}e^{i\beta_2\Phi(r')}\right>_{latt}\approx
\left({a\over L}\right)^{(\beta_1+\beta_2)^2/4\pi}\left({|r-r'|\over a}
\right)^{\beta_1\beta_2/2\pi}
\endeq
The well known observation follows that the correlator vanishes
unless charge neutrality is satisfied, that is $\beta_1+\beta_2=0$. 
We then have, where $r,r'$ are now arbitrary points in the continuum,
\begeq
\left< e^{i\beta\Phi(r)}e^{-i\beta\Phi(r')}\right>\approx
\left({a\over |r-r'|}
\right)^{\beta^2/2\pi}
\endeq
In the following we will sometimes, but not always, set $a=1$. 

\subsection{Normal ordering and OPE}

We now introduce complex coordinates $z=x_1+ix_2,\zbar=x_1-ix_2$. 
We have
$\partial={1\over 2}(\partial_1-i\partial_2),\partialbar
={1\over 2}(\partial_1+i\partial_2)$,
 $d^2z=2dx_1dx_2$, and we define the delta function by 
$\int d^2z\delta^2(z,\zbar)=1$, so $\delta^2(z,\zbar)={1\over 2}\delta^2(x_1,x_2)$.
The action reads now
\begeq
S=\int d^2z\partial\Phi\partialbar\Phi,
\endeq
and the laplacian 
\begeq
\Delta=2\partial\partialbar+2\partialbar\partial
\endeq
Of crucial importance is the result \footnote{A physical way to prove this is 
to observe that, in practice, ${1\over z}$ has to be regulated with a short
 distance cut-off which does introduce a $z$ dependence, as for instance in
${H(z\zbar -a^2)\over z}$, $H$ the Heavyside function.} 
\begeq
\partial\left({1\over \zbar}\right)=2\pi\delta^2(z,\zbar)\label{basicid}
\endeq
from which it follows that\footnote{Note that the two derivatives cannot be interchanged
on singular functions, that is why $\Delta\ln r=2\pi\delta^2(x_1,x_2)$, 
and not twice as much.}
\begeq
\partial\partialbar \ln|z|^2=\partial
\partialbar \ln \zbar=2\pi\delta^2(z,\zbar)
\endeq
It is customary to write the basic correlator as\footnote{Of course the notation is somewhat
redundant, since the value of $z$ determines
$x_1$ and $x_2$; but in what follows, we will reserve the notation $f(z)$ 
for analytic functions.} 
\begeq
\left<\Phi(z,\zbar)\Phi(z',\zbar')\right>=-{1\over 4\pi}\ln|z-z'|^2
\endeq
Note that, in this expression, we have completely 
discarded the $L$ dependence that occurs in the lattice system. A reason
for doing so is that $\Phi$ is not a ``good'' field anyway, and that we will
usually consider rather derivatives of $\Phi$, for which this ambiguity does not
matter. The $L$ dependence is however  crucial for exponentials of the 
field $\Phi$. When we discard it, we have to remember that, at the end of the 
day, only correlators which have vanishing charge are non zero.

Now, still using  Wick's theorem it follows  that
\begeq
\partial\partialbar\left<\Phi(z,\zbar)\Phi(z',\zbar')\ldots\right>
=-{1\over 2}\left< \delta^2(z-z',\zbar-\zbar')\ldots\right>,
\endeq
where the dots stand for any other insertions in the path integrals,
that involve no field either at $z$ or $z'$. A relation that holds in  this
sense is simply rewritten
\begeq
\partial\partialbar\Phi(z,\zbar)\Phi(z',\zbar')
=- {1\over 2}\delta^2(z-z',\zbar-\zbar')\label{queqofmot}.
\endeq
This is the first example of equations we will write quite often
between ``operators'' in the theory - the word operator here 
occurs naturally when one splits open the path integral
to obtain a hamiltonian description, see later. 

Recall that the equations
of motion for the field $\Phi$ read,  on the other hand
\begeq
\partial\partialbar\Phi(z,\zbar)=0
\endeq
It follows that the product $\Phi(z,\zbar)\Phi(z',{\zbar}')$ obeys the equation of motion
except at coincident points.

In the sequel, we shall constantly use the concept of {\sl normal ordering}. 
We define the normal ordering of the product of two bosonic fields by\footnote{
If one wishes to keep the $a$ and $L$ factors, the normal ordering formula reads
$:\Phi(z,\zbar)\Phi(z',\zbar'):\equiv \Phi(z,\zbar)\Phi(z',\zbar') +{1\over 4\pi}
\ln\left(|z-z'|/a\right)^2$. One has then $\left<:\Phi^2:\right>=\ln L/a$.}
\begeq
:\Phi(z,\zbar)\Phi(z',\zbar'):\equiv \Phi(z,\zbar)\Phi(z',\zbar') +{1\over 4\pi}
\ln|z-z'|^2
\endeq
This definition is such that the normal ordered product of fields now {\sl does}
satisfy the equation of motion even at coincident points
\begeq
\partial\partialbar:\Phi(z,\zbar)\Phi(z',\zbar'):=0
\endeq
As a result of this,  the normal product is (locally) the sum of an analytic
and antianalytic function, and can be expanded in powers of $z$. Thus,
for instance
\begeq
\Phi(z,\zbar)\Phi(0,0)=-{1\over 4\pi}\ln|z|^2+:\Phi^2(0,0):+z:\partial\Phi\Phi(0,0):
+\zbar:\bar{\partial}\Phi\Phi(0,0):+\ldots\label{fifi}
\endeq
This is the first example of an operator product expansion (OPE). Like equation 
(\ref{queqofmot}),
its precise meaning is that it holds once inserted inside a correlation
 function. OPEs in conformal field theories are not asymptotic, but rather
convergent expansions; their radius of convergence is given by the distance to the 
nearest other operator in the correlation functions of interest. The right 
hand side of (\ref{fifi}) involves products of fields at coincident
points, which turn out to be well defined in this theory. Notice that
 $:\Phi^2(0,0):$
in (\ref{fifi}) could be defined equally well by point splitting, as will be discussed 
later.

For a product of more than two bosonic operators, the definition of normal
order can be extended iteratively
\begeqar
:\Phi(z,\zbar)\Phi(z_1,\zbar_1)\ldots\Phi(z_n,\zbar_n):=
\Phi(z,\zbar):\Phi(z_1,\zbar_1)\ldots\Phi(z_n,\zbar_n):\nonumber\\
+{1\over 4\pi}\left(\ln|z-z_1|^2:\Phi(z_2,\zbar_2)\ldots\Phi(z_n,\zbar_n):+
\hbox{ permutations}\right)
\endeqar
such that the classical equations of motion are still satisfied (a quick way to understand 
normal order,
as is clear from the previous formula, is that quantities inside double dots are not contracted
with one another when one computes correlators). 

Of crucial importance is the normal ordered exponential $:e^{i\beta\Phi}:$. It
is a good exercise  to recover the OPE
\begeq
:e^{i\beta_1\Phi(z,\zbar)}:\ :e^{i\beta_2\Phi(0,0)}:=|z|^{\beta_1\beta_2\over 2\pi}
:e^{i\beta_1\Phi(z,\zbar)+i\beta_2\Phi(0,0)}:
\endeq
and
\begeq
\partial\Phi(z,\zbar):e^{i\beta\Phi(0,0)}:={-1\over 4\pi z}:e^{i\beta\Phi(0,0)}:
+:\partial\Phi(z,\zbar)e^{i\beta\Phi(0,0)}:\label{derexp}
\endeq
The quantities inside the normal ordering symbols can now be
expanded  in powers of $z,\zbar$ like 
an ordinary function. 

\smallskip
\noindent{\bf Exercise}: Show that the ``quantum Pythagoras'' theorem holds:
$$
\left.\left(:\cos 2\sqrt{\pi}\Phi:\right)^2+\left(:\sin 2\sqrt{\pi}\Phi:\right)^2\right|_{finite}=
-4\pi\partial\Phi\partialbar\Phi
$$
\smallskip

Notice that in (\ref{derexp}), we could have 
treated  $\partial\Phi$ as an  analytic function: a perfectly legitimate thing
to do when one computes correlators. Hidden here is, in fact,
 the very convenient  decomposition 
of the field $\Phi$ itself into the sum of an analytic and antianalytic 
component (one has to be very careful however when one writes
$\Phi(z,\zbar)=\phi(z)+\phibar(\zbar)$; first, 
because such a decomposition does not hold for the general fields summed
 over in the path integral,
and second, because  the field $\Phi$ does not obey the equations of 
motion at coincident points). Pushing that line of thought a bit further however, one has
\begeqar
\left<\phi(z)\phi(z')\right>=-{1\over 4\pi}\ln(z-z')\nonumber\\
\left<\phibar(\zbar)\phibar'(\zbar')\right>=-{1\over 4\pi}\ln(\zbar-\zbar')
\endeqar
this up to phases due to the branch cuts. We will also use in the following the
dual field
\begeq
\tilde{\Phi}=\phi-\bar{\phi}
\endeq

The exponentials are scalar operators, ie they  are invariant under
 rotations. More general operators also have a 
spin, so their two point function reads
\begeq
\left< O(z,\zbar)O(0,0)\right>={1\over z^{2h} \zbar^{2\bar{h}}}
\label{hhbar}
\endeq
The simplest example is provided by $O=\partial\Phi$ which has $h=1,\bar{h}=0$. 
In general, single valuedness of physical correlators requires $h-\bar{h}$ 
to be an integer. The numbers $h$ and $\bar{h}$ are called usually
right and left conformal dimensions. The dimension of $O$ is $d=h+\bar{h}$,
while its spin is $s=h-\bar{h}$.

\subsection{The stress energy tensor}

The stress energy tensor is defined in the classical theory as 
follows. Consider a coordinate transformation $x_\mu\to x_\mu+\epsilon_\mu$ 
(that is,
changing the arguments of the fields in the action from 
$x_\mu\to x_\mu+\epsilon_\mu$) 
. The variation of the action reads, to lowest order \footnote{The factor
of $2\pi$ is peculiar to the conformal field theory literature.},
\begeq
\delta S=-{1\over 2\pi}\int \partial_\mu\epsilon_\nu T_{\mu\nu} dx_1dx_2\label{actchange}
\endeq
Elementary calculation shows that 
\begeq
T_{\mu\nu}=-2\pi\partial_\mu\Phi\partial_\nu\Phi+\pi\delta_{\mu\nu}\partial_\rho\Phi
\partial_\rho\Phi\label{classt}
\endeq
The stress energy tensor in the quantum theory is defined 
through Ward identities: the  end result is the same formula as (\ref{classt}),
but where products of fields are normal ordered. It enjoys some very important
properties: the symmetry $T_{12}=T_{21}$  as a result of rotational invariance,
 and the
 tracelessness $T_{11}+T_{22}=0$ as a result of scale invariance. In addition, the 
stress energy tensor is always conserved, ie the operator
equation  $\partial_\mu T_{\mu\nu}=0$ holds. This can be checked explicitely
for the free boson using (\ref{classt}) and the classical equations of motion,
which we recall are satisfied by normal ordered products \footnote{More 
generally, that $T$ is classically conserved follows simply from the fact that
the action is stationary as a consequence of the classical equations of motion,
so $\delta S$ must vanish for arbitrary $\epsilon$.}.
In general, one introduces complex components
\begeqar
T_{zz}=&{1\over 4}\left(T_{11}-T_{22}-2iT_{12}\right)\nonumber\\
T_{\zbar\zbar}=&{1\over 4}\left(T_{11}-T_{22}+2iT_{12}\right)\nonumber\\
T_{z\zbar}=T_{\zbar z}=&{1\over 4}\left(T_{11}+T_{22}\right).
\endeqar
which, in the free boson case, read simply
\begeqar
T_{zz}=&-2\pi:\left(\partial\phi\right)^2:\\
T_{\zbar\zbar}=&
-2\pi:\left(\partialbar\phibar\right)^2:\\
 T_{z\zbar}=&0.\\
\endeqar
Using the equations of motion, $T_{zz}$ is analytic (again, 
in the special sense that, when inserted in correlation functions,
the dependence is analytic away from the arguments of the other operators)
and will simply be denoted $T(z)$ in what follows; similarly, $T_{\zbar\zbar}$
is antianalytic. It is easy to chek that these properties
extend to other models: a remarkable consequence of  locality 
(so there is a stress energy tensor to begin with) and masslessness is thus the existence 
of a conserved current, a field of dimensions $(2,0)$. Currents provide a powerful tool
to classify the fields of the theory, as we will see shortly. 

Plugging back our results into (\ref{actchange}), one checks  that the variation of the action
$\delta S$ vanishes exactly for a conformal transformation $z\to z+\epsilon(z)$.
This is the celebrated conformal invariance, which we will discuss in more 
details below.

The short distance expansion of $T$ with itself reads
\begeq
T(z)T(0)={1\over 2z^4}+{2\over z^2}T(0)+{1\over z}\partial T(0)+\hbox{analytic}
\endeq
The coefficient of the $1/2z^4$ term is uniquely determined once the 
normalization of $T$ has been chosen. For other massless relativistic
field theories, this coefficient takes the value $c/2$ where 
$c$ is a number known as the {\sl central charge}. For a free boson we 
see that $c=1$. For $n$ independent free bosons, $c=n$. For a
free Majorana fermion
$c={1\over 2}$. The relative normalization of the two other factors
is fixed by the requirement that $T(z)T(0)=T(0)T(z)$.

\smallskip
\noindent {\bf Exercise}: show this.
\smallskip

\footnotesize

The fact that $T$ is analytic except when its argument coincides
with the one of some other field inside a correlator has
an interesting consequence for the trace of the stress energy tensor. 
Indeed, $\partialbar \langle T(z) T(0)\rangle$ is now a derivative
of the delta function! Using the conservation equation $\partial_\mu T_{\mu\nu}$,
which reads in complex components, $\partialbar T+ {1\over 4}\partial (T_11+T_22)=0$,
if follows that the two point function of the trace of the stress energy tensor 
is non zero:
\begeq
\langle [T_{11}+T_22](x_1,x_2)[T_{11}+T_{22}](0,0)\rangle ={\pi c\over 3}
 \Delta \delta^2(x_1,x_2) \label{anoma}
\endeq
This is a simple example of an anomaly, a quantity which is  zero classically,
but non zero quantum mechanically. It is a bit dangerous to give too much meaning
to (\ref{anoma}) however, since the trace is not really an independent 
object - it is much safer to remember again that classical equations of motion
hold except at coincident points. 

\normalsize

Remarkably, the stress energy tensor was introduced in  a statistical
mechanics long ago by Kadanoff and Ceva \cite{KC}. These authors were interested in 
the way Ising correlators change under shear and scaling  transformations. They 
recognized that, in a critical theory, rescaling in the $x_1$ and $x_2$
directions was equivalent to changing the horizontal and vertical
couplings, and thus that the effect of shear and scaling could be 
taken into account by introducing an operator ``conjugated'' to these changes
in the correlators,
just like, say a change in temperature could be taken into 
account by introducing the total energy in the correlators. It is thus
possible to physically identify $T$, and to wonder how its continuum limit
behaves, and how the various algebraic properties we are going to 
derive emerge.

\subsection{Conformal in(co)variance}

To fix ideas, let us now consider the free bosonic theory on a cylinder
of circumference ${1\over T}$ and length $L$. Introducing
the complex coordinate $w=x+iy$ such that the imaginary axis is parallel to 
the cylinder's length (see figure 4), the two point function
of the field $\Phi$ in that geometry is easily found to be 
\begeq
\left< \Phi(w,\wbar)\Phi(w',\wbar')\right>_{cylinder}=-{1\over 4\pi}\ln\left|{1\over \pi T}
\sin\pi T(w-w')\right|^2.
\endeq

\begin{figure}[tbh]
\centerline{\psfig{figure=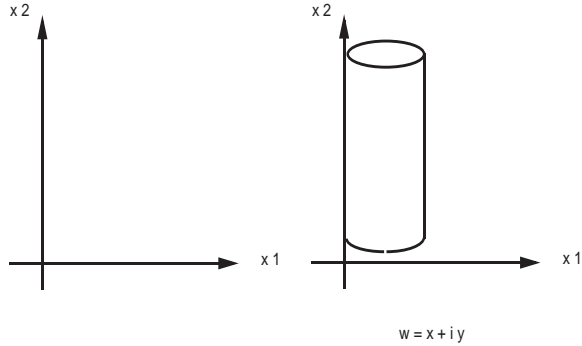,height=1.8in,width=3.in}}
\caption{Some of the geometries used in the text.}
\end{figure}

From this, it follows that
\begeq
\left< :e^{i\beta\Phi(w,\wbar)}:\ :e^{-i\beta\Phi(w',\wbar')}:\right>_{cylinder}
= \left({\pi T\over |\sin \pi T(w-w')|}\right)^{\beta^2/ 2\pi}
\endeq
Let us focus on the holomorphic
part
\begeq
\left< :e^{i\beta\phi(w)}:\ :e^{-i\beta\phi(w')}:\right>_{cylinder}
= \left({\pi T\over \sin \pi T(w-w')}\right)^{\beta^2/ 4\pi}
\endeq
This can be shown to be, equivalently,
\begeq
\left< :e^{i\beta\phi(w)}:\ :e^{-i\beta\phi(w')}:\right>_{cylinder}
=\left({dz\over dw}\right)^{h}\left({dz'\over w'}\right)^h 
\left< :e^{i\beta\Phi(z)}:\ :e^{-i\beta\Phi(z')}:\right>_{plane},
\label{twopointconf}
\endeq
where we used the mapping $z=e^{-2i\pi Tw}$. The latter formula expresses the {\sl covariance} of the two point function
under the conformal transformation. Another example of such covariance is provided by the
derivative of the field
\begeq
\left< \partial_w\phi(w)\partial_{w'}\phi(w')\right>
=\left({dz\over dw}\right)\left({dz'\over dw'}\right)\left< \partial_z\phi(z)
\partial_{z'}\phi(z')
\right>, \label{twopointconfi}
\endeq
where we suppressed mention of the geometry, which is implicit
in the variables used. 

Relations like (\ref{twopointconf},\ref{twopointconfi}) are well expected, since
the gaussian action is, in fact, conformal invariant; this
 follows, as discussed above,  from the properties of the stress energy tensor,
and thus is expected to generalize to other local massless field theories.
More directly, this invariance is easily established for the free boson,
since upon  changing
 the argument of the field from $z\to w$ in the action, $S$ is invariant,
the Jacobian cancelling the term coming from the partial derivatives.

Of course,
one has to be quite careful in using the conformal invariance of the action, since the correlators
are not invariant - ie, one has for instance, 
$\left< \Phi(w,\wbar)
\Phi(w',\wbar')\right>\neq\left< \Phi(z,\zbar)\Phi(z',\zbar')\right>$,
while the naive change of variables in the action would suggest
the propagators to map straightforwardly, and thus the equality 
to hold. The reason for this 
discrepancy comes from the cut-off, which is also modified in a 
conformal transformation. We, on the other hand, wish to use the same
regularization whatever the geometry, ie, for instance, use a square lattice
of constant $a$ to regularize both the problem in the plane and on the 
cylinder; hence, there is an ``anomaly''.  

Fields obeying the general covariance relation (and a similar one for the antiholomorphic part)
\begeq
\left< O(w)O(w')\right> =\left({dz\over dw}\right)^h\left({dz'\over dw'}\right)^h
\left< O(z)O(z')\right>
\endeq
are called {\sl primary fields}. The field $\phi$ itself is not primary,
though, in a way, it satisfies the equivalent of the previous relation
with $h=0$, since
\begeq
\left< \phi(w)\phi(w')\right>=\left< \phi(z)\phi(z')\right>+{1\over 8\pi}
\ln \left({dz\over dw}
{dz'\over dw'}\right)
\endeq
Fields which are not primary exhibit in general more complicated covariance
relations. An example is provided by the second derivative of $\phi$, which 
we leave to the reader to work out. A more interesting example is
furnished by the stress energy tensor. Though we have defined it so 
far by normal ordering, it is clear that an equally good definition
is obtained by point splitting, ie \footnote{Here, it does not matter how and by what
amount the two points are split of course, provided they both tend to $z$
at the end}
\begeq
T(z)=-2\pi\lim_{d\to 0}\left[\partial\phi(z+d/2)\partial\phi(z-d/2)+{1\over 4\pi
d^2}\right]\label{tdef}
\endeq
We have thus, from the change of variables
$$
T(z)=-2\pi\lim_{d\to 0}\left[w'(z+d/2)w'(z-d/2)\partial_w\phi(w(z+d/2))\partial
\phi(w(z-d/2))
-{1\over 4\pi d^2}\right]
$$
To define the stress energy tensor on the cylinder, we use the same
definition (\ref{tdef}), but with $z$ replaced by $w$ \footnote{That is, normal ordering is 
always defined
by subtracting the short distance, geometry independent, divergences.}. Therefore
\begeq
T(z)=\left[w'(z)\right]^2T(w)+{1\over 12}\{w,z\}\label{strestransf},
\endeq
where the added term is 
$$
{1\over 12}\{w,z\}=\lim_{d\to 0}{1\over 2}{w'(z+d/2)w'(z-d/2)\over [w(z+d/2)-w(z-d/2)]^2}-{1\over 2d^2}
$$
This is known under the name of Schwartzian derivative, and reads
\begeq
\{w,z\}={2w'''w'-3w'^4\over 2w'^2}
\endeq
It enjoys nice properties under the composition of successive 
conformal transformations, that we leave to the reader to investigate.
An important property following from  (\ref{strestransf}) is that $T$ acquires a finite
expectation value on the cylinder, while it did not have one in the plane
\begeq
\left< T\right>_{cylinder}=-{c\over 24}\left(2\pi T\right)^2\label{tcylin}
\endeq

\smallskip
\noindent {\bf Exercise}: show this by using the propagators on the plane and the cylinder,
together with appropriate definitions of normal ordering.
\smallskip 

The OPE of the stress energy tensor with any field of the theory 
has the general form
\begeq
T(z)O(0,0)=\ldots+{h\over z^2}O(0,0)+{1\over z}\partial O(0,0)+\ldots\label{primexp}
\endeq
The two terms explicitely written follow from the fact that $O$ 
has dimension $h$ and the use of the Ward identity (\ref{ward}). It can be shown
that $1/z^2$ is the highest singularity if $O$ is primary.

\smallskip
\noindent {\bf Exercise}: check this for the free boson by considering various examples.
\smallskip

\footnotesize

\subsection{Some remarks on Ward identities in QFT.}

Suppose in general that there is a transformation of the field $\Phi'(x_1,x_2)=
\Phi(x_1,x_2)+\delta\Phi(x_1,x_2)$ that 
leaves the 
product of the path integral measure and the Boltzmann weight $e^{-S}$ 
invariant. Examples of such
transformations are provided for instance by translations or rotations in ordinary
isotropic homogeneous physical systems. Consider then a transformation 
$\Phi'(x_1,x_2)=
\Phi(x_1,x_2)+\rho(x_1,x_2)\delta\Phi(x_1,x_2)$. For general $\rho$,
this is not a symmetry of the problem anymore. On the other hand, we can 
always change variables in the functional integral and reevaluate any correlator
in terms of the new field $\Phi'$. This means that we have the identity
\begeq
0=\int [d\Phi']e^{-S'}-\int [d\Phi]e^{-S}\label{changofv}
\endeq
where in $S'$ one maybe had to add up terms coming from the 
change of variables in the path integral. On the other hand, one can expand the 
right hand side of this equation to first order in the change 
of fields assumed small. Since for $\rho$ 
a constant the product of the measure and the weight would be invariant,
 this means that the right hand side of (\ref{changofv}) must
depend on the gradient of $\rho$ only, ie one has
\begeq
\hbox{rhs= }
{i\over 2\pi}\int [d\Phi]e^{-S}\int j_\mu\partial_\mu\ \rho dx_1dx_2
\label{noeth}
\endeq
The quantity $j_\mu$ is called a Noether current. Since it comes from local
manipulations, it must 
be a local quantity. Now, that (\ref{noeth})
vanishes is something that  must hold for
any reasonably smooth function $\rho$.  Let us
choose $\rho$ to be  equal to unity inside a disk of radius $R_1$, and to
vanish
on and outside of $R_2$, while it is arbitrary in between.
Integrating (\ref{noeth}) by parts, we get an integral of $j_\mu$ on the circle $R_1$,
together with an integral on the annulus between $R_1$ and $R_2$ of 
$\partial_\mu j_\mu$. Since the functions $\rho$ is quite arbitrary there,
it  follows that the current has
to be conserved, that is 
\begeq
\partial_\mu j_\mu=0\label{conserv}
\endeq
This result would still hold of course with fields
inserted far from $R_1$ and $R_2$, so (\ref{conserv}) truly holds as
an operator equation, in the sense explained above.

As an application, consider  a translation $x_\mu\to x_\mu+\epsilon_\mu$, 
where $\epsilon_\mu$ is
small and constant: we obtain a current
which, in the classical case, coincides with $ij_\mu=T_{\mu\nu}\epsilon_\nu$. 
The 
foregoing procedure is a generalization to the quantum theory, and the 
conservation equation follows from Noether's theorem.

Now consider some field $O$ inside the circles, say right at the origin.
Under the transformation $\Phi\to\Phi'$, this field 
becomes $O'=O+\delta O$ (the change is expanded to
 first order  as before,
but of course $O'$ might as well depend on the derivatives of $\rho$ in general). 
We now have, since all we are doing is changing variables in the path integral
\begeq
0=\int [d\Phi'] O'\ e^{-S'}\ldots -\int [d\Phi] O\ e^{-S}\ldots
\endeq
Expand this to first order. This time, the integration by part gives
\begeq
-i\delta O={1\over 2\pi}\int (dx_2 j_1-dx_1j_2)O(0,0)=
{1\over 2i\pi} \int(dz j-d\zbar\bar{j})O(0,0)\label{changeofop}
\endeq

As an application, consider a transformation $z\to z+\alpha(z)$. where $\alpha$ is 
small.
The Noether current associated with it is given by $j=i\alpha(z)T(z)$ and 
$\bar{j}=i\bar{\alpha}(\zbar)\bar{T}(\zbar)$. For a transformation that
is conformal inside a contour $C$, and differentiably connected to a (necessary 
non conformal) transformation vanishing at large distances, one finds
from (\ref{changeofop}) the key result
\begeq
\delta O={1\over 2i\pi}\int_C\alpha(z)T(z)O(0)dz-{1\over 2i\pi}\int_{\overline{C}}
 \bar{\alpha}(\zbar)
\bar{T}(\zbar)O(0)d\zbar\label{ward} 
\endeq

\smallskip
{\bf Exercise}: derive from this (\ref{primexp}) and the 
fact that there are no singularities stronger than $1/z^2$ 
 for a primary operator.
\smallskip

Notice finally that for the free boson, the expression of the stress energy tensor 
is almost the classical one, up to normal ordering, and it appears as if the 
integration measure essentially plays no role in the construction of the 
Ward identities. That one can forget about the behaviour of the measure in
conformal transformations is justified a posteriori, by the fact that the quantum
currents are indeed conserved. The measure would play a more subtle 
role for theories defined on curved two-dimensional
manifolds.

\normalsize

\subsection{The Virasoro algebra: intuitive introduction}

As noticed before, the main consequence of conformal invariance is the existence of 
a conserved current, the stress energy tensor $T$. In general, one sets
\begeq
T(z)=\sum_{n=-\infty}^\infty {L_n\over z^{n+2}}
\endeq
that is, plugging this expansion into the OPE provides a definition
of what the field $L_nO$ actually is: for instance $L_0 O=hO$, 
$L_{-1}O=\partial O$, $L_{2}T=1$, etc. In general, one does not expect
 fields
with negative dimensions to appear, or at least not fields
 with arbitrarily large
negative dimensions (weird things can occur in non unitary theories adapted to 
disordered systems in particular though). This means that for every field,
 $L_n O$ must
vanish for $n$ positive large enough. Of particular interest are the primary
fields, for which
the highest singularity is $1/z^2$, ie they satisfy $L_n O=0,n>0$.

For the moment, we can contend
 ourselves with the   intuitively
reasonable notion that the $L_n$ are ``operators'' acting on the space of
 fields of the theory - here, exponentials  multiplied by
 normal ordered polynomials in derivatives
of the field $\Phi$ - so the $L_n$ are not  unlike differential operators
acting on functions (in fact, the $(L_{-1})^n$ are just that). 

 It is then
tempting to ask oneself what the algebra of these operators is, that is
how do $L_n(L_m O)$ compare with $L_m(L_n O)$? This is easily done by using 
contour integration together with the short distance expansions. The commutator
$[L_n,L_m]O$ can be computed as follows. We have 
\begeq
L_nL_m O=\int_{C'}{dz\over 2i\pi}z^{n+1}T(z)\int_{C}{dw\over 2i\pi} w^{m+1}T(w) O(0)
\endeq
where the contours encircle the origin and $C$ is inside $C'$ (see figure 5).

\begin{figure}[tbh]
\centerline{\psfig{figure=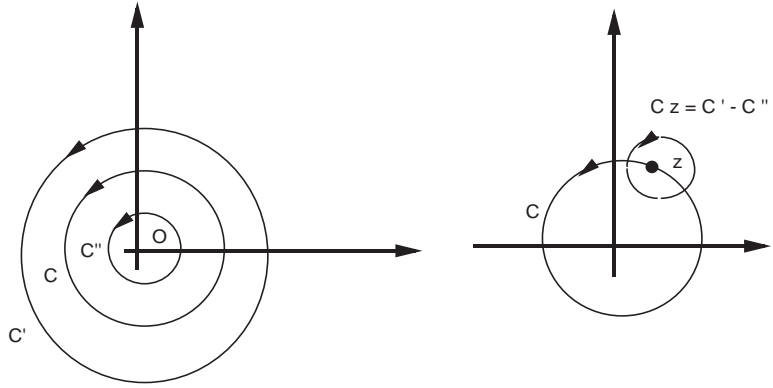,height=2.in,width=4.in}}
\caption{The contour manipulations that lead to the definition
of a commutator in radial quantization.}
\end{figure}

 Indeed, imagine writing the
OPE of the integrand. First we expand $T(w)O(0)$ to extract the field $L_mO$, 
on which the action of $L_n$ is then obtained by the second integration. That the 
countour $C$ is inside $C'$ is natural from the point of view of ``radial quantization'' which,
as we will see later, gives a precise operatorial definition to the $L_n$'s.
It is also necessary if one wishes to use the OPE in the order we just said
for convergence reasons. The product $L_mL_nO$ is computed in the same fashion 
with this time with a contour $C''$ inside $C$: this forces one  to expand first the product $T(z)O$,
resulting in the opposite order for the operators. Comparing the two, and forgetting
the operator $O$ itself, we see that
\begeq
[L_n,L_m]=\int_C{dz\over 2i\pi}\int_{C_z}{dw\over 2i\pi} z^{n+1}w^{m+1}T(z)T(w)
\endeq
and a computation using the OPE of $T$ with itself gives
\begeq
[L_n,L_m]=(n-m)L_{n+m}+{c\over 12}(n^3-n)\delta_{n+m}
\endeq
This is the celebrated Virasoro algebra, the $L_n$ being called Virasoro 
generators. It is an infinite dimensional Lie algebra. 

\smallskip
\noindent {\bf Exercise}: compute the action of the Virasoro generators of say derivatives of the 
field $\phi$, and directly check the Virasoro commutation relations.
\smallskip 

A very important use of this algebra is to provide one with a natural 
structure
to organize and recognize the fields in a theory. Of course, one does not quite
need this powerful tool for the free boson, whose fields are 
easily built ``by hand'', but for more complex theories, this is really
very useful. Given a lattice model with microscopic variables, 
arbitrary combinations of neighbour variables can be built, whose scaling limit
may or may not give rise to new scaling fields:
 which are truly new, which are nothing 
but ``$L_n$'''s (descendents) of others? What happens is that a theory
has a certain number of primary fields, which is very often finite (eg, three
for the Ising model), and all the other fields are just descendents of these ones. 
The whole set of fields is thus organized into
products of representations of the left and right Virasoro algebras, for which 
the primary fields are heighest weight states. This can be expressed by the compact form
\begeq
{\cal H}=\sum_{h,\bar{h}} Vir_h\otimes \overline{Vir}_{\bar{h}}
\endeq
The situation
is quite similar to the case of angular momentum in ordinary quantum mechanics, where
the space of say the possible electronic states of some atom can be 
organized in terms of representations of the angular momentum algebra. Of course,
here we have an algebra with an infinite number of generators, instead of three
for angular momentum in three space dimensions. Qualitatively,
there is an infinite number of Virasoro generators because there are an infinite
number of elementary conformal transformations, one
for each power of $z$: $z^n$.

As in the theory of angular momentum, unitarity contrains the 
quantum numbers, that is the values of conformal weights for a given 
central charge. This in turn gives rise  to strong constraints 
for multipoint correlation functions; this is beyond the scope of 
these lectures, but not by far. In particular, correlations at strong
coupling fixed points in the Kondo model can be computed just by using that 
technique \cite{ALgreens}. 

The reason why we focused on the algebra of the $L_n$'s is because of the 
special
role of the stress energy tensor in conformal transformations. Of course, we 
could define other operators and other algebras associated with any field that
 has 
an integer dimension (so the contour integrals can be closed in the complex 
plane.
 Generalizations also occur for fields with non integer, rational dimensions,
and a cut plane, but this is more complicated). A natural candidate in
 condensed matter is provided by currents,
for instance $\partial\phi$. Set therefore
\begeq
i\partial\phi={1\over\sqrt{4\pi}}\sum_n {\alpha_n\over z^{n+1}},
\ i\partialbar\phibar={1\over\sqrt{4\pi}}\sum_n {\bar{\alpha}_n
\over \zbar^{n+1}}\label{currentmodeexp}
\endeq
From 
\begeq
\left< \partial\phi(z)\partial\phi(w)\right> ={1\over 4\pi}{1\over (z-w)^2},
\endeq
it follows that
\begeq
[\alpha_n,\alpha_m]=n\delta_{m+n}
\endeq
Here we recognize the oscillator algebra standard in the quantization of the free boson,
and maybe it is time to discuss more what the mode expansion
has to do with hamiltonian quantization.

\subsection{Cylinders}

I shall mostly discuss what happens in the case of
the cylinder.
 The key idea here is to remember  that path integrals
are scalar products of states. 
If we insert a field $O$ at $y=-\infty$ on the cylinder, this corresponds to
having prepared the system in a state (an ``in'' state) $\left|O\right>$, to which
 is associated, in
 the
field representation a wave function $\Psi_O[\Phi_{S^1}]$, result
of a partial path integration 
\begeq
\Psi_O[\Phi_{S^1}]=\int_{\Phi_{S^1}\ fixed} [d\Phi_\Omega] O(-\infty)\ e^{-S_\Omega}
\endeq
where the integral is taken over all configurations of the 
field in the bottom part of the cylinder $\Omega=(-\infty,0]\times S^1$,
 the values at the boundary $S^1$ 
 being held
to $\Phi_{S^1}$ \footnote{I am not being too careful here about what
happens at $-\infty$.}. Similarly if we insert a field $O'$ at $\infty$, this corresponds
to projecting the system on an ``out'' state $\left|O'\right>$, to which 
is associated
a wave fucntion $\Psi_{O'}(\Phi_{S^1})$. The scalar product of
 these two states is then
\begeq
\left< O|O'\right>=\int [d\Phi_{S^1}] \Psi_{O'}^*(\Phi_{S^1})\Psi_O(\Phi_{S^1}),\label{fieldrepcor}
\endeq
and this is essentially the correlation function of the two fields $O,O'$. 
Of course, by translation invariance on the cylinder, this does not depend on the 
particular place where we have cut open the path integral.

To make things concrete, let us discuss an example we will use explicitly later,
with $O=O'=I$, the identity operator - ie, nothing is 
actually inserted at $\pm \infty$. In this case, (\ref{fieldrepcor}) is just the partition function $Z$
of the problem.
To find the wave functions, let us split open the path integral
at $y=0$, and let us Fourier decompose 
\begeq
\Phi_{S^1}(x)=\sum_n \Phi_n e^{i\omega_nx},
\endeq
where $\omega_n=2\pi nT$. Introduce then the solution of the Laplace equation $\Delta\Phi_0=0$
subject to the constraint $\Phi_0(x,y=0)=\Phi(x)$. One finds easily
\begeq
\Phi_0(x,y)=\sum_n \Phi_n e^{-|\omega_n y|}e^{i\omega_nx}
\endeq
We can now split the field in the path integral into $\Phi=\Phi_0+\Phi'$
where $\Phi'$ vanishes at $y=0$. Because of this, together with the
 fact that $\Phi_0$ solves Laplace equation, integration by parts shows that
the path integral factorizes into the partition fucntions of two half cylinders
with Dirichlet boundary conditions (we will get back to these later; the point
 here is that they are independent
of $\Phi_{S^1}$), and an interesting term
\begeq
Z=Z_D^2 \int [d\Phi_{S^1}] \exp\left(-{1\over T}\sum_n |\omega_n| |\Phi_n|^2\right)
\label{parti}
\endeq
Comparing with (\ref{fieldrepcor}), it follows that, up to a phase
\begeq
\Psi_I\left[\Phi_{S^1}\right]\propto \exp\left[-{\pi T^2\over 2} \int_{S^1} \left({\Phi(x)-\Phi(x')
\over\sin \pi T(x-x')}\right)^2\right]
\endeq
where we Fourier transformed back the integrand in  (\ref{parti}). 

We notice here as a side remark that the issue of finding wave functions 
is more than formal: the computation above was carried out for instance by people
interested in finding the wave function of the Thirring model in terms of the
original fermions \cite{Stone}.

In this point of view, we have a Hilbert space made up of states
in one to one correspondence with the various fields of the theory. For operators
other than the identity, one has to be a little bit careful. While the in state
is always obtained by inserting $O$ at $-\infty$, the out state is obtained 
actually by inserting $\left(e^{-2i\pi Tw}\right)^{2h}
\left(e^{-2i\pi T\wbar}\right)^{2\bar{h}} O$ at $+\infty$.

The same 
analysis can be carried out in the plane in the framework of ``radial
quantization'', where time is $\ln|z|$. This is why the expansions we used earlier were called 
OPE. To be correct however, they do have a meaning as OPE's only when the
operators are radially ordered, since recall that, to an Euclidian
Green function computed with a path integral, there corresponds a {\sl time
ordered} Green function in the quantum field theory. Of course, other hamiltonian
descriptions (for instance, the standard one where
imaginary time runs say along $x_2$) 
could be obtained by splitting open the path integrals differently.

The remarkable thing now is, that the  hamiltonian on the cylinder
has a very simple spectrum. Indeed, first observe that 
for primary operators, using the formula (\ref{twopointconf}), the two point function
on the cylinder decreases at large distance along the cylinder as 
\begeq
\left< O(w)O(w')\right>\approx \exp\left[-{2\pi T}(h+\bar{h})(y-y')\right]
\exp\left[-2i\pi T(h-\bar{h})(x-x')\right]\label{decay}
\endeq
For non primary operators, the same can be shown to hold, the more complicated
terms in the covariance formula decreasing more quickly. 

On the other hand, suppose we want to describe the theory on the 
cylinder in a hamiltonian formalism with imaginary time along the $y$ axis. As
is well known, the rate of decay of correlation functions is given by the gaps 
of the hamiltonian, and the decay (\ref{decay}) indicates that there
is an eigenstate of the hamiltonian whose eigenenergy is $2\pi T(h+\bar{h})$
over the ground state. More generally, the computation of any physical
property of the theory boils down to evaluating correlators, which all 
obey (\ref{decay}): therefore, the whole space of the quantum field theory
must be organized in states associated with the various observables, such that
their eigenenergy is $2\pi T(h+\bar{h})$
over the ground state! This is exactly what we expected from the 
hamiltonian formalism described before, with one 
additional piece of information: the spectrum of $H$. 

In addition, it is important to stress that the stress tensor acquires a non
vanishing expectation value on the cylinder, due to the schwartzian derivative
(\ref{tcylin}) . As a result,
the hamiltonian on the cylinder reads
\begeq
H=2\pi T\left(L_0+\bar{L}_0-{c\over 12}\right)\label{cylindhamil}
\endeq
and the momentum
\begeq
P=2\pi T \left(L_0-\bar{L}_0\right)
\endeq
where $L_0$ has eigenvalues $h$, $\bar{L}_0$ eigenvalues $\bar{h}$. 

We can of course define the whole set of Virasoro generators on the cylinder
by
\begeq
T(w)=(2\pi T)^2\left(\sum_n L_n e^{-2i\pi nT w}
-{c\over 12}\right)
\endeq
 Now we have a precise meaning to give the $L_n$
as operators, and their commutator can be computed, of course giving rise to the
same Virasoro algebra derived more intuitively before. Notice the identity
\begeq
H={1\over 2\pi}\int_0^{1/T} (T+\bar{T})dx\label{hperio}
\endeq
This is independent of $x$, a result of analyticity.

It should be clear that the whole conformal invariance analysis could be
written within the hamiltonian formulation. For instance, the OPE of $T$
with itself corresponds to the commutator
\begeq
{1\over 2i\pi}\left[T(x),T(x')\right]=\delta(x-x')T'(x)-2\delta'(x-x')T(x)+{c\over 6}\delta'''(x-x'),
\endeq
and the OPE of $T$ with a primary field
\begeq
{i\over 2\pi}\left[T(x),O(x')\right]=\delta(x-x')\partial_x O-\delta'(x-x')hO
\endeq

\subsection{The free boson via hamiltonians}

We now discuss the hamiltonian formalism more specifically
 for the free boson. The Lagrangian
is 
\begeq
L={1\over 2}\int_0^{1/T} dx\left[\left(\partial_t\Phi\right)^2-\left
(\partial_x\Phi\right)^2\right]
\endeq
from which the momentum follows 
\begeq
\Pi=\partial_t\Phi
\endeq
and the standard hamiltonian
\begeq
H={1\over 2}\int_0^{1/T} \left[\Pi^2+\left(\partial_x\Phi\right)^2\right]\label{stanham}
\endeq
with the canonical equal time commutation relations
\begeq
\left[\Phi(x,t),\Pi(x',t)\right]=i\delta(x-x')
\endeq
The field is periodic in the space direction, that is $\Phi(x,t)\equiv
\Phi(x+1/T,t)$. We chose to compactifiy the field
on a circle of radius $r$, that is we identify $\Phi\equiv\Phi+2\pi r$. 
The mode expansion of the field reads then (see eg \cite{GSW} for many more details on this)
\begeq
\Phi(x,t)=\xhat+T\phat t+2\pi T rw x+{i\over\sqrt{4\pi}}\sum_{n\neq 0}
{1\over n}\left(\alpha_n e^{2i\pi T n(x-t)}-\bar{\alpha}_{-n}
e^{2i\pi Tn(x+t)}\right),
\endeq
where  $\xhat=T\int \Phi(x,t)dx=\Phi_0$ is the boson zero mode,
while $\phat=\int \Pi(x,t)dx=\Pi_0$ is the total momentum. $w$ in an
 integer (the winding number); $\phat$ is 
quantized such that $r\phat=k$ is also an integer. The commutation relations
of the operators are $[\xhat,\phat]=i$ and
\begeq
\left[\alpha_n,\alpha_m\right]=
\left[\bar{\alpha}_n,
\bar{\alpha}_m\right]=n\delta_{n+m},\ 
\left[{\alpha}_n,\bar{\alpha}_m\right]=0
\endeq
The operators $\alpha_n,\bar{\alpha}_n$ are related with the
usual creation and annihilation operators of the free boson harmonic oscillators
by
\begeq
\alpha_n=-i\sqrt{n}a_n,n>0,\ \alpha_n=i\sqrt{-n}a_{-n}^\dagger, n<0
\endeq
and
\begeq
\bar{\alpha}_n=-i\sqrt{n}\abar_{-n},n>0,\ \bar{\alpha}_n
=i\sqrt{-n}\abar_{n}^\dagger, n<0
\endeq
Note that if we go to euclidian space time, replacing $t$ by $-iy$
and then use the conformal coordinates $z=e^{2\pi T(y-ix)}$ ($z=e^{-2i\pi Tw}$,
 $\zbar=
e^{2\pi T(y+ix)}$), we obtain the expansion
\begeq
\Phi(z,\zbar)=\Phi_0-i\left({\phat\over 4\pi}+{wr\over 2}\right)\ln z
-i\left({\phat\over 4\pi}-{wr\over 2}\right)\ln \zbar+{i\over \sqrt{4\pi}}
\sum_{n\neq 0}{1\over n}\left(\alpha_n z^{-n}+\bar{\alpha}_n \zbar^{-n}\right)
\endeq
When the winding number is 
non zero, the field is not periodic around the origin; rather, a ``vortex''
is inserted there. When $w=0$, we can set
 $\alpha_0=\bar{\alpha}_0={\phat\over \sqrt{4\pi}}$,
one recovers the expansion (\ref{currentmodeexp}) for $i\partial\phi$. In 
general, we will set
\begeq
\alpha_0={\phat\over \sqrt{4\pi}}+wr\sqrt{\pi},\ \bar{\alpha}_0=
{\phat\over \sqrt{4\pi}}-wr\sqrt{\pi}
\endeq
The hamiltonian (\ref{stanham}) reads , before regularization
\begeq
H=2\pi T\left[\pi(wr)^2+{\phat^2\over 4\pi}+{1\over 2}\sum_{n\neq 0} \left(
 \alpha_{-n}\alpha_n
+  \bar{\alpha}_{-n}\bar{\alpha}_n\right)\right]
\endeq

A question that arises now is the relation between the normal ordering
defined in the field theory  and the normal ordering in the usual sense
of ordering free bosonic operators in quadratic expressions:
$:\alpha_n\alpha_m:=\alpha_{inf(n,m)}\alpha_{sup(n,m)}$. 
The two {\sl might} differ by a constant; in the present  case actually,
they coincide provided one uses zeta regularization.
Indeed, by ordering $H$,  we encounter a divergent 
term $\sum n$, which we can regularize by (for results on the zeta function, see \cite{zeta})
\begeq
\sum_1^\infty  n=\zeta(-1)=-{1\over 12}
\endeq
With this prescription, the hamiltonian with the vacuum energy divergence
subtracted  reads as it should (\ref{cylindhamil}), with 
the Virasoro generators 
\begeq
L_n={1\over 2}\sum_m \alpha_{n-m}\alpha_m,\ 
\bar{L}_n={1\over 2}\sum_m \bar{\alpha}_{n-m}\bar{\alpha}_m
\endeq
together with
\begeq
L_0={1\over 2}\alpha_0^2+\sum_{n=1}^\infty \alpha_{-n}\alpha_n,\ 
\bar{L}_0={1\over 2}\bar{\alpha}_0^2+\sum_{n=1}^\infty \bar{\alpha}_{-n}
\bar{\alpha}_n
\endeq
The modes $\alpha_n$ and $\bar{\alpha}_n$ are annihilation operators for $n>0$
and creation operators for $n<0$. The whole space of fields is thus obtained
starting from highest weight states $\left|w,k\right>$ which are annihilated
by the annihilation operators and are eigenstates of the zero modes, and applying
creation operators to them. Schematically, one has
\begeq
{\cal H}=\sum_{w,k} Heis_{w,k}\otimes \overline{Heis}_{w,k}
\endeq
Of course, the $\left|w,k\right>$ states are primary, and thus
highest weight of the Virasoro algebra. Accordingly, one could as well
build the whole space of fields by acting on them with $L_n's$. This would be more complicated;
for instance, the field $\partial\phi$, which is simply the result of $\alpha_{-1}\left|0,0\right>$,
is not obtained from the action of $L_{-1}$ on that state at all. This means in general
that more primary fields are necessary than the $\left|w,k\right>$ in the Virasoro description.

\subsection{Modular invariance}

A convenient way of encoding the field content of the theory is
to write the torus partition function, that is, the partition function
when one imposes periodic boundary conditions in the imaginary time
direction, too. One has, using (\ref{cylindhamil})
\begeq
Z=\hbox{Tr }\exp\left[-{2\pi TL}\left(L_0+\bar{L}_0-{c\over 12}\right)\right]
\endeq
Using the mode decomposition, one finds easily
\begeq
Z={1\over \eta(q)\bar{\eta} (\bar{q})}\sum_{w,k} q^{
h_{wk}}\bar{q}^{\bar{h}_{wk}}\label{partfct}
\endeq
where $q=e^{-2\pi TL}=\bar{q}$ (the notation allows consideration
of more complicated parallelograms),
\begeq
\eta(q)=q^{1\over 24}\prod_{n=1}^\infty (1-q^n)
\endeq
and 
\begeq
h_{wk}=2\pi\left({k\over 4\pi r}+{wr\over 2}\right)^2,\ \bar{h}_{wk}= 
2\pi\left({k\over 4\pi r}-{wr\over 2}\right)^2
\endeq

An important property of this partition function is that it is {\sl modular invariant}. 
What this means is, suppose one considers quantization of the free boson
with time in the $x$ instead of $y$ direction. The radius being the same, 
this will lead to the same expression  as (\ref{partfct}) but with $L$ and $1/T$ 
exchanged, that is
\begeq
Z={1\over \eta(q')\bar{\eta} (\bar{q}')}\sum_{w,k} \left(q'\right)^{
h_{wk}}\left(\bar{q}'\right)^{\bar{h}_{wk}}\label{partfctmod}
\endeq
where $q'=e^{-2\pi/TL}$. The expressions (\ref{partfct})
and (\ref{partfctmod}) do turn out to be equal thanks to some
elliptic functions identities (see next section). They ought to be, of course,
since they represent the same physical object from two different points of view. 

For more sophisticated theories, the partition function cannot be computed 
a priori, but it is possible to determine it by imposing that it does not
depend on the description, ie is {\sl  modular invariant}. See \cite{philippe} and references 
therein for more 
details.

\section{Conformal invariance analysis of quantum impurity fixed points}

\subsection{Boundary conformal field theory}

An excellent reference for this  part is the original work of J. Cardy (\cite{johnbdr}).
 Consider now a field theory defined only on the half plane $x_2>0$ (figure 6) - it might 
be for instance the continuum limit of a 2D statistical mechanics model which is
at its critical point in the bulk, that is $T=T_c$, the usual 
 critical temperature of the system. Various situations could occur at the
boundary depending on whether the coupling there is enhanced,  or whether 
some quantum boundary degrees of freedom have been added.

\begin{figure}[tbh]
\centerline{\psfig{figure=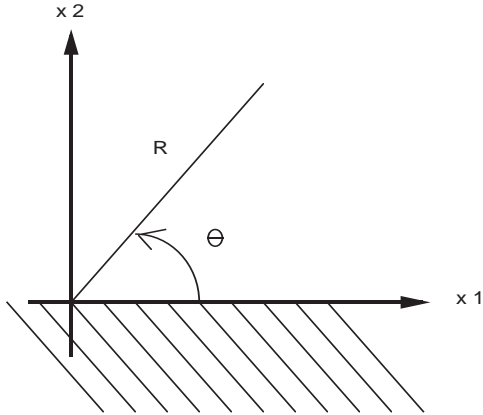,height=2.15in,width=2.5in}}
\caption{The geometry for boundary conformal field theory.}
\end{figure}

 Consider,
to fix ideas, the simplest case where 
the statistical mechanics model would have the same couplings in the
 bulk and the boundary (
the so called ``ordinary transition''). Intuitively, one expects the 
system to still be invariant under global rotations, dilations and translations that preserve
the boundary, and that this invariance should be promoted to a local one, ie conformal invariance
in the presence of the boundary. 

Physical fields are now characterized both by a bulk and a boundary anomalous
dimension. If both fields are taken deep inside the system, they behave 
as in the bulk case. On the other hand, if they are near the boundary,
one has, for example,
\begeq
\left< O(x_1,x_2)O(x'_1,x'_2)\right>\approx {1\over |x_1-x'_1|^{2d_s}},~~~ |x_1-x'_1|>>x_2,x_2'
\endeq
ie the large  distance behaviour of the correlators 
parallel to the surface is determined by the boundary dimension.  We quote also the formula
\begeq
\left< O(x_1,x_2)O(x'_1,x_2')\right>\approx {1\over R^{d+d_s}}
\left(\cos\theta\right)^{d_s-d}
\endeq

A condition of boundary conformal invariance is that  $T_{12}=T_{21}=0$ when $x_2=0$,
which means physically that there is no flux of energy through the boundary. As a result,
the left and right components of the stress tensor are not independent anymore, but 
$T=\bar{T}$ for $Im z=0$; this is expected, since the theory is invariant only under the transformations
that preserve this boundary, that is satisfy $w=\bar{w}$ for $Im z=0$. As a result however,
one can define  formally the stress tensor in the region $Im z<0$ by setting
\begeq
T(z)=\bar{T}(z),\ Im z<0\label{conti}
\endeq
Instead of having a half plane with left and right movers, we can thus equivalently 
describe the problem with only right movers on the full plane.For instance, the two point 
correlation function in the half plane is related with the four point
correlation function in the full plane.  Also radial quantization
corresponds to propagating outwards from the origin in the upper half plane, with 
hamiltonian (see 
figure 7)
\begeq
\int_C T(z)dz+ cc
\endeq

\begin{figure}[tbh]
\centerline{\psfig{figure=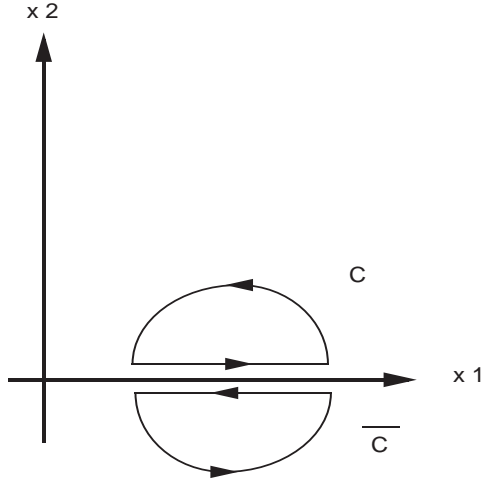,height=2.5in,width=2.5in}}
\caption{Geometry of the contours for  the boundary case.}
\end{figure}

Using the continuation (\ref{conti}), this becomes a closed contour integral of $T$ only:
thus, the Hilbert space of the theory with boundary is described by a sum of representations
of a single Virasoro algebra this time:
\begeq
{\cal H}=\sum_h Vir_h\label{hilbertopen}
\endeq

The natural mapping in this problem is $w=-{L\over i\pi} \ln z$, which maps
the half plane onto a strip of width $L$ \footnote{Notice that here I have put $L$ in the mapping,
instead of $T$. In the periodic case, we could have used the mapping $w=-{L\over 2i\pi}\ln z$
to produce a similar result. This is all equivalent, but I prefer the present choice, where $1/T$
is always the periodic direction in the problem.} with the same boundary conditions
on both sides \footnote{A strip with different boundary conditions on either side
would correspond to a half plane with different boundary conditions $x<0$ and $x>0$,
with a ``boundary conditions changing operator'' inserted right at $x=0$.}. 
The hamiltonian now reads
\begeq
H={\pi\over L}\left(L_0-{c\over 24}\right)\label{hopen}
\endeq
Note that there are, roughly, two factors of two differing from the periodic
hamiltonian: the prefactor has a $\pi$ instead of $2\pi$, and there is 
a single Virasoro generator in the bracket. The space onto which the
periodic hamiltonian (\ref{cylindhamil}) acts is uniquely defined
by the (bulk) theory one is dealing with, say the Ising model - as we discussed, this specification
amounts to giving the various representations of $Vir\ \otimes\ \overline{Vir}$
defining the model. For (\ref{hopen}), the space depends on the boundary conditions;
it is specified by a set of representations of a single Virasoro algebra. By a careful study of the 
two point function in the plane, together with the conformal transformation (where the jacobians
still involve the bulk dimension), one can show that the gaps of $H$ are given by the 
corresponding surface dimensions. 

It is important to stress again that the same physical obervable will be associated with 
different representations of the Virasoro algebra in the bulk and boundary cases. For instance,
the spin in the Ising model coresponds, in the bulk, to $Vir_{1/16}\otimes\overline{Vir}_{1/16}$,
while with free boundary conditions, it corresponds to $Vir_{1/2}$ (with fixed boundary conditions,
the spin is the same as the identiy operator).

\subsection{Partition functions and boundary states}

To classify boundary conditions, it is extremely useful to 
deal with partition functions a bit. We consider thus a cylinder
with a periodic direction of length $1/T$ and a non periodic one of length
$L$: on either side, boundary conditions of type $a,b$ have been imposed. 
We can describe the situation in two ways (see figure 8): either imaginary time
runs in the direction parallel to the boundary (``open channel''),
in which case we can write the partition function as 
\begeq
Z= Tr e^{-H_{ab}/T}
\endeq
where $H_{ab}$ is the 
hamiltonian  (\ref{hopen}) with boundary conditions $a$ and $b$, or imaginary time
can run in the direction perpendicular to the boundary (``closed channel''), in which
case
\begeq
Z=\left< B_a|e^{-LH}|B_b\right>\label{partbdr}
\endeq
where $\left|B_a\right>,\left|B_b\right>$ are boundary states, and $H$ is the 
periodic hamiltonian (\ref{cylindhamil}).

\begin{figure}[tbh]
\centerline{\psfig{figure=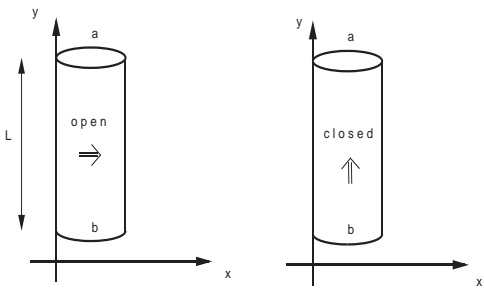,height=1.5in,width=2.5in}}
\caption{The open and closed channel geometries when boundaries are present.}
\end{figure}

Observe that the the boundary states
are not normalized: they  are entirely determined, including their norm, 
by the condition
that (\ref{partbdr}) gives the right partition function. To make things more concrete,
fixed boundary conditions in the Ising model for instance are represented, in the 
microscopic Hilbert space, by the state $\left|B\right>_{fixed}=\prod_i \left|+\right>$, 
while for free boundary conditions one has  $\left|B\right>_{free}=\prod_i \left(
\left|+\right>+\left|-\right>\right)$.

Here the boundary states are states in the Hilbert space of the bulk theory, ie
in $Vir\ \otimes\ \overline{Vir}$. Conformal invariance at the 
boundary requires
\begeq
\left(L_n-\bar{L}_{-n}\right)\left|B\right>=0 \label{cfbdr}
\endeq
A solution to this 
equation is provided by so called Ishibashi states \cite{Ishibashi} 
\begeq
\left|h\right>=\sum_n \left|h,n\right>\otimes \left|\overline{h,n}\right>
\endeq
where $\left|h,n\right>$ denotes an orthonormal basis of the representation $Vir_h$,
and $\left|\overline{h,n}\right>$ the corresponding basis of $\overline{Vir}_h$. 

In the case of the free boson, a boundary state will satisfy (\ref{cfbdr})
if it satisfies a stronger constraint
\begeq
\left(\alpha_n\pm \bar{\alpha}_{-m}\right)\left|B\right>=0\label{modecond}
\endeq
This in fact corresponds to Neumann and Dirichlet boundary conditions,
for which $T_{12}\propto \partial_1\Phi\partial_2\Phi=0$.  The
negative sign in (\ref{modecond}) is solved by
\begeq
\left|B\right>\propto \exp\left[-\sum_{n=1}^\infty 
{\alpha_{-n}\bar{\alpha}_{-n}\over n}
\right]
\left|0,k\right>
\endeq
Therefore, we can build boundary states by
\begeq
\sum_{k} c_k \exp\left[-\sum_{n=1}^\infty {\alpha_{-n}\bar{\alpha}_{-n}
\over n}\right]
\left|0,k\right>
\endeq
The question of interest is to determine the coefficients
$c_k$. A quick way to proceed \footnote{This topic  goes back to the 
early days of open string theory. A nice recent paper on the subject is \cite{IanMasaki},
where the following computations are carried out in many more details.}
is to recognize here a Dirichlet state: indeed, suppose we act with $\Phi(x,t=0)$
on the boundary state. Because of the condition (\ref{modecond}),
the oscillator part just does not contribute;
what does contribute is only the $\xhat$ part, which acts as
$\xhat=i{\partial\over\partial p}$. Therefore, we have
\begeq
\left|B_D(\Phi_0)\right>= {\cal N}_D \sum_{k=-\infty}^\infty e^{-ik\Phi_0/r} 
\exp\left[-\sum_{n=1}^\infty {\alpha_{-n}\bar{\alpha}_{-n}\over n}\right]\left|0,k\right>
\endeq
The last question, which is actually of key
importance for what follows, is the determination of the overall factor 
${\cal N}$:
in other words, what is the overall normalization of boundary states? This is 
where the consideration of partition functions is useful.

To answer this, we observe that, if we compute the partition function
with height $\Phi_0$ on both sides, the identity representation
should appear once and only once. On the other hand,
the partition function is easily computed in the 
closed channel from the boundary states: one finds, for more general
pair of values at the boundary
\begeq
Z=\left< B_D(\Phi_0)|e^{-LH}|B_D(\Phi'_0)\right>={\cal N}_D^2 {1\over \eta(\tilde{q})}
\sum_{k=-\infty}^\infty \tilde{q}^{k^2/8\pi r^2}e^{ki(\Phi_0-\Phi'_0)/r}
\endeq
where $\tilde{q}=e^{-4\pi TL}$. We now perform a modular transformation
to reexpress this partition function in terms of the other parameter 
$q=e^{-\pi/LT}$. One has (the proof of this is a bit intricate. See eg \cite{Apostol},
chapter 3.)
\begeq
\eta(\tilde{q})={1\over\sqrt{2TL}}\eta(q)
\endeq
and, 
by using Poisson resummation formula for the infinite sum, 
\begeq
\sum_{n} \exp\left(-\pi an^2+bn\right)={1\over \sqrt{a}}
\sum_k\exp-{\pi\over a}\left(k+{b\over 2i\pi}\right)^2
\endeq
 one finds
\begeq
Z_{DD}=\sqrt{\pi} 2r{{\cal N}_D^2\over \eta(q)}
\sum_n q^{{1\over 2\pi}\left(\Phi_0-\Phi'_0+2\pi nr\right)^2}\label{dirich}
\endeq
This expression has a simple interpretation: one sums over all the 
sectors where the difference of heights between the two sides
of the cylinder is $\Phi_0-\Phi'_0+2\pi rn$. For each such sector,
the partition function is the product of a basic partition function
corresponding to heights equal (without the $2\pi r$ identification)
on both sides, times the exponential of a classical action. The latter
is easily obtained: the classical field is $\Phi={\Phi_0-\Phi'_0+2\pi nr \over L}
y$, whose classical action is 
$$
\exp\left[-{1\over 2LT} (\Phi_0-\Phi'_0+2\pi nr)^2\right]
$$
Consider now (\ref{dirich}). We know that the partition function must write as a sum of 
characters (that is, $Tr_{Vir_h} q^{L_0-c/24}$, as follows from (\ref{hilbertopen}) and 
(\ref{hopen})) 
of the Virasoro algebra with integer coefficients; even though I will not spend time 
discussing what the characters at $c=1$ are ($ q^h/\eta$ for generic $h$),
it is easy to see that this implies that the prefactor in (\ref{dirich}) has to be an integer. Since 
we do not expect the normalization of the boundary states to change discontinuously
with $\Phi_0$, this integer is actually a constant, whatever $\Phi_0,\Phi_0'$. We can in particular 
choose $\Phi_0=\Phi_0'$, for which the identity representation $Vir_{h=0}$ appears in
the spectrum; of course it should appear only once, and therefore
\begeq
{\cal N}_D={1\over \sqrt{2r\sqrt{\pi}}}
\endeq

The other condition corresponds to Neumann boundary conditions, 
or, equivalently, Dirichlet boundary conditions on the dual field 
$\tilde{\Phi}=\tilde{\Phi}_0$. 
One finds the boundary state
\begeq
\left|B_N(\tilde{\Phi}_0)\right>={\cal N}_N\sum_{w=-\infty}^\infty
e^{-2i\pi rw\tilde{\Phi}_0}\exp\left[\sum_{n=1}^\infty {\alpha_{-n}
\bar{\alpha}_{-n}\over n}\right]
\left|w,0\right>
\endeq
The Neumann Neumann partition function reads then
\begeq
Z_{NN}={1\over \eta(q)}\sum_n 
q^{{1\over 2\pi}\left(\tilde{\Phi}_0-\tilde{\Phi'}_0+n/r\right)^2},
\endeq
and one has
\begeq
{\cal N}_N={1\over 2}\sqrt{2r\sqrt{\pi}}
\endeq

The Neumann Dirichlet partition function is actually independent
of the values of $\Phi_0,\tilde{\Phi}_0$, since then
the field cannot wind in any direction. 

\smallskip
\noindent{\bf Exercise}: show the following 
\begeq
Z_{ND}={1\over 2\eta(q)}\sum_n q^{{1\over 4}(n-1/2)^2}.
\endeq

The consideration of  boundary states is extremely 
powerful to find out and study boundary fixed points. A general strategy is, knowing 
the Virasoro algebra symmetry of the model at hand, to try to find out 
combinations of Ishibashi states that are acceptable boundary states. Solving this 
problem involves rather complicated constraints. For instance, if one has several possible candidates 
$\left|B_i\right>$, the partition function with boundary conditions $ij$ can easily 
be evaluated in the closed channel; after modular transformation to the open channel,
it should expand as a sum of characters of the Virasoro algebra with {\sl integer coefficients}.
Another constraint is that the identity representation should appear at most once in
all open channel partition functions.
Clearly, this becomes  a rather technical subject; more details can be found in 
the paper of J. Cardy \cite{Johnbdrst}. Questions like the completeness of boundary states (ie 
whether {\sl all} the boundary fixed points of a given bulk problem are known) are still
open in most cases.

\subsection{Boundary entropy}

Let us now suppose that we have a one dimensional 
quantum field theory defined on a segment of length $L$, with some boundary conditions
at $x=0$ and $x=L$. As is well known, the partition function at temperature $T$
of this theory will be given by the same expression as the partition function
of the two dimensional systems considered previously; notice however that I have
changed conventions calling now $x$ (resp. $y$) what was $y$ (resp. $x$) 
previously (see figure 9) \footnote{This is to match as much as possible with the literature; in any case, there is 
no perfect notation that would be convenient all the way through.}.
\begin{figure}[tbh]
\centerline{\psfig{figure=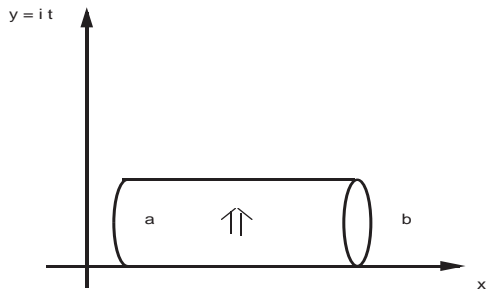,height=1.5in,width=2.5in}}
\caption{The geometry for defining boundary entropy.}
\end{figure}

 In terms 
of the parameter $q$, this partition function is expressed as a sum
of terms ${q^{h}\over \eta}$ with integer coefficients: 
the spectrum of $H_{ab}$ is discrete,
and its ground state has integer degeneracy, nothing very exciting.
 In the limit $L\to \infty$,
the spectrum becomes gapless however, and one has to be more careful about
the concept of degeneracy. If we take this limit, 
the free energy of the quantum field  theory behaves as
\begeq
F=-Lf+f_a+f_b
\endeq
where $f$ is a free energy per unit length, $f_a,f_b$ are boundary contributions. These
contributions will involve, as $T\to 0$, a boundary energy that is non universal,
but also a boundary entropy. 
It is easy to see what this entropy will be by using a modular transformation.
The same partition function expresses  then as  a sum of ${\tilde{q}^h\over \eta(\tilde{q})}$
with some non integer coefficients that come from the Poisson resummation formula
(in general, from the modular $S$ matrix). In the large $L$ limit, $\tilde{q}\to 0$.
From the fact that
\begeq
F=-T\ln Z
\endeq
we see that, as $T\to 0$, $f= O(T^2)$ (the
ground state energy 
of $H_{ab}$ is set to zero in this approach; for the exact dependence 
of $f$ on $T^2$ see section 6), while $f_a$ and $f_b$ 
are of the form 
\begeq
f_a=-T\ln g_a, f_b=-T\ln g_b
\endeq
where \cite{ALprldegeneracy}:
\begeq
g_a=\left<B_a|0\right>,\ g_b=\left<0|B_b\right>
\endeq
A one dimensional massless quantum field theory defined on a line with 
boundary conditions (or boundary degrees of freedom as we will see next) therefore has
a non trivial zero temperature boundary entropy, or ground state
degeneracy. 

\smallskip
\noindent{\bf Exercise}: Show
that the precise meaning of this degeneracy is related with the behaviour
of the density of states 
\begeq
D(n)\approx {g_ag_b\over 2} \left({c\over 6n^3}\right)^{1/4} \exp\left( 2\pi 
\sqrt{cn\over 6}\right)\label{precise}
\endeq
where we parametrized the excitation energies of $H_{ab}$ by 
$e_n={n\pi\over L}$, $n,L$ large (when computing the partition function
and its logarithm, do not forget to integrate the fluctuations around the saddle
point!).
\smallskip

As we have seen in the previous subsection, some boundary conditions
have a degeneracy $g<1$, ie a negative boundary entropy. This is a bit shocking,
but of course we should remember, first, that $g$ is more a prefactor 
in an asymptotic formula for degeneracies (\ref{precise}) 
than a true ground state degeneracy (at $L=\infty$, there is no gap),
and second,   that we are dealing with quantum field theories and that 
this is only a finite, properly  regularized ``entropy''. The same remark applies, somehow,
to $g$ being non integer. However, it is perfectly possible to have non integer
degeneracies for semi-classical systems involving kinks \cite{Paulkondo}. 

\vfill
\eject

\Large \noindent {\bf Intermezzo}

\noindent \huge {\bf Perturbation near the fixed points}

\normalsize

\bigskip
\bigskip

A scale-invariant boundary condition
is a RG fixed point
(recall that the bulk is always critical in
the type of systems that we are considering here).
As  with any RG fixed point, there
is a set of {\it relevant/marginal/irrelevant}
boundary operators (and couplings) associated with each
scale-invariant boundary condition.
These operators have support only  at the
boundary, i.e. at one point in position space (at
the  position of the impurity).

If  no relevant boundary operators
are allowed, then  the scale
invariant boundary condition represents a stable
fixed point (the zero temperature fixed point, describing
the Kondo model at strong coupling,  is an example; so is the Dirichlet fixed point
 in the tunneling
problem, to which we will get back soon).
Irrelevant boundary operators give
perturbatively calculable corrections  to physical properties
evaluated  at the RG  fixed point.  Many important
physical features of the Kondo model are actually
due to the effect of the leading (dominant)
irrelevant boundary operator \cite{ALgreens}.

Adding a {\sl relevant} boundary operator
to the Hamiltonian   describing a particular
scale-invariant boundary condition,  destroys that
boundary condition, and causes   crossover to
a new, scale-invariant boundary condition at large
distances  and low temperatures (in the infrared).
 In other words,  we have a (boundary)
 RG flow, describing the crossover
 from the initial  scale-invariant
 boundary  condition  (in the ultraviolet, i.e.
 at short distances or high temperatures) to a new scale
 invariant boundary condition (in the infrared, i.e.
 at large distances and low temperature).

Note that at every stage of
 this  flow, the bulk remains always critical and unchanged;
 the only action is at the boundary.
 An interesting observation concerning general boundary
 RG flows was made in \cite{ALprldegeneracy}:
the zero-temperature boundary entropies ($s=\ln g$ in the previous section) generally obey
$$
s_{UV} > s_{IR}, \qquad {\rm (decrease \ of \ boundary \ entropy)}
$$
This may be viewed as a boundary analogue of
the well known $c$-theorem of bulk conformal
field theory \cite{cthm}.  (Note, however, that
the universal numbers
$s_{UV}$ and $ s_{IR}$ do
 not seem to be obviously related
 to a dynamical  quantity, in contrast with the central charge, 
which is related to
the stress tensor of CFT).

A well known  example is the one-channel
Kondo model. Initially,
at weak coupling  (at high temperature, in the ultraviolet),
we have a quantum mechanical spin
 decoupled from the electron degrees
of freedom of the metal.  An isolated
($s=1/2$) spin has a zero-temperature
entropy of $s_{UV} = \ln 2$.
At strong coupling (at low temperature, in the infrared),
this impurity spin is completely
screened by the conduction
electrons. This means that  no
dynamical degrees of freedom are left,
and thus we have $s_{IR}= 0$.

\section{The boundary sine-Gordon model: general results}

\subsection{The model and the flow}

We consider now the model we had decided to tackle in the introduction
\begeq
S={1\over 2}\int_{-\infty}^0 dx\int dy \left[\left(\partial_x\Phi\right)^2
+ \left(\partial_y\Phi\right)^2\right]+\lambda\int dy\cos{\beta\over 2}\Phi(0,y).\label{bsg}
\endeq
This model is called the boundary sine-Gordon model since it has a sine-Gordon type interaction,
but at the boundary. In more general terms than those of the edge states tunneling, the physics of this model is rather clear. The limits $\lambda=0$ 
and $\lambda=\infty$ are fixed points, corresponding 
to conformal invariant boundary conditions, respectively of Neumann and Dirichlet 
types.  Away from these limits, the model
is not scale invariant because of the boundary interaction. In the vicinity of $\lambda=0$,
the RG equation is
\begeq
{d\lambda\over db}=\left(1-g\right)\lambda+O(\lambda^3),
\endeq
where we have set $g=\nu={\beta^2\over 8\pi}$. It is natural to expect that $\lambda$ 
flows all the way from $0$ to $\infty$ under
renormalization. Equivalently, the boundary conditions look like Neumann at very
 high energy (UV)
but like Dirichlet at low energy (IR) - the dimension of the physical coupling is 
$[\lambda]=L^{g-1}$, so the  typical energy scale for the cross over between UV and IR 
behaviours is  $T_B\propto \lambda^{1\over g-1}$. Equivalently also, 
 the field $\Phi$ feels
 Neumann boundary conditions
close to the boundary, but feels Dirichlet boundary conditions 
instead far from it, with a cross over distance $1/T_B$. 

Notice that the boundary entropies of the UV and IR fixed points are different. To compute 
them, we can use the results of the previous section after having identified
the radius of the boson. In the shift $\Phi\to\Phi+2\pi r$, the interaction 
$\cos{\beta\over 2}\Phi$
must be unchanged, which requires
\begeq
r={2\over\beta}
\endeq
It follows that
\begeq
g_N=\left({\beta^2\over 4\pi}\right)^{-1/4},\ g_D=\left({\beta^2\over 16\pi}\right)^{1/4} 
\endeq
Notice the ratio
\begeq
{g_N\over g_D}=\left({\beta^2\over 8\pi}\right)^{-1/2}
\endeq
For the case of a relevant perturbation we are considering here,
this ratio is larger than one: the boundary entropy is greater 
in the UV than it is in the IR. This is in agreement with the intuitive
idea that degrees of freedom disappear under the renormalization
group, leading to a loss of information. There is a well known conjecture
stating that for any allowed flow in a unitary system (that is, roughly,
a system with real, local hamiltonian), $g_{UV}>g_{IR}$. For the case of irrelevant
perturbation, one finds $g_N<g_D$, so according to this the flow should not
be possible, which is indeed the case: since the operator is irrelevant, it does not
generate any flow, and one should observe N boundary conditions both at
small and large distance.

\subsection{Perturbation near the UV fixed point}

The first question we will be interested in is  the calculation of the 
boundary free energy at any temperature $T$ and coupling $\lambda$. This can 
be represented by a Coulomb gas expansion as follows. First, by using  a conformal 
mapping, one finds the two point function of the free boson with Neumann boundary conditions
on the half cylinder
\begeq
\left< \Phi(y)\Phi(y')\right> = -g\ln\left|{\sin\pi T(y-y')\over \pi T}\right|
\endeq
\smallskip
\noindent{\bf Exercise}: derive this, by first computing the two point
function on the half plane.
\smallskip

We can then evaluate the ratio of partition functions with and without boundary
interaction as follows
\begeq
{Z(\lambda)\over Z(\lambda=0)}=1+\sum_{n=0}^\infty {1\over (2n)!}
\lambda^{2n} \int_0^{1/T} dy_1\ldots dy_{2n}
\left< \cos{\beta\over 2}\Phi(y_1)\ldots\cos{\beta\over 2}\Phi(y_{2n})\right>.
\endeq
Of course, only electrically neutral configurations with $n$ positive and $n$ 
negative charges contribute. After some rescaling, one finds 
\begeq
{Z(\lambda)\over Z(\lambda=0)}=1+\sum_{n=1}^\infty (\tilde{\lambda})^{2n} I_{2n},
\endeq
where the dimensionless coupling is 
\begeq
\tilde{\lambda}={\lambda\over 2T}(2\pi T)^g,
\endeq
and the integrals are
\begeq
I_{2n}={1\over (n!)^2} \int_0^{2\pi}du_1\ldots \int_0^{2\pi}du'_{2n}\left|
{\prod_{i<j} 4\sin{u_i-u_j\over 2}\sin{u'_i-u'_j\over 2}
\over\prod_{i,j}2\sin{u_i-u'_j\over 2}}\right|^{2g}.
\endeq
This is the partition function of a classical  Coulomb gas in two space dimensions,
with the charges moving on a circle of unit radius (see figure 10).

\begin{figure}[tbh]
\centerline{\psfig{figure=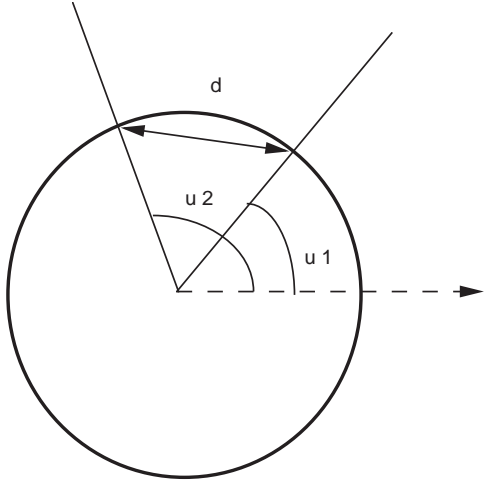,height=2.5in,width=2.5in}}
\caption{Charges of a two dimensional Coulomb gas that move on a circle $d=2\sin{u_2-u_1\over 2}$.}
\end{figure}

The integrands have small distance behaviours $1/u^{2g}$. It follows that there is no
short distance divergence, and the integrals
 are all
finite for $g<{1\over 2}$ (there are never large distance divergences
 here since we have a temperature). 
When $g>{1\over 2}$, the integrals have divergences. In the sequel, I will 
always regularize integrals dimensionally,  not by introducing a cut-off. To explain
 what this means,
consider the case $n=1$, which can be done by elementary computations
\begeq
I_2={\Gamma(1-2g)\over \Gamma^2(1-g)}.
\endeq
This can then be continued beyond $g={1\over 2}$ simply by using the known continuation
of $\Gamma$ to negative arguments. How to do this in the case of arbitrary $n$ is a bit more
tricky. A way to do it relies on the remarkable fact 
that the integrals $I_{2n}$ can be expressed in an almost closed
form by appealing to techniques of Jack polynomials \cite{FLeS,Jack}. I will only give the result here
\begeq
I_{2n}={1\over [\Gamma(g)]^{2n}}
\sum_m\prod_{i=1}^n \left({\Gamma[m_i+g(n-i+1)]\over
\Gamma[m_i+g(n-i)+1]}\right)^2
\endeq
where the sum is over all sets (Young tableaux) $m=(m_1,\ldots,m_n)$
with integers $m_i$ obeying $m_1\geq m_2\ldots \geq m_n\geq 0$. This expression can be used
to compute the $I_{2n}$ numerically to high values of $n$, or, more fundamentally,
to perform the analytic continuation in $g$. I will not discuss this further,
and get back now to the physics of this model. 

For $g<{1\over 2}$ at least, the perturbative expansion is well defined, giving a series
in $\lambda$ with positive coefficients.  This series will presumably have a finite radius 
of convergence - although one does not expect the appearance of a singularity on the 
positive real axis (this would correspond to the existence of a phase transition on the one dimensional
boundary). Beyond this radius, some other technique has to be used to 
understand quantitatively what happens. It is possible to argue what the 
leading behaviour of the partition function
at large $\lambda$ should be. Indeed, $Z$ actually depends only on the ratio 
of the two energy scales $T_B$ and $T$, so large $\lambda$ is like
small temperature. But small temperature corresponds, going back to an euclidean
description, to a cylinder of large diameter. In this limit, the partition function
per unit length of the boundary should have a well defined, ``thermodynamic'' limit,
so  $Z$ should go as $Z\propto\exp\left({T_B\over T}\right)$. This means our perturbative 
series has to go as $\exp(cst \tilde{\lambda}^{1/1-g})$. 

\subsection{Perturbation near the 
IR fixed point}

A natural idea to find out what happens beyond the radius of convergence 
is to think of the problem from a ``dual'' point of view, ie around the $\lambda=\infty$
infra red fixed point. The first question one may ask is along which 
irrelevant operator this fixed point is approached. There are several, equally interesting 
ways to answer this question. 
 
The first one starts by considering the 
model where the bulk degrees of freedom have been integrated out, leading to the 
action (at zero temperature, which makes the formulas a bit simpler)
\begeq
S={1\over 2\pi}\int dydy'\left[{\Phi(y)-\Phi(y')\over y-y'}\right]^2
+\lambda\int dy\cos{\beta\over 2}\Phi(y)+{m\over 2}\int dy\left(\partial_y\Phi\right)^2
\endeq
where we have added an irrelevant mass term to make some integrals finite. 

We can find kinks interpolating between adjacent vacua 
and satisfying the equations of motion \cite{schmid}
\begeq
m\partial^2_y\Phi=-{\lambda\beta\over 2}\sin{\beta\over 2}\Phi
\endeq
A simple solution of this equation is indeed 
\begeq
\Phi\equiv f_{ins}(y)={2\pi\over\beta}+{8\over \beta}\tan^{-1}\left[\exp\left({\beta\over 2}
\sqrt{\lambda\over m}y\right)\right]
\endeq
The energy of this kink is infinite, but can be made finite by subtracting 
a constant term from the action, replacing the $\cos{\beta\over 2}\Phi$ by 
$\cos{\beta\over 2}\Phi-1$. If we then consider a configuration of the field $\Phi$ 
made of a superposition of far apart instantons and anti-instantons,  
\begeq
\Phi=\sum \epsilon_j f_{ins}(y-y_j)
\endeq
the kinetic term of the action can be conveniently 
evaluated by Fourier transform
\begeq
S_{kin}=\int |\Phi(\omega)|^2 |\omega|{d\omega\over 2\pi}
\endeq
At large distances, one finds 
\begeq
S_{kin}\approx {16\pi\over \beta^2}\sum_{j<k} \epsilon_j\epsilon_k \ln|y_j-y_k|
\endeq
This is in exact correspondence with the Coulomb gas expansion discussed previously,
but with the exchange ${\beta^2\over 8\pi}\to {8\pi\over \beta^2}$. It follows
that the IR action reads, at  leading order 
\begeq
S\approx {1\over 2}\int_{-\infty}^0 dx\int dy \left[\left(\partial_x\tilde{\Phi}
\right)^2
- \left(\partial_y\tilde{\Phi}\right)^2\right]+\lambda_d\int dy\cos{4\pi\over \beta}
\tilde{\Phi}(0,y),\label{bsgdual}
\endeq
where we recall that $\tilde{\Phi}$ is the dual of the boson $\Phi$. It follows
that the IR fixed point is approached along an operator of
 dimension $h={8\pi\over \beta^2}={1\over g}$. One also checks that $\lambda_d\propto 
\lambda^{-{1\over g}}$. 

It is important to stress now that, while the flow away from the UV fixed point
is fully specified by a single perturbing term, the situation is very different
for the approach towards the IR fixed point. Of course, one is 
free if one wishes to perturb the D boundary conditions by a single
irrelevant operator as represented in (\ref{bsgdual}), though of course one has to be especially
careful in defining the theory because of the strong short distance divergences in the integrals.
The point is, that there is only {\sl one} particular way of approaching the IR
fixed point that corresponds to the trajectory originating at our UV fixed point. This means that
the large $\lambda$ behaviour of the series we are interested in would be 
computable from the knowledge of an action of the form
\begeq
S={1\over 2}\int_{-\infty}^0 dx\int dy \left[\left(\partial_x\tilde{\Phi}
\right)^2
+ \left(\partial_y\tilde{\Phi}\right)^2\right]+\lambda_d\int dy\cos{4\pi\over \beta}
\tilde{\Phi}(0,y)+\sum_k \lambda^{h_k-1\over g-1} O_k\label{horridir},
\endeq
where $O_k$ belong to a very large class of operators allowed by symmetry: there are 
for instance all the $\cos{n\over 2\beta}\tilde{\Phi}$, $\left(\partial\tilde{\Phi}\right)^2$,
and many others. Since all these operators come with appropriately 
scaled powers of the coupling constant, they all give contributions 
to physical properties that depend on our single scaling variable $\tilde{\lambda}$,
and no operator can be discarded (let me stress that an expansion such  as (\ref{horridir}) 
does not make
much sense until one specifies the regularization procedure employed).

Here of course the reader should ask: but why didn't we add that collection of
operators near the UV fixed point as well? The point is that we had control
of what we wanted to do near the UV fixed point, and only a maniac would want to 
use such an irrealistically finely tuned combination of operators to perturb a fixed point.
However we have {\sl no} control about the way the IR fixed point is approached: 
this is entirely determined by the  dynamics of the quantum  field theory, and it turns out
to be quite complicated. It is important in particular to realize that, starting 
from (\ref{horridir}) and trying to go against the renormalization
group flow, there is, most probably,  only {\sl one} choice of IR perturbation that would get back to
our UV fixed point.

It seems very hard to
push the instanton expansion beyond the first non trivial order to
try to get at (\ref{horridir}), or get the expansion of physical quantities for
large $\lambda$: this has for a long time made IR perturbation theory 
impossible to carry out beyond the first trivial order. 

Remarkably however,(\ref{horridir}) can be entirely
determined using ideas of integrability. 
In a scheme where everything
is dimensionally regularized,  the only  vertex 
operator that is present 
near the IR fixed point is $\cos{8\pi\over\beta}\tilde{\Phi}$: none of the other
harmonics actually appear!
 There are also very strong constraints on the 
other operators. 

In any case, the non-perturbative
 region of large $\lambda$ is very hard to access quantitatively using perturbation of
the IR fixed point. Fortunately, the problem can be tackled 
 by using ideas of integrability, a topic to which we will turn soon. 

\subsection{An alternative to the instanton expansion: the conformal invariance analysis}

Clearly, the instanton expansion, if physically appealing, is a bit laborious,
especially when one considers how little information it finally provides. 
The conformal invariance analysis gives an alternative way, usually more 
reliable, to know which operators are present near the IR fixed point. Indeed, 
this information is encoded in the partition function $Z_{DD}$ (\ref{dirich}) for 
$\Phi_0=\Phi'_0$: the modes in the sum correspond to operators with dimension
$h={8\pi n^2\over \beta^2}$, the cosines identified previously, while the 
other terms obtained by expanding the eta function correspond to powers of 
derivatives of the field.  

\vfill
\eject

\part{Integrability and the complete flow}

The constraint of integrability has been used with much
success to study crossover scaling in  bulk 2D
theories exactly.
However, it is often objected that integrable models
are not so relevant for experimentally observable physics
for at least two reasons:

\noindent (a): In order to achieve integrability, extensive
fine-tuning of parameters is often required.
Therefore, it is often believed that  exact
predictions made by studying an exactly integrable
model  might often not be generic and therefore
difficult to observe experimentally. For example integrable spin
chains with spin greater than $1/2$ are gapless, while the generic
even-spin spin chain and those observed have gaps in the spectrum.

\vskip .2cm

\noindent (b): A very important set of experimentally
accessible observables are transport properties.
Amongst those is the conductance which is usually
computed from  (equilibrium) Green's functions using
the Kubo-formula.   It is usually very difficult
or impossible to compute exact Green's functions,
even when the system is  (Bethe-Ansatz)
integrable.  Therefore, before the progress
made in the last couple of years, integrability was largely restricted to
the computation of thermodynamic quantities, excluding
transport properties at finite temperature.

\vskip .5cm

\noindent  In fact, the situation is quite  different for integrable quantum impurity problems.
 For these problems, exact transport properties can
be computed (even out of equilibrium) and
integrability can  answer directly
experimentally important questions.

\noindent In particular:

\vskip .2cm
\noindent (i):  In order to achieve integrability in
quantum impurity problems, one often needs to adjust
very  few parameters -- sometimes none!
For instance, both the Kondo effect and the point contacts
in fractional quantum Hall Effect devices,
 provide an experimental realization
of an integrable system without any fine-tuning. Similarly, while integrable higher spin quantum
spin chains are non generic, the higher spin Kondo problem, is integrable.

\vskip .2cm
\noindent (ii):  Exact transport properties (at non-zero
temperature)  can be computed from integrability \cite{FLSprli}.
The linear response conductance for the quantum Hall point contact, for instance,
agrees quantitatively with recent
experiments by Milliken/Webb/Umbach \cite{exper}: this is the quantity we will discuss
in what follows. 
Many other properties can also be computed, and I will discuss them briefly at the end.

\bigskip

The method  to obtain those exact results
is a bit unconventional: it relies crucially
on a judicious choice of basis of the Hilbert space
of the system. We use a basis that is natural
from the point of view of integrability. It
is simply the  basis in which all the infinite
number of  conservation laws (that exist since
we have an integrable system) are diagonal.
This  basis turns out to  have a ``Fock-space'' like
structure, i.e. it is spanned by ``quasiparticles''.
It is in this basis that the  quantum impurity
interaction becomes tractable.
In order to compute transport properties, we
use a kinetic equation for those quasiparticles of
the Bethe-Ansatz. This is non-trivial,
since we are really describing a fully
interacting system, where a single-particle  (Fermi-liquid)
concept such as a kinetic equation seems out of place,
at first sight.
The single particle kinetic
equation  would fail to produce exact results
in interacting systems
due to the existence
of particle production processes in the single-particle
basis. However, the  quasiparticle basis
dictated to us by integrability
is precisely characterized by the absence
of quasiparticle production (and ``factorized scattering'').
This particular
and special feature of an integrable
theory,  allows us to
use a kinetic equation to compute
transport properties exactly from  integrability.

A historical note is necessary here. 
The point of view I will use 
is  different from the original works on integrable quantum impurity problems.
In the latter works, the authors started from a ``bare'' theory, and proved by hand that
there were simple eigenstates obtained by making a two body scattering ansatz, 
 the Bethe ansatz (there won't be much about the Bethe ansatz per se in these lectures).
 They then proceeded to build the ``physical'' theory by filling up
the ground state, and studying excitations above it. Nowadays, it has become
customary to start directly with the physical theory, and prove its integrability
using a very different and powerful tool that I will introduce below, perturbed conformal
theory. The spectrum of excitations and the S matrix are then deduced (some would say guessed)
by using symmetry arguments; in particular by analyzing non local currents (for this aspect, 
see eg \cite{andre}). The approach gives rise somehow naturally to the computation
of transport properties, in particular by making physical sense of massless scattering,
and {\sl that} has definitely been a progress. Another key advantage is that, for a given ``physical''
theory, there are many possible ``bare'' choices, that is many different possible regularizations.
Usually only one of them is integrable, and not always the 
obvious one; for instance, S. Zamolodchikov showed in his pioneering 
work that the Ising model at $T_c$ with a magnetic field is an integrable quantum field theory,
but it is well known that  its standard regularized square lattice version is {\sl not}
 integrable (in the case we are interested in, the boundary sine-Gordon model,
I am actually not aware of any simple integrable bare hamiltonian).
 All 
 this is not to diminish the beauty and
 astonishing insight
of the pioneering works  about quantum impurity problems \cite{andrei,wiegmann}, nor the 
huge body of work on the Bethe ansatz and Yang Baxter equation that it is impossible 
to even start to acknowledge here.

\section{Search for integrability: classical analysis}

As emphasized in the first sections, one of the main uses of conformal
invariance is to provide a convenient basis to the Hilbert space
of observables in terms of representations of the infinite 
dimensional symmetry. For the free boson, the basis furnished by 
irreducible representations of the Virasoro algebra is just one of many 
choices: the basis furnished instead by representations of the Heisenberg
algebra is also possible, and sometimes more convenient. 

When one wishes to study the problem with a boundary interaction, the question arises,
of which basis will be the most convenient to work. It turns out it is still
 a third choice, provided by a ``massless scattering theory''. To understand what this
 means,
it is good to first consider the classical case.

In the classical limit, one can scale the parameter $\beta$ off the action. Going
to real time, one obtains
a classical scalar field $\Phi(x,t)$ satisfying the
Klein-Gordon equation in the bulk $x\in [-\infty,0)$:
\begeq
\partial_t^2 \Phi - \partial_x^2 \Phi =0\label{KGeqn}
\endeq
together with the  boundary conditions (where $\lambda$ is also rescaled):
\begeq
\left.\partial_x \Phi \right|_{x=0} = \left.\lambda \sin\left(  
{1\over 2}\Phi
\right) \right|_{x=0} \label{intbcs}
\endeq

Now, the point is that the ``most  natural'' basis of solutions for the bulk problem,
that  is plane waves, behaves  badly with respect to the boundary interaction: if 
a plane wave is sent towards the boundary, what bounces back is a complicated 
superposition. Is it possible to find a better basis made of wave packets that
will bounce nicely?

To find such a basis, we make a detour through 
a more complicated problem which everybody knows is integrable, the massive sine-Gordon
model. That is, we want to think of our Klein-Gordon problem as the 
$\Lambda \to 0$ limit of the sine-Gordon equation:
\begeq
\partial_t^2 \Phi - \partial_x^2 \Phi = -\Lambda
\sin(\Phi)\label{SGeqn}
\endeq
It turns out  that this  massive model, in the presence of the boundary
interaction, is still  integrable. We will discuss this point in more details below,
and start instead by considering 
how things look like in the massless limit. 

There are two types of finite-energy solutions of the classical  
sine-Gordon
equation: solitons, which are time-independent and topologically
non-trivial, and breathers, which are time-dependent and topologically  
trivial.
Intuitively, a breather can be thought of as a bound state of a kink  
and an
antikink oscillating in and out (i.e. breathing).
Here,  we will discuss only the solitons; the analysis for  
the
breathers follows analogously.

A major triumph of the theory of non-linear partial differential  
equations was
the construction of explicit solutions of (\ref{SGeqn}) for any number of  
moving
solitons. 
The solitons' energies and momenta
are conveniently expressed
in terms of rapidities $\alpha_j$, defined by
$E_j=M\cosh\alpha_j$ and $P_j=M\sinh\alpha_j$, $M=\Lambda^{1/2}$. The velocity of each is  
thus
given by $\tanh \alpha_j$ (positive for a right-moving soliton).
We have set the speed of ``light'' to be 1.

Consider now a two-soliton solution of (\ref{SGeqn}) on
$(-\infty,\infty)$.  This solution is usually expressed as:
\begeq
\Phi(x,t) = 4 ~arg (\tau) \equiv 4~ \arctan\left(
{{\cal I}m(\tau)  \over {\cal R}e (\tau) }  \right) \label{sgsoln}
\endeq
where the $\tau$-function solution is given by:
\begeqar
\tau = 1 - & \epsilon_1 \epsilon_2
\left( \tanh{\alpha_1-\alpha_2 \over 2}\right)^2 ~e^{ -
E_1 (x-a) - E_2 (x-b) +P_1 t +P_2 t}  \nonumber\\
{}+ & i \left[  \epsilon_1 e^{- E_1 (x-a)+P_1 t}
{}+ \epsilon_2 e^{- E_2 (x-b)+P_2 t} \right]\label{twosols}
\endeqar
The constants
$a$ and $b$ represent the initial positions of the two solitons, and
$\epsilon_j = +1$ if the $j^{\rm th}$ soliton is a kink, while
$\epsilon_j = -1$ if it is an anti-kink.

What happens if we try to take the massless limit of this solution?
For a wavepacket to have finite energy in the massless limit $m\to 0$,
the rapidity $|\alpha|$ must go to infinity. We thus define
$\alpha\equiv A+\theta$, and let $A\to\infty$ such that the  
parameter
$m\equiv {1\over 2} Me^{A}$ remains finite. The energy and momentum  
of a
right-moving ``massless'' soliton then reads
\begeq
E=P=m e^\theta\label{dispi}
\endeq
For a left mover, $\alpha\equiv -A+\theta$, and its  
energy
and momentum read 
\begeq
E=-P=m e^{-\theta}.
\endeq

Suppose that both of these solitons are right-moving. Then the massless
limit yields:
\begeq
\tau = 1 - \epsilon_1 \epsilon_2
e^{-\Delta}e^{ -E_1(\eta - a) -E_2(\eta  - b) }
{}+  i \left[ ~ \epsilon_1 e^{- {E_1 (\eta - a)} }
{}+ \epsilon_2 e^{- { E_2 (\eta - b)}} ~\right]\label{infrapid}
\endeq
where $\eta = (x-t)$ and
$$\Delta \equiv  - \log\left[(\tanh(\theta_1 - \theta_2))^2\right].$$
This leads to  an a priori strangely complicated solution of the
Klein-Gordon equation. Observe that:
\begeqar
arg \left\{ 1 - \epsilon_1 \epsilon_2
 ~e^{ - E_1(\eta - a) - E_2(\eta  - b) }
{}+  i \left[ ~ \epsilon_1 e^{- E_1 (\eta - a) }
{}+ \epsilon_2  e^{-  E_2 (\eta - b)} ~\right] \right\} \nonumber\\
{}= arg \left[ 1 + i \epsilon_1 e^{- E_1 (\eta - a) }  \right]
{}+ arg \left[ 1 + i \epsilon_2 e^{- E_2 (\eta - b) }  \right]
{}=\tan^{-1}\left[\epsilon_1 e^{- E_1 (\eta - a) }\right]
+\tan^{-1}\left[\epsilon_2 e^{- E_2 (\eta - b) }\right]
\label{superpos}
\endeqar
This is easily checked to be the sum of two one-soliton solutions;
 the factor $\Delta$ thus measures
the extent to which the two-soliton solution is {\sl not} a superposition
of one-soliton solutions.

More precisely, consider the limit $a \to \infty, \eta \to \infty$
so that $E_1(\eta - a)$ is finite.  This corresponds to moving the  
first
kink off to $x=+\infty$ and following it.  The $\tau$ function  
collapses
to the one-kink form $\tau = 1 + i\epsilon_1 e^{- E_1 (\eta - a)} $.
Moving this soliton through the second one corresponds to taking it
to $x = -\infty$, or taking the limit $a \to - \infty, \eta \to  
-\infty$
(with  $E_1(\eta - a)$ finite).  Discarding an overall multiplicative
factor (which is irrelevant in the computation of $\phi = 4  
\arg(\tau)$),
we see that in this limit, $\tau = 1 + i\epsilon_1 e^{- E_1 (\eta -
a) -\Delta}$.
Thus these preferred Klein-Gordon wave packets exhibit
{\sl non-trivial monodromy}.  The foregoing time delay $\Delta$ is precisely
the classical form of a massless scattering matrix.

One obtains the same $\Delta$ for two left-moving
solitons.   For a left-moving and a right-moving soliton colliding
one easily sees that the massless limit of $\left(
\tanh {1\over 2}(\alpha_1-\alpha_2)\right)^2$ is unity.  The solution  
collapses
to the superposition of a left-moving wave packet and a right-moving
 wave packet exactly as in (\ref{superpos}), with no time delay.  This
is the classical manifestation of the fact that the left-right quantum
scattering matrix $S_{LR}$ elements are at most rapidity-independent  
phase
shifts.

Consider now the Klein-Gordon equation on $[-\infty,0]$ with the  
boundary
condition (\ref{intbcs}). 
Here is a   direct way of seeing the integrability of the Klein-Gordon (and
indeed, sine-Gordon) equation with boundary conditions (\ref{intbcs}).
The idea is to show that the method of images can
be used on $(-\infty,\infty)$ even in the non-linear system, so as to
replicate the boundary conditions (\ref{intbcs}) on $[-\infty,0]$.
The scattering of a kink, or anti-kink, from the boundary can be
described  by a three-soliton solution on $(-\infty,\infty)$. These
three solitons consist of the incoming soliton, its mirror image
with equal but opposite velocity, and
a stationary soliton at the origin, to adjust the boundary conditions 
 (see \cite{SSW} for more details).
If one takes the infinite rapidity limit of this three-soliton solution  
then
the stationary soliton simply collapses to an overall shift of $\phi$
by a constant, while the mirror images (since they are moving in  
opposite
directions) reduce to a superposition of two wave packets.  One thus
obtains:
\begeq
\Phi = \Phi_0 + 4~ arg \big[ 1 + i \epsilon_1 e^{-
E (\xi - a) }  \big] + 4~ arg \big[ 1 + i \epsilon_2  
e^{- E (\eta - b) }  \big] \ \label{bndryscatt}
\endeq
where $\xi = x+t$, $\eta = x - t$.  By direct computation
one finds that this solution satifies (\ref{intbcs}) with:
\begeq
 e^{E(a+b)}
= - \epsilon_1 \epsilon_2 ~ {(2E+\lambda) \over (2E-\lambda)} \label{twosolbc}
\endeq
The constant $\Delta_B\equiv -E(a+b)$ represents 
the delay of the reflected pulse.
If one defines the classical boundary scale $\theta_B$ via $g = 2m  
e^{-\theta_B}$,
then this delay
may be written as
\begeq
\Delta_B = \log \left[- \epsilon_1 \epsilon_2
 \tanh {1\over 2} (\theta - \theta_B)\right] \label{refdelay}
\endeq
Note that the sign  $ \epsilon_2$ is to be chosen so as to make
the argument of the logarithm real.  This determines whether the
reflection of a kink will be a kink or an anti-kink. Thus we see that  
$\theta_B$ is
the scale at which behavior crosses over from the region of the Neumann
critical point (where the classical boundary scattering is completely
off-diagonal) to the Dirichlet boundary critical point (where classical  
boundary
scattering is diagonal).

The conclusion of this section is clear: there are classical wave packets
that scatter very nicely at the boundary. The price to pay is that they are
considerably more complex than plane waves, and ``weakly interacting'' - that is,
they scatter non trivially through one another. 

The natural attitude after having established such results classically
is to see whether they are preserved quantum mechanically. 
One can for instance establish integrability order by order 
in a loop expansion (which here amounts to an expansion in powers 
of $\beta$). Here I want to show a more direct way to proceed, that
generalizes to all sorts of theories.

\section{Quantum integrability}

\subsection{Conformal perturbation theory}

As  in the previous section, we start by considering 
theories with a bulk interaction. All what follows is based on the very insightful 
work of A.B.  Zamolodchikov \cite{sashaising}.
 Consider therefore
the usual sine-Gordon model with action\footnote{It is safer to assume here that
 ${\beta^2\over 8\pi}<{1\over 2}$,
so no counter term is necessary to define the perturbed action. Only a finite number of such counter terms
would be required in general anyway, because of the ``super-renormalizability'' common
to most perturbed conformal field theories. }
\begeq
S={1\over 2}\int dxdy \left[\left(\partial_x\Phi\right)^2+\left(\partial_y\Phi\right)^2\right]
+\Lambda\int dxdy\cos\beta\Phi
\endeq
As in the classical case, quantum integrability is established 
by proving the existence of non trivial integrals of motion.  Rather than doing perturbation 
around $\beta=0$, what one can do instead is perturbation around the conformal limit. This 
requires first the understanding that a conformal theory is integrable \footnote{At least
partly integrable - this subtlety does not seem to matter for most problem
occuring in condensed matter.}, a rather straightforward property.

Indeed,  let us  try to build a set of conserved quantities
for a quantum field theory. 
We consider Euclidian space with imaginary time in the $y$ - direction:
a quantity will thus be conserved if its integral  along two horizontal
contours at different values of $y$ gives the same result. Using complex coordinates, 
this will
occur if we have a pair of quantities, say $T_{n}$ and $\Theta_n$ such that $\partialbar T_n=
\partial\Theta_n$. Right at the conformal point, by analyticity,
$T$, all its derivatives and (regularized) powers, do provide conserved quantities.

To clarify this a little, let us consider the classical case. Going to 
imaginary time and complex 
coordinates, the equation of motion is
\begeq
\partial\partialbar\Phi={\Lambda\over 4}\sin\Phi
\endeq

\smallskip
\noindent{\bf Exercise:} Show that the first pairs leading to conserved quantities are
\begeqar
T_2=\left(\partial\Phi\right)^2,~~~\Theta_2=-{\Lambda\over 2}\cos\Phi\nonumber\\
T_4=\left(\partial^2\Phi\right)^2-{1\over 4}\left(\partial\Phi\right)^4, \Theta_4
={\Lambda\over 4}\left(\partial\Phi\right)^2\cos\Phi,
\endeqar
that is, the relation $\partialbar T_{2n}=\partial\Theta_{2n}$ holds. Go back to real time,
and find out which quantity is, indeed, conserved by time evolution.
\smallskip

Away from the conformal
point, what will happen is that there will sometimes be a deformation
of (some of) these quantities that is still  conserved. To see that,
let us start by looking at the stress energy tensor, and see what $\partialbar T$ 
becomes with a perturbation. To make sense of this question, we have to insert
$T$ inside a correlator, as usual. The difference with the conformal case is that now
the action reads, quite generally
\begeq
S=S_{cft}+\Lambda \int dxdy O
\endeq
so we have to expand the Boltzmann weight in powers of $\Lambda$. This gives an infinity
 of terms,
each of which has now a Boltzmann weight with a pure $S_{cft}$, so the results right
at the conformal point can be used for them. But then it seems that $T$ being analytic
at the conformal point, nothing will make it non analytic away from it! That is not true
because of what happens at coincident points. The integral of the perturbing field 
will affect only the behaviour near $z$, so we can use the OPE of $T$ with the
 perturbation
\begeq
T(z)O(z',\zbar')={hO(z',\zbar')\over (z-z')^2}+{\partial O(z',\zbar')\over z-z'}+\ldots
\endeq
Now, using the identity (\ref{basicid}) it follows that 
\begeq
\partialbar T=\pi(1-h) \Lambda\partial O
\endeq
Hence, for $T_2=T$, $\Theta_2=-\pi\Lambda(1-h) O$, and we have a conserved quantity - 
to first order in $\Lambda$ that is. Before wondering about higher orders,
let us stress what ensured the existence of a conserved quantity: the fact that
the residue of the 
simple pole of the OPE of $T$ with the perturbation was a total derivative. One can then try to see whether there are other quantities
for which a similar thing holds. Let us consider therefore the combination
$T_4=4\pi^2 :\left(\partial\phi\right)^2:+A :\left(\partial^2\phi\right)^2:$. After laborious 
computation, one finds that the residue of the simple pole, in the sine-Gordon case of
 interest,
is 
$$
\left({Ai\beta\over 2\pi}+i{\beta^3\over 8\pi}\right):\partial^3\phi e^{i\beta\phi}:
-3\beta^2:\partial\phi\partial^2\phi e^{i\beta\phi}:
-4i\pi\beta :\left(\partial\phi\right)^3 e^{i\beta\phi}:
$$
which is a total derivative when (notice this would still hold  with $\beta$ subsituted
 with ${8\pi\over\beta}$)
$$
A=2\pi\left(3-{\beta^2\over 8\pi}-{8\pi\over \beta^2}\right)
$$
The same sort of argument can be built to show that a conserved 
quantity can be obtained for every even $n$, which is the integral of a local field 
of dimension $n$.

We now have to discuss what happens beyond first order. Suppose we carry out the computation
to order $n$; a priori, we expect the result to be something like
\begeq
\partialbar T_4=\partial \Theta_4 +O(\Lambda^n)\label{conservi}
\endeq
where $\Theta_4$ is of order one in $\Lambda$. The left hand side has dimensions $(4,1)$,
 while $\Lambda^n$ has 
dimensions $\left(n(1-{\beta^2\over 8\pi}), n(1-{\beta^2\over 8\pi})\right)$. This means
that a local field with dimensions $\left(4-n+n{\beta^2\over 8\pi}, 1-n+n{\beta^2\over 8\pi}\right)$
has to appear multiplying the $\Lambda^n$ term.  Since $\beta^2<8\pi$, only a finite
number of cases allow a positive set of dimensions, and for each of these except $n=1$,
and for $\beta^2$ generic, one
checks that there is no field with these dimensions. Hence, the conservation at
lowest order extends, generically, to conservation at arbitrary order \footnote{Non generic cases could be more 
complicated; consider the example of $h=1$ for instance, where now arbitrary orders
are allowed in the right hand side of (\ref{conservi})!}. This proves 
quantum integrability, perturbatively that is.

Finally, these conserved quantities also turn out to be in involution, ie they
define mutually commuting operators. The proof is based on the Jacobi identity, which implies
that, if two conserved quantities do not commute, then their commutator
is also a conserved quantity - a little more thinking then establishes this is
not possible. I also would like to remark that in the early literature about quantum
integrable models, there is the implicit ``suspicion'' that quantization might
destroy integrability, ie reduce the classical symmetry of the theory. It is 
important to realize that such a thing is not always true. In fact, in two dimensions
at least, the quantum
theory often has more symmetry than the classical theory.

\subsection{S-matrices}

The next step in the analysis of a massive integrable quantum field theory 
requires going to a scattering description \cite{ZZ}. Let us assume quite generally 
that 
we have massive particles distinguished by some label $a$,  with mass $M_a$. 
We write their two momentum $p^\mu$ in terms of a rapidity variable $\alpha$:
 $E=M\cosh\alpha, P=M\sinh\alpha$.

We now consider scattering processes. There are the 
 ``in-states'', corresponding physically to a bunch of particles arranged on
the x-axis by decreasing order of rapidities, which we describe formally 
by a ket  $\left|\alpha_1,
\ldots,\alpha_N\right>_{in}^{a_1,\ldots,  a_n}$. At large time,
they give rise to  ``out-states'', formally described by $\left|\alpha'_1,
\ldots,\alpha'_{n'}\right>_{out}^{a'_1,\ldots,  a'_{n'}}$, with an a priori
 different set of particles  arranged
on the x-axis by increasing order of, a priori different,  rapidities (see figure 11). 

\begin{figure}[tbh]
\centerline{\psfig{figure=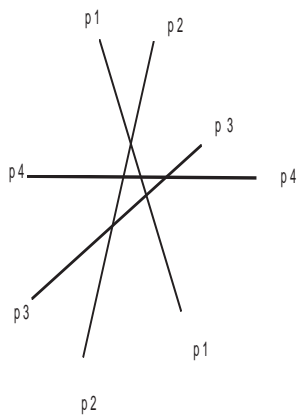,height=2.15in,width=1.5in}}
\caption{Scattering of quasiparticles in an integrable, $1+1$
quantum field theory.}
\end{figure}

Both the in and the out states are a complete set of states in a local quantum field theory,
and they are connected by the S-matrix. 

Now, the  existence of infinitely many 
conserved quantities has  very drastic consequences on the scattering of these 
particles. Indeed, the conserved quantities have to act
simply on the multiparticles states - they are in fact  proportional
to the sums of odd powers of the momenutm. As a result, it follows that, 
in the scattering 
process: 

\noindent $(i)$ The number of particles is conserved; in fact, the number of particles
of the same mass is conserved

\noindent $(ii)$  The final set of two-momenta coincides with the initial set of two-momenta

From this in turn, it follows that the S matrix factorizes into a product of 
2-body scattering processes. To see this, consider for instance conjugating 
the S matrix by the operator $e^{i p_1^\mu x_\mu}$. Since the S matrix
conserves momenta, it actually does commute with this operator, so we 
don't change anything. On the other hand, the operator has a non trivial physical
action: it changes the space time coordinates of particle 1. If we chose 
$p_1$ appropriately, we can arrange for particle $1$ to scatter with the other
particles only after they are very highly separated, so the scattering of this particle
is a succession of two particle scatterings. By proceeding inductively,
one deduces that indeed, the S matrix factorizes. Moreover, for the whole thing 
to be consistent, the scattering must be ``associative''; that is,
the scattering of three particles can be decomposed into three
pairwise scatterings,  with a result independent of which particular decomposition is
used. This is illustrated graphically
 in figure 12

\begin{figure}[tbh]
\centerline{\psfig{figure=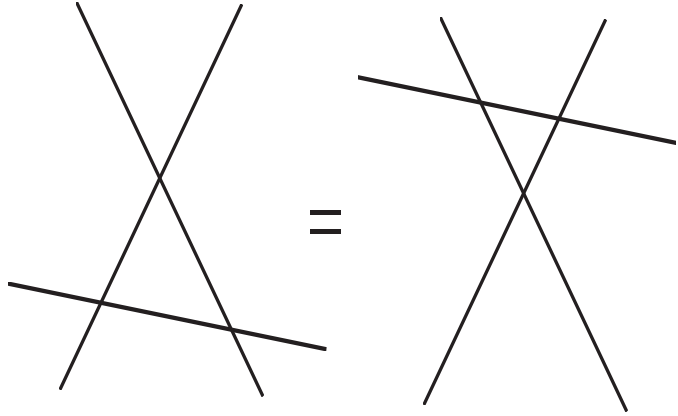,height=2.15in,width=3.5in}}
\caption{Factorization of the scattering.}
\end{figure}

 This constraint, the so called Yang Baxter equation,
is the pillar of the algebraic approach to integrable quantum field theories (for a review see for instance 
\cite{YBreview}).

Notice that the YB equation is trivial if all the particles have different masses.
It becomes more interesting in the case where several particles 
have the same mass, but differ by some other quantum number, eg the charge. In general,
we define the S matrix elements by the relation (figure 13)
\begeq
\left|\alpha_1,\alpha_2\right>^{in}_{a_1,a_2}=S_{a_1a_2}^{a'_1,a'_2}
\left|\alpha_1,\alpha_2\right>^{out}
_{a'_1,a'_2}\label{smatdef}
\endeq

\begin{figure}[tbh]
\centerline{\psfig{figure=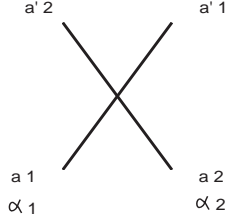,height=1.15in,width=1.15in}}
\caption{The matrix element $S_{a_1a_2}^{a'_1,a'_2}$ corresponds to 
the  process illustrated here.}
\end{figure}

Relativistic invariance contrains the S matrix to depend on the difference
of rapidities $\alpha_1-\alpha_2$.

 In the case of the sine-Gordon model, perturbation in $\beta$, 
 considerations of (quantum affine) symmetry, minimality and consistency
assumptions, lead to the following results. The spectrum is made up of 
 the kink and antikink of mass $M\propto  \Lambda^{1\over 2-2g} $, together with
 breathers. For $n-1< {1\over g}\leq n$, there are $n-2$ such breather states. 
Their masses are $M_k=2M \sin\left[k\pi g/2(1-g)\right]$.

The kink S-matrix is 
closely related to the matrix of Boltzmann weights in the 6-vertex model.
There are three key amplitudes
\begeqar
a(\a)&=\sin[\gamma(\pi+i\a)]Z(\a)\nonumber\\
b(\a)&=-\sin(i\gamma\a) Z(\a)\nonumber\\
c(\a)&=\sin(\gamma\pi) Z(\a),\label{sixvrmat}
\endeqar
where $\gamma={1\over g}-1$. 

\begin{figure}[tbh]
\centerline{\psfig{figure=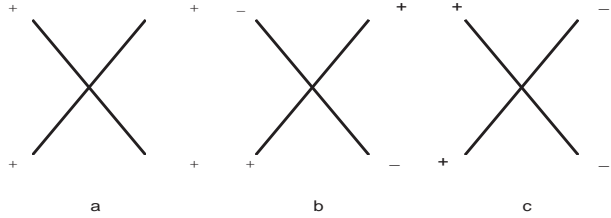,height=1.15in,width=3.15in}}
\caption{The three types of possible processes involving kink and antikink.}
\end{figure}

The element $a(\a_1-\a_2)$ describes the process
$\left|\a_1\a_2\right>_{++} \rightarrow \left|\a_1\a_2\right>_{++}$,
as well as
$\left|\a_1\a_2\right>_{--} \rightarrow \left|\a_1\a_2\right>_{--}$,
$b$ describes $+-\to +-$,
$c$ describes the
non-diagonal process $+-\to -+$ (see figure 14), and there is a symmetry under
interchange of
kink to antikink (corresponding to $\Phi\to -\Phi$).

\smallskip
\noindent{\bf Exercise:} Show that (\ref{sixvrmat}) gives rise, indeed,
to a solution of the Yang Baxter equation. Hint: proceed graphically 
as sketched in figure 15 - to represent the matrix multiplications
of YB, simply draw all the physical processes that connect a given pair 
of  initial and final
states, and add up their amplitudes. 
\smallskip

\begin{figure}[tbh]
\centerline{\psfig{figure=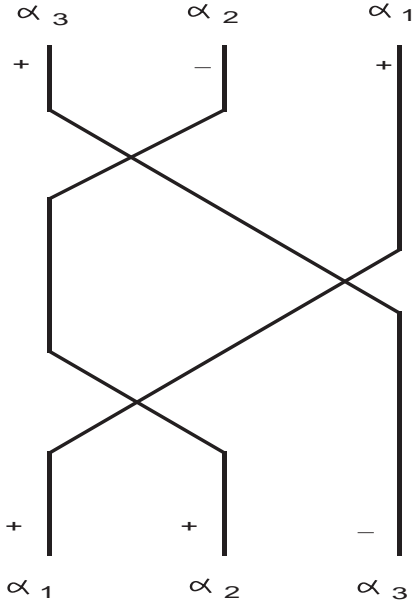,height=3.15in,width=2.15in}}
\caption{The three types of possible processes involving kink and antikink.}
\end{figure}

The
function
$Z(\a)$ is a normalization factor, which can be written as
$$Z(\a)={-1\over \sin[\gamma(\pi +i\a)]}\exp\left(i
\int_{-\infty}^{\infty} {dy\over 2y}
\sin {2\a\gamma y \over\pi} {\sinh [(\gamma-1)y]
 \over \sinh y\cosh(\gamma y)} \right).$$
The breather-kink and
breather-breather $S$ matrices are well known
; we do not write them down here (see below for some examples,
and in the appendix).

These S matrices  of course are not only characterized by 
the fact that they must solve the Yang-Baxter equation:
there are several other physical requirements,
like unitarity and crossing symmetry. In the formula above, these translate into
the relations 
\begeqar
a(\alpha)=b(i\pi-\alpha)\nonumber\\
c(\alpha)=c(i\pi-\alpha)
\endeqar
form which unitarity $S(\alpha)S(-\alpha)$ follows.

 In addition, one must have a 
``closed bootstrap'': for instance,
breathers appear as poles in the kink antikink scattering, and their
S matrix can be computed using this fact (for a review see \cite{giuseppephysrep}, \cite{patrickbook}). 

When $1/g$ is an integer, the bulk scattering is
diagonal ($c$ vanishes) and $a=\pm b$.
Therefore, the only allowed
processes are transmissions: particles go through  one another without 
exchanging quantum numbers. This is the simplest case, to which we will
restrict in what follows.

It is reasonable to think of the S matrix, together with the mass spectrum, as the
 quantum
equivalent of the knowledge of the $\tau$ function in the classical case
\footnote{There are more accurate formulations of this statement.}. We can then
take the massless limit exactly as we did before, by letting the mass parameter 
$M\to 0$, and 
at the same time boosting the rapidities. One obtains then a collection
 of left and right
moving kinks, antikinks and breathers.  The LL and RR scattering 
matrices  take exactly the same expressions as in the massive case in terms of the 
new rapidities $\t$, while the LR S matrices go to constants, which can in most cases 
just be forgotten. This provides an alternative description of the free boson in terms 
of ``massless scattering''. We will see in a little while how quantities of the 
conformal field theory can be recovered, if one wishes, within that description.

It is fair to stress here that the RR or LL scattering are hard to make sense of 
in the context of a true, physical scattering process. If particles have the label R
say,
this means they are  moving at the speed of light in the right direction, so, 
for instance,
all the lines in the figure 11 which illustrates the Yang Baxter equation become parallel! 
The point is that the scattering has to be interpreted in the massless limit as
a set of commutation relations for creation operators, an idea which we will discuss 
below. The idea of massless scattering appeared a bit weird at the beginning \cite{FT},
\cite{ZZmassless},\cite{FStrieste},
but its predictive power and favorable comparison with experiments
gained it respectability quickly. A more rigorous approach along the lines
of lattice regularizations is proposed in \cite{RS}.

\subsection{Back to the boundary sine-Gordon model}

For the moment, we finish following
 the logic of the classical analysis in the quantum case. 
First, one can prove that the quantum sine-Gordon model with a bulk {\sl and} boundary
interaction 
\begeq
S={1\over 2}\int_{-\infty}^0 dx\int dy \left[\left(\partial_x\Phi\right)^2
+ \left(\partial_y\Phi\right)^2+\Lambda \cos\beta\Phi(x,y)
\right]+\lambda\int dy\cos{\beta\over 2}\Phi(0,y)
.\label{sgbsg}
\endeq
allows the existence of conserved quantities as well \cite{GZ}. The proof proceeds in the same 
spirit as for the bulk case, and we are not going to reproduce it here,
though it is an excellent  exercise for the diligent reader.

\smallskip
\noindent{\bf Exercise:} Show that, in the classical theory, the pair $T_4,\Theta_2$
still leads to a conserved quantity provided there exists still another local 
quantiy $\theta_3$ such that
\begeq
\left.T_4+\bar{\Theta}_2-\bar{T}_4-\Theta_2\right|_{x=0}={d\over dy}\theta_3
\endeq
Show then, that for the  sine-Gordon model with a potential $V[\Phi(x=0,y)]$
at the boundary, only the choice $V\propto \cos{1\over 2}(\Phi+\Phi_0)$
allows the existence of the integral of motion.
\smallskip

 Integrability means 
that the particles also have to scatter nicely at the boundary,
ie scatter one by one, without particle production, in a way that is compatible
with the bulk scattering. This latter condition can simply be expressed 
graphically as shown on figure 16, and corresponds to the ``boundary Yang-Baxter''
equation. 

\begin{figure}[tbh]
\centerline{\psfig{figure=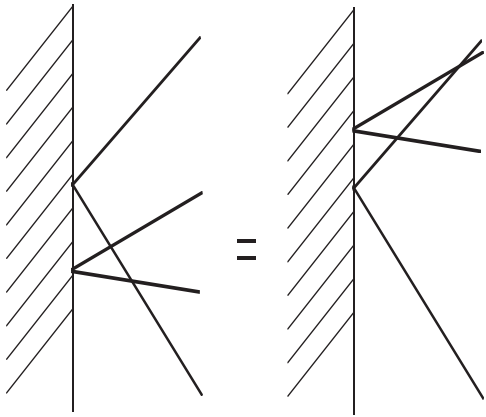,height=2.15in,width=2.5in}}
\caption{The boundary Yang-Baxter equation.}
\end{figure}

The problem of determining the reflection matrix is purely technical. It 
does have quite a simple answer in the massless limit. Introduce
\begeqar
R_{+-}(\theta)=R_{-+}(\theta)={-i\exp(\gamma\t)\over 1-i\exp(\gamma\t)}\exp\left[i\chi_g(\t)\right]\nonumber\\
R_{++}(\theta)=R_{--}(\theta)={1\over 1-i\exp(\gamma\t)}\exp\left[i\chi_g(\t)\right]\nonumber
\endeqar
where $\chi_g$ is a phase that will disappear at the end of the computations, 
$\gamma={1\over g}-1$. The reflection 
matrix for coupling $\lambda$ is then  given by $R(\theta-\theta_B)$ where $T_B=e^{\theta_B}\propto
 \lambda^{1/(1-g)}$ (the exact correspondence depends on the regularization scheme;
 it is given in \cite{FLeS} in the case of dimensional regularization) is the equivalent of 
the Kondo temperature. As $\theta\to\infty$, $R_{++}\to 0$ (the 
scattering is completely off-diagonal) corresponding to Neumann boundary conditions,
while as $\theta\to -\infty$, $R_{+-}\to 0$ (the scattering is 
completely diagonal) corresponding to Dirichlet 
boundary conditions. Notice also the 
unitarity condition $\left|R_{+-}\right|^2+\left|R_{++}\right|^2=1$.

\section{The thermodynamic Bethe-ansatz: the gas of particles 
with ``Yang-Baxter statistics''.}

The thermodynamic Bethe 
ansatz was probably written first in \cite{YY} in the context of the XXZ model. Its use
in quantum field theory, in particular to compute the central charge and study RG flows,
was pioneered by a beautiful series of papers of Al. Zamolodchikov; see for instance
 \cite{Alyoshatba}. A useful and pedagogical review on many of these topics can be found 
in \cite{vladimir}.

\subsection{Zamolodchikov Fateev algebra}

It is convenient to think of the particles in terms of creation
and annihilation operators. For this, let us introduce, still denoting 
the type of the particles by a label $a$, operators $Z_a(\theta)$ and
$Z_a^\dagger(\theta)$ satisfying the relations
\begeqar
Z_{a_1}(\theta_1)Z_{a_2}(\theta_2)&=&S_{a_1a_2}(\theta_1-\theta_2)Z_{a_2}
(\theta_2)Z_{a_1}(\theta_1)\nonumber\\
Z^\dagger_{a_1}(\theta_1)Z^\dagger_{a_2}(\theta_2)&=&S_{a_1a_2}(\theta_1-\theta_2)
Z^\dagger_{a_2}
(\theta_2)Z^\dagger_{a_1}(\theta_1)\nonumber\\
Z_{a_1}(\theta_1)Z_{a_2}^\dagger(\theta_2)&=&S_{a_1a_2}(\theta_1-\theta_2)Z^\dagger_{a_2}
(\theta_2)Z_{a_1}(\theta_1)+2\pi\delta_{a_1a_2}\delta(\theta_1-\theta_2)\label{zfdef}
\endeqar
Here, we restricted to the case of diagonal scattering. Note that the compatibility 
between the first two  relations uses unitarity in the form 
$S^\dagger (\theta)=S^{-1}(\theta)=S(-\theta)$. The space of states
is generated by the kets 
\begeq
\left|\theta_1,\ldots,\theta_n\right>_{a_1,\ldots,a_n}=Z_{a_1}^\dagger(\theta_1)
\ldots Z_{a_n}^\dagger(\theta_n)\left|0\right>,
\endeq
where $\left|0\right>$ denotes the physical vacuum. Similarly, the dual space is 
generated by the bras
\begeq
\left< \theta_n,\ldots,\theta_1\right|=\left< 0\right|Z_{a_n}(\theta_n)\ldots
Z_{a_1}(\theta_1)
\endeq
The metric is, from (\ref{zfdef}), induced by
\begeq
{}_{a_1}\left< \theta_1|\theta_2\right>_{a_2}=2\pi\delta_{a_1a_2}\delta(\theta_1-\theta_2)
\endeq
If for instance $\theta_1>\theta_2$, then the in and out states are, respectively
\begeqar
\left|\theta_1,\theta_2\right>_{a_1a_2}^{in}=\left|\theta_1,\theta_2\right>_{a_1a_2}\nonumber\\
\left|\theta_1,\theta_2\right>_{a_1a_2}^{out}=\left|\theta_2,\theta_1\right>_{a_1a_2}
\endeqar
When the rapidity sets are not ordered, one obtains states which are neither in nor out;
or course they are related to either of these by products of S matrix elements. 

To make things more concrete, let us  discuss briefly wave functions in coordinate 
representation, restricting for simplicity to two particles.
 To satisfy the relations (\ref{zfdef}), it is easy to 
see that the wave function must have a singularity at coincident coordinates, and be
 of the form
\begeq
\left|\theta_1\theta_2\right>_{a_1a_2}\propto 
\int_{x_1<x_2} dx_1dx_2\  e^{i(P_1x_1+P_2x_2)}\left|x_1,x_2\right>_{a_1a_2}
-S_{a_1a_2}(\theta_1-\theta_2)\int_{x_1>x_2} dx_1dx_2\  e^{i(P_1x_1+P_2x_2)}\left|x_1,x_2\right>_{a_1a_2}\label{wavefucnt}.
\endeq
where we assumed that the particles are fermions, $S(0)=-1$. Equivalently, one has
\begeq
\left|x_1x_2\right>_{a_1a_2}\propto \int_{\theta_1>\theta_2}d\theta_1d\theta_2\ 
\left[e^{i(P_1x_1+P_2x_2)}+S_{a_1a_2}(\theta_1-\theta_2)
e^{i(P_1x_2+P_2x_1)}\right]\left|\theta_1,\theta_2\right>_{a_1a_2}^{in}
\endeq
where we see the appearance of the well known Bethe wave function \cite{vladimir} .

\subsection{The TBA}

The next step is to get a handle on the massless scattering description. 
The latter turns out to be quite convenient to discuss thermodynamic properties, 
and this is what we shall start with.  

As a simple example we consider a hypothetical theory made up of a 
 single type of massless
particle, say right-moving, with energy and momentum parametrized as
in (\ref{dispi}).  The scattering is described by a single $S$-matrix element
$S_{RR}$.  Quantizing a gas of such particles on  a circle of length $L$
requires the momentum of the $i$th particle to obey (we have set $\hbar=1$)
\begeq
\exp\left(im e^{\theta_i}L\right)\prod_{j\neq i}
S_{RR}(\theta_i-\theta_j)=1\label{quan}. 
\endeq
One can think of this intuitively as bringing the particle around the
world through the other particles; one obtains a product of
two-particle $S$-matrix elements because the scattering is
factorizable. A bit more rigorously, one can deduce 
this from the wave function as in (\ref{wavefucnt})

Going to the $L\rightarrow\infty$ limit, we introduce the density of
rapidities indeed occupied by particles $\rho(\theta)$ and the density
of holes $\tilde\rho$. A hole is a state which is allowed by the
quantization condition (\ref{quan}) but which is not occupied, so that the
density of possible rapidities is $\rho(\theta)
+\rho^h(\theta)$.  Taking the derivative of the log of  (\ref{quan})
yields
\begeq
2\pi[\rho(\theta)+\rho^h(\theta)]=m L
e^\theta+\int_{-\infty}^\infty K(\theta-\theta')\rho(\theta')d\theta'\label{logquan},
\endeq
where 
$$K(\theta)={1\over i}{d\over d\theta}\ln S(\theta).$$
To determine which fraction of the levels is occupied we do the
thermodynamics, following the pioneering work of Yang and Yang. The energy is
$${\cal E}=\int_{-\infty}^\infty \rho(\theta)m
e^{\theta} d\theta,$$
and the entropy is
$${\cal S}=\int_{-\infty}^\infty\left[
(\rho+\rho^h)\ln (\rho+\rho^h)
-\rho\ln(\rho)-\rho^h\ln(\rho^h)\right]d\theta.$$
\smallskip
\noindent{\bf Exercise}: derive this relation by using Stirling's formula $\Gamma(z)
\approx z^{z-{1\over 2}}e^{-z}\sqrt{2\pi}$.
\smallskip

The free energy  ${\cal F}=({\cal E}-T{\cal S})$ is found by
minimizing it with respect to $\rho$.  The variations of ${\cal E}$ and 
${\cal S}$ are
\begin{eqnarray}
\delta{\cal E}=&\int_{-\infty}^\infty 
\delta\rho m e^\t d\t\nonumber\\
\delta{\cal S}=&\int_{-\infty}^\infty\left[(\delta\rho+
\delta\rho^h)\ln (\rho+
\rho^h)-\delta\rho\ln(\rho)-\delta\rho^h\ln(\rho^h)\right]
d\theta\nonumber
\end{eqnarray}
It is convenient to parametrize
\begeq
{\rho(\theta)\over\rho^h(\theta)}=\exp\left(-{\epsilon\over T}\right)
\endeq
giving
$$\delta {\cal S}=\int_{-\infty}^\infty\left[\delta\rho
\ln\left(1+e^{\epsilon/T}\right)+\delta\rho^h\ln
\left(1+e^{-\epsilon/T}\right)\right]d\theta.$$
Using (\ref{logquan}) allows us to find $\tilde\rho$ in terms of $\rho$.
Denoting convolution by $\star$, this gives $2\pi(\delta\rho+
\delta\rho^h)=K\star\delta\rho$
so
$$\delta {\cal S}=\int_{-\infty}^\infty\left[
{\epsilon\over T}+{K\over 2\pi}\star\ln\left(1+e^{-\epsilon/T}\right)
\right]\delta\rho d\theta.$$
Hence the extremum of ${\cal F}$ occurs for
\begeq
m e^\theta=\epsilon+T{K\over 2\pi}\star\ln\left
(1+e^{-\epsilon/T}\right).\label{tba}
\endeq
and one has then, expressing $\rho^h$ from (\ref{logquan}) 
and using (\ref{tba}) 
\begeq
{\cal F}=-LT^2{m \over 2\pi T}\int_{-\infty}^\infty
 e^\theta\ln\left(1+e^{-\epsilon/T}\right)d\theta.\label{free}
\endeq

It is a simple exercise to show that this formula, together with  (\ref{tba}), 
generalizes to a theory with several species of particles, provided the scattering is 
diagonal. This corresponds to the case ${\beta^2\over 8\pi}=g={1\over t}$, $t$ an integer,
to which we restrict in what follows. In that case, recall that we have 
a kink and antikink of mass parameter  $m$, and breathers of mass 
parameter $m_k=2m\sin{k\pi\over 2(t-1)}$, 
with $k=1,\ldots,t-2$. We will also allow for different chemical potentials 
$\mu_k$ for the various particles. Defining now $\epsilon$'s through
\begeq
{\rho_j(\theta)\over\rho^h_j(\theta)}=\exp\left({\mu_j-\epsilon_j\over T}\right)
\endeq
the equivalent of  (\ref{logquan}) is now
\begeq
2\pi[\rho_j(\theta)+\rho^h_j(\theta)]=m_j L
e^\theta+\sum_k\int_{-\infty}^\infty K_{jk}(\theta-\theta')\rho_k(\theta')d\theta'\label{logquani},
\endeq
and the equivalent of (\ref{tba}) 
\begeq
m_je^\theta=\epsilon_j+T\sum_k{K_{jk}\over 2\pi}\star\ln\left
(1+e^{\mu_k-\epsilon_k\over T}\right).\label{gentba}
\endeq
The equivalent of (\ref{free}) is, in turn:
\begeq
{\cal G}=
{\cal E}-T{\cal S}-\sum_k \mu_k{\cal N}_k=-LT^2\sum_k {m_k\over 2\pi T}\int_{-\infty}^\infty
 e^\theta\ln\left(1+e^{\mu_j-\epsilon_k\over T}\right)d\theta.\label{genefree}
\endeq
For the case $g={1\over 3}$ for instance, one has
\begeqar
K_{bb}=2K_{++}=2K_{+-}=-{2\over\cosh\t}\nonumber\\
K_{b+}=K_{+b}=-2\sqrt{2}{\cosh\t\over\cosh2\t}\label{explicit}
\endeqar

It has become common in the literature to reformulate the TBA in a convenient
form by using  simple diagrams. It is a laborious but straightforward exercise
to demonstrate, using the kernels given in the appendix, that (\ref{gentba}) is equivalent
to the following simple system \footnote{The case where $g$ is not of the simple 
form $1/integer$ can also be handled of course. It is technically more difficult
because the scattering is non diagonal, so an additional Bethe ansatz
is necessary to diagonalize the scattering to start with, before the periodicity of the 
wave function can be imposed \cite{PaulKen}.}
\begeq
\epsilon_j=T\sum_k N_{jk} {s\over 2\pi}\star \ln\left(1+e^{\epsilon_k-\mu_k\over T}
\right)\label{universal}
\endeq
Here, $s(\theta)={(t-1)\over\cosh(t-1)}(\theta)$, $N_{jk}=1$ if the nodes $j$ and $k$ 
are neighbours on the following diagram, $0$ otherwise

\bigskip
\noindent
\centerline{\hbox{\rlap{\raise28pt\hbox{$\hskip5.5cm\bigcirc\hskip.25cm +$}}
\rlap{\lower27pt\hbox{$\hskip5.4cm\bigcirc\hskip.3cm -$}}
\rlap{\raise15pt\hbox{$\hskip5.1cm\Big/$}}
\rlap{\lower14pt\hbox{$\hskip5.0cm\Big\backslash$}}
\rlap{\raise15pt\hbox{$1\hskip1cm 2\hskip1.3cm s\hskip.8cm t-3$}}
$\bigcirc$------$\bigcirc$-- -- --
--$\bigcirc$-- -- --$\bigcirc$------$\bigcirc$\hskip.5cm $t-2$ }}

\bigskip
\noindent {\bf Exercise}: establish this for the case $g={1\over 3}$.
\bigskip

The equations (\ref{universal}) have to be supplemented by the boundary conditions
\begeq
\epsilon_j\approx m_j e^\theta,\ \theta>>1
\endeq

\footnotesize

\subsection{A standard computation: the central charge}

The thermodynamics of a chiral theory like the one we just studied is not so exciting;
this is because, after all, the $1+1$ theory is conformal invariant, 
so the results at different temperatures are essentially equivalent. This can easily
be seen on the TBA equations: a change in $T$, or the mass scale $m$, 
can be fully absorbed by a boost of the particles,
ie a shift of rapidities, exactly like for the changes in mass scale encountered before.
As a result, we see that the integrals in (\ref{free}) are independent of the temperature,
so ${\cal F}\propto LT^2$. 

That this is so, and the coefficent of proportionality, are directly 
related with considerations from the beginning of these lectures. 
Indeed,  we have ${\cal F}=-T\ln Z$, where $Z$ is the
partition function of the one dimensional quantum field theory at
temperature $T$. In Euclidean formalism, this corresponds to a theory on
a torus with finite size in time direction $R=1/T$. By modular
invariance, identical results should be obtained if one quantizes the
theory with $R$ as the space coordinate . For large $L$,
$Z=e^{-E(R)L}$, where $E(R)$ is the ground-state (Casimir) energy with
space a circle of length $R$. Thus ${\cal F}=LE(R)/R$. Conformal
invariance requires that at a fixed point this Casimir energy is
$E(R)=-{\pi c\over 6R}$, where $c$ is the central charge. Going
back to the thermal point of view, ${\cal F}=-{L\pi cT^2\over 6}$ and
the specific heat is\footnote{With massive particles 
or with nontrivial
left-right massless scattering, ${\cal F}$ does depend on
$M/T$, giving a running central charge.}  ${\cal C}={L\pi cT\over 3}$. 

It is possible to  analytically find this central charge from (\ref{tba}). This 
is a bit technical, but worth studying, since it is a crucial {\sl a posteriori} test
of the whole thing. We take the
derivative of (\ref{tba}) with respect to $\theta$ and solve for $e^{\theta}$.
Substituting this in (\ref{free}), we have 
\begin{eqnarray}
{\cal F}&=-{TL\over 2\pi}\int d\t
\left[{d\e\over d\t} \ln(1+e^{-\e/T}) -\int {d\t'\over 2\pi} \ln(1+e^{-\e(\t)/T})
K(\t-\t') {d\e\over d\t'}{1\over 1+e^{\e(\t')/T}}\right]\nonumber\\
&=-{TL\over 2\pi}\int d\t {d\e\over d\t}\left[ \ln(1+e^{-\e/T}) + 
{\e-m e^\t\over T}{1\over 1+e^{\e(\t)/T}}\right]\nonumber\\
&=-{\cal F}-
{TL\over 2\pi}\int d\t {d\e\over d\t} \left[\ln(1+e^{-\e/T})+{\e/T\over
1+e^{\e/T}}\right],
\end{eqnarray}
where we use (\ref{tba}) again to get to the second line. We can replace the
integral over $\t$ with one over $\epsilon$, giving an ordinary integral
$$
{\cal F}=-{TL\over 4\pi}\int_{\e(-\infty)}^{\infty}
 d\e \left[\ln(1+e^{-\e/T})+{\e/T\over
1+e^{\e(\t)/T}}\right] ,$$ 
A change of variables gives
\begeq
{\cal F}=-{T^2L\over 2\pi}{\cal L}\left({1\over 1+x_0}\right),
\endeq
where ${\cal L}(x)$ is the Rogers dilogarithm function
$${\cal L}(x)=-{1\over 2}\int_0^x\left({\ln(1-y)\over y}+
{\ln y\over 1-y}\right) dy,$$
and $x_0\equiv\exp[\e(-\infty)/T]$ is obtained from (\ref{tba}) as 
\begeq
{1\over x_0}=\left(1+{1\over x_0}\right)^I\label{xoo},
\endeq
with $I={1\over 2\pi}\int K$. 

For example, when the $S$ matrix is a constant, $K=0$, $x_0=1$ and
\begeq
{\cal F}=-{LT^2\pi\over 24},
\endeq
where we used $L(1/2)={\pi^2\over 12}$.  Here we find $c_L={1\over 4}$.
In a left-right-symmetric quantum field theory, the right sector makes
the same contribution, giving the total central charge $c={1\over 2}$
required for free fermions. 

Similar computations can be carried out for more complicated theories, leading
to beautiful expressions of central charges in terms of sums 
of dilogarithms (see eg \cite{dilogs}). In the case of interest, one finds of course $c=1$. 

\normalsize

\subsection{Thermodynamics of the flow between N and D fixed points}

We now wish to do the thermodynamics in the presence of the boundary, to obtain
the boundary free energy, and the associated flow of boundary entropies. 
To start, it is better 
to map the problem onto 
a line of length $2L$ ($-L<x<L$)
by considering the left movers to be right movers with $ x> 0$.
Thus we have only
R movers scattering among themselves and off the boundary, which  
can now
be thought of as an impurity (a particle with rapidity $\t_B$). The reflection matrix  becomes
 a transmission matrix, with appropriate relabellings, for instance $R_{+-}\to T_{++}$, etc. (This  
trick is
the same than what we did  for boundary conformal field theory, and can only be used in the  
massless
limit.). For simplicity, we put periodic boundary conditions on the  
system;
these do not change the boundary effects at $x=0$.

Recall we consider only the 
case  $\gamma={1\over g}-1$ a positive integer, where the bulk  
scattering
is diagonal. The  impurity scattering still is not, but we can redefine  
our
states to be $\left|1,2\right>\equiv (\left|+\right>\pm \left|-\right>)/\sqrt{2}$ 
so that the impurity scattering is now diagonal:
\footnote{If $\gamma$ is even, this actually makes the bulk scattering  
completely
off-diagonal (e.g.\ $\left|11\right>$ scatters to $\left|22\right>$), but the TBA equations  
turn
out the same.}
\begeqar
T_{11}(\t)=R_{++}+R_{+-}&=\exp\left[i\chi_g(\theta)\right]\nonumber\\
T_{22}(\t)=R_{++}-R_{+-}&=\tanh\left({\gamma\t\over 2}
-{i\pi\over 4}\right) \exp\left[i\chi_g(\theta)\right].
\endeqar

We can now write the Bethe equations. These differ from the bulk ones only
by the presence of the additional impurity scattering 
\begeq
2\pi(\rho_j(\t)+\rho_j^h(\t))= m_j e^{\t} + \sum_k
K_{jk}\star\rho_k(\t)+ {1\over 2L}
\kappa_j(\t-\t_B)\label{bdrtba}
\endeq
where 
\begeqar
K_{jk}(\t)&=&{1\over i} {d\over d\t}
\ln S_{jk}(\t)\nonumber\\
\kappa_{j}(\t-\t_B)&=&{1\over i} {d\over d\t} \ln R_{j}(\t-\t_B).
\endeqar
The effect of the boundary is seen in the last piece of  (\ref{bdrtba})
proportional to $1/L$. 

The minimization equations are independent of the boundary terms,
since these do not appear directly in ${\cal E}$ or ${\cal S}$, 
and they disappear when one takes a variation of (\ref{bdrtba}). Thus equations 
(\ref{gentba}) still hold. 

Boundary terms {\sl do} enter the free energy or the grand potential
however when one rewrites it 
in terms of the
$\epsilon$'s. One finds
\begeq
{\cal F}={\cal F}_{bulk}-
T\int {d\t\over 2\pi}\sum_{j=1}^{t}
\kappa_{j}(\t-\ln(T/T_B))\ln(1+e^{-\epsilon_j(\t)}).\label{totalf}
\endeq
As discussed before,  ${\cal F}_{bulk}=
-{\pi c\over 6}T^2L$ in a massless bulk theory, 
where $c$ is the central charge of the
conformal field theory, $c=1$ here. The second term in (\ref{totalf}) is the 
boundary free energy.

Although the equations (\ref{gentba}) for $\epsilon(\t)$ cannot 
be solved explicitly  
for all temperatures, the free energy is
easy to evaluate as $T\to 0$ and $T\to\infty$,  
as we
will show next. Moreover, one can extract the  
analytic
values of critical exponents by looking at the form of the expansions  
around
these fixed points. Also, they are straightforward to solve numerically  
for any
$T$.

Several notes of caution
are necessary. At the order we are working, the formula  for the entropy is not quite  
correct, because
there are $1/L$ corrections to the Stirling formula used in its  
derivation.
Also,
at this order, the logarithm of the partition function is not ${\cal  
E}-T{\cal
S}$: it depends not only on the saddle point
value of the sum over all states, but also on fluctuations. Their net
effect is that we cannot compute the $g$ factors from ${\cal F}$ alone.
However, both of these corrections are subleading contributions to the  
bulk
free energy, and do not depend on the boundary conditions. Therefore   we can still
 compute {\bf differences} of $g$  
factors from
${\cal F}$; the corrections are independent of the boundary scale  
$\t_B$ and
cancel out of the difference.

We can evaluate the impurity free energy explicitly in several limits.
In the IR limit $T/T_B\rightarrow 0$ the integral is dominated by
$\t\rightarrow\infty$
where the source terms in (\ref{gentba}) become very big. Hence
$\epsilon_r(\infty)=\infty$
and the impurity free energy vanishes in this limit.
In the UV limit $T/T_B\rightarrow \infty$ the integrals
are dominated by the region where $-\t$ is large
so that the source terms disappear
in (\ref{gentba}) and the $\epsilon_r$ go to constants. These are found by using the 
alternative form (\ref{universal}), which reads here, denoting $x_j=e^{\epsilon_j/T}$
\begeq
x_j=\prod_k (1+x_k)^{N_{jk}/2}
\endeq
One finds
\begeq
x_n\equiv e^{\epsilon_n(-\infty)/T}=(n+1)^2-1 ;\qquad \quad 
x_{1,2}=\gamma\label{xsols}
\endeq
Therefore we obtain
\begeqar
\ln{g_N\over g_D}=&\left.
{-{\cal F}_{imp}\over T}\right|_{UV}-\left.{-{\cal F}_{imp}
\over T}\right|_{IR}\nonumber\\
=&\sum_{n=1}^{t-2}
I^{(n)}\ln(1+1/x_n)+(I^{(+)}+I^{(-)})\ln(1+1/x_\pm)
\endeqar
where
$$I^{(r)}\equiv\int {d\t\over 2\pi}\kappa_r(\t)=\tilde\kappa(0).$$
From results given in the appendix, one finds 
 $I^{(n)}=n/2$ and $I^{(+)}+I^{(-)}=\gamma/2$, and thus
\begeqar
\ln{g_N\over g_D}&=&{\gamma\over 2}\ln{\gamma+1\over
\gamma}+\sum_{n=1}^{\gamma-1}{n\over 2}\ln{(n+1)^2\over n(n+2)}\nonumber\\
&=&{1\over 2}\ln (\gamma+1)={1\over 2}\ln t.
\endeqar
This is in agreement with the ratio calculated from conformal field  
theory.

We can also find the dimension of the perturbing operators.
{}From the equations one deduces the following expansions  
for
$T/T_B$ large:
$$Y_r(\t)=e^{\epsilon_r(\t)}=\sum_j Y_r^{(j)}e^{-2j\gamma\t/(\gamma+1)}.$$
As a result it is straightforward to see that near $\lambda=0$,
${\cal F}$ can be expanded in powers
of $(T_B/T)^{2\gamma/(\gamma+1)}$.
On the other hand we expect ${\cal F}$ to be an analytic
function of $\lambda^2$. Hence
\begeq
\lambda\propto \left(m e^{\t_B}\right)^{\gamma/(\gamma+1)}.
\endeq
This agrees with the conformal result that the perturbing
operator $\cos[ \beta\Phi(0)/2]$ has boundary dimension
$d=1/(\gamma+1)=\beta^2/8\pi$.
In the IR limit of $T/T_B$ small, one can expand out the kernels  
$\kappa_r$ in
powers of $\exp(\t_b-\t)$. This leads to the fact 
that the irrelevant operator which
perturbs the Dirichlet boundary conditions has dimension $d$=2. This is  
the
energy-momentum tensor. (Recall that there is another irrelevant  
operator in
the spectrum with dimension $d=\gamma+1$, which for $0\le \gamma <1$ is the  
appropriate
perturbing operator.)

\section{Using the TBA to compute static transport properties}

Let us pause for a moment to compare the gas of Yang-Baxter interacting quasi particles
to say free fermions. Within the TBA, the interactions have been fully encoded into
non trivial pseudo energies $\epsilon_j(\theta)$: that is, at temperature $T$,
the filling fractions of the various species are not independent, but correlated
via the coupled integral equations discussed previously.  This has some striking consequences.
For instance, we see from (\ref{xsols}) that the filling fraction of kinks or antikinks at
 rapidity $-\infty$ (ie at vanishing bare energy) is $f={1\over t}$. Except for $t=2$ (which is a free
fermion theory) there is no symmetry between particles and holes.
It is important to realize that
the interactions would have other effects, in general, for other questions asked. For instance,
in the case of free fermions, 
 the total density $n=\rho+\rho^h$, $\rho=nf$, the {\sl
fluctuations} also depend on the $\epsilon_j$ through the well known formula 
$\overline{\left(\Delta \rho
\right)^2}=
n f(1-f)$. Such a formula does not hold in the present case: the fluctuations of the various
species are correlated - their computation plays an important role in the DC noise at non 
vanishing temperature
and voltage, see \cite{FSnoise}. Similarly, physical operators  have complicated 
matrix elements in the multiparticle basis; the current for instance is able to create {\sl any}
neutral configuration of  quasiparticles by acting on the vacuum. There is thus a somewhat deceptive
simplicity in what we have done so far. However, for the DC conductance, it turns out that
the knowledge of the distribution functions is all that is necessary,
so for that particular aspect, our quasiparticles are not so far from free ones.

\subsection{Tunneling in the FQHE}

At this stage, it is useful to recall the tunneling problem of the introduction: we had L and R moving 
electrons that were backscattered by the impurity. Certainly if a R moving particle bounces back on the
gate voltage to become a left one, the charge $Q_R+Q_L$ is conserved. Now $Q_L+Q_R$ is essentially
 the charge of the
even field in the manipulations discussed in the introduction, which we found has no dynamic indeed.
Now when a R mover bounces back as a L mover, there is a change in the non conserved
charge $Q\equiv Q_R-Q_L$; this one is proportional to the charge of the odd
 field, which has a non trivial 
dynamics. Now, following carefully the formulas for bosonization, one finds the simple
result that a right moving kink, for which ${\beta\over 2\pi}\int \partial_x\Phi=1$, also has 
physical charge $Q=1$, and similarly for antikinks and left moving particles. Therefore, the non conservation
of the physical charge due to backscattering is the same as the non conservation of charge in 
the boundary sine-Gordon model. More precisely, when a kink comes in and bounces back as an antikink, 
as happens most of the time near the UV fixed point (Neumann boundary conditions), the charge $Q$ is 
conserved in the original problem. On the other hand, when a kink bounces back as a kink, as happens 
near the IR fixed point (Dirichlet boundary conditions), the charge in the original
problem is not conserved; rather, $\Delta Q=-2$. Let me stress here that the kink in the boundary sine-Gordon
theory however would look horribly complicated in the original problem, because the changes of variables we have performed
are non local. Only the conserved {\sl charge} is easy to follow \footnote{Charge is like current here,
where we have set the Fermi velocity equal to one.}.

\subsection{Conductance without impurity}

In the absence of impurity, that is with Neumann boundary conditions
in the original boundary problem, charge is straightforwardly transported. A right 
moving kink or antikink just goes through. Of course, if there are as many
particles of each specie, no current is transported overall. If however,
a voltage $V$ is applied, kink and antikink are at a different chemical potential,
$\mu=\pm {V\over 2}$ - this follows since the $U(1)$ charge in the boundary sine-Gordon model
is nothing but the physical charge $Q$. The current that flows through the system is thus
\begeq
I=\int_{-\infty}^\infty (\rho_+-\rho_-)(\theta)d\theta
\endeq
We can use our TBA to evaluate this expression quickly. First, we introduce the filling 
fractions
\begeq
f_{\pm}={1\over 1+e^{(\epsilon_\pm\mp{V\over 2})/T}}
\endeq
Second, we observe that  the very convenient identity 
$n_j=\rho_j+\rho_j^h={1\over 2\pi}{d\epsilon_j\over d\theta}$ holds, and that,
moreover, $\epsilon_+=\epsilon_-\equiv \epsilon$.

\smallskip
\noindent {\bf Exercise}: Prove these  two statements by staring at the TBA equations.
\smallskip

It thus follows that
\begeq
I={1\over 2\pi}\int_{-\infty}^\infty\left(f_+-f_-\right){d\epsilon\over d\theta}d\theta
\endeq
and thus
\begeq
I={T\over 2\pi}
\int_{-\infty}^\infty d\theta {d\over d\theta}\ln\left[
{1+e^{-V/2T}e^{-\epsilon/T}\over 1+e^{V/2T}e^{-\epsilon/T}}\right]
\endeq
The current is thus entirely determined by the values of $\epsilon$ at 
$\pm\infty$, exactly like for the central charge. As before, $\epsilon(\infty)=\infty$,
but the value of $\epsilon(-\infty)$ now does depend on the voltage. One finds
in fact, solving again (\ref{universal}) but with a voltage, 
\begeq
e^{\epsilon_n(-\infty)/T}=\left[{\sinh(n+1)V/2tT\over\sinh V/2tT}\right]^2-1,
~~e^{\epsilon_\pm(-\infty)/T}={\sinh (t-1)V/2tT\over\sinh V/2tT}
\endeq
(observe one recovers the result (\ref{xsols}) as $V\to 0$) 
from which an elementary computation shows the simple result (recall $g={1\over t}$)
\begeq
I={gV\over 2\pi}
\endeq
The bizarre factor of $2\pi$ occurs here because we have set $e=\hbar=1$ (recall 
that in physical units, $I=g{e^2\over h}V$).

\smallskip
\noindent {\bf Exercise}: Prove the last two formulas.
\smallskip

This is what one expected, and of course there are quicker ways to derive this result.
The point however, is that the same computation carries over without much additional
difficulty to the case where the impurity is present. 

\subsection{Conductance with impurity}

In the general case, we will write the source drain current as $I=I_0+I_B$
where $I_0={gV\over 2\pi}$ is the current in the absence of backscattering, 
and $I_B$ is the backscattered current. In the original problem, $I_B$ is 
for instance the rate at which the charge of the right moving edge is depleted. 
Of course, $\partial_t Q_L=-\partial_t Q_R$ in each hopping event, so $I_B=\partial_t{Q\over 2}$. 
In the steady state this rate is constant. When for instance $V$ is positive, 
there are more kinks than antikinks injected with a thermal distribution
into the system from their respective infinite reservoirs; it is assumed that these reservoirs 
are so big that the backscattering does not change 
their properties. 

We now derive an analytic expression for this
backscattering current using  a kinetic rate
equation for  quasiparticles
of the Bethe ansatz.
It is possible to compute the rate of change of $\Delta Q/2$
in the basis of the Bethe ansatz quasiparticles, since
each scattering event of a kink (antikink) into an antikink (kink)
changes the physical charge $\Delta Q/2$ by $-1$ (by $ (+1)$).
This kinetic equation is of course very familiar.
However, in general there would be no reason
why it should be applicable to an
interacting system.
But it is exact in the case that we are considering,
even though the system is interacting.
The reason for this lies in the constraints
of integrability: as discussed above, in the very
special quasiparticle  basis of the  Bethe ansatz,
these quasiparticles scatter off of the point contact independently
(``one-by-one''), and all quasiparticle production processes
are absent \footnote{Some more detailed justifications are available; see \cite{LSdual}
and references therein.}.

This allows us to express the rate of
change in $\Delta Q$, in terms of the transition probability
$|T_{+-}|^2$ (recall that in the unfolded point of view $T_{+-}=R_{++}$)
 and the number of  kinks and antikinks
(carriers of charge $\Delta Q =\pm 1$) in the  rapidity
range between $\theta$ and $\theta + d\theta$
$$
n_\pm(\theta) f_{\pm} (\theta) d\theta,
$$
where $n_\pm$ is the density of states and $f_{\pm}$ are the filling
fractions.
\noindent The number of kinks of rapidity $\theta$
that scatter into antikinks per unit time is
\begeq
|T_{+-}(\theta)|^2 \rho_{+-} d \theta \label{rateplusminus}
\endeq
where
$\rho_{+-}$ is the probability that the initial kink state
is filled and the final antikink state is empty (in all these quantities, there is also
a $V$ dependence, which we keep implicit here).
For a system of free fermions we would have
$$\rho_{+-} = f_+ ( 1-f_-)
$$
but in our interacting system we only
have
$$\rho_{+-} = f_+ - f_{+-}
$$
where $f_{+-}$ is the probability that
 {\it both}, the kink and the antikink states
are filled.  For the number of antikinks of rapidity $\theta$
that scatter into kinks per unit time
one finds a formula similar to (\ref{rateplusminus}),
 with $\rho_{+-} \to \rho_{-+}$.
In the final rate equation, only the difference between
these two probabilities
$$
\rho_{+-}-\rho_{-+}=  f_+ - f_-$$
 appears. Notice that the unknown $f_{+-}=f_{-+}$
has cancelled out (it can in fact be determined by
techniques more elaborate than the TBA \cite{FSnoise}).
Therefore, the backscattering current
is
\begeq
I_B(V) =- \int d  \theta~ n (\theta)
|T_{+-}(\theta-\theta_B)|^2 [f_+(\theta) - f_-(\theta)]\label{finalresult}
\endeq
All ingredients in this formula are exactly known:
the scattering matrix has a simple analytic form
and the occupation factors and densities
of state are obtained exactly from the thermodynamic
Bethe ansatz (TBA).
Notice that this equation is valid for any value of the
driving voltage $V$. It thus automatically describes
non-equilibrium transport.

By the same manipulations as before, it then follows that 
\begeq
I={1\over 2\pi}\int_{-\infty}^\infty \left(f_+-f_-\right)\left|T_{++}\right|^2 {d\epsilon\over d\theta}d\theta
={T\over 2\pi}\int_{-\infty}^\infty d\theta 
{1\over e^{2(t-1)(\theta-\theta_B)}}{d\over d\theta}\ln\left[
{1+e^{-V/2T}e^{-\epsilon/T}\over 1+e^{V/2T}e^{-\epsilon/T}}\right]
\endeq
Of special interest is the linear conductance, which we obtain
by taking a derivative at $V=0$; this gives, after reinserting the factor $2\pi$,
\begeq
G={(t-1)\over 2}\int_{-\infty}^\infty d\theta{1\over 1+e^{\epsilon/T}}
{1\over \cosh^2\left[(t-1)(\theta-\ln(T_B/T))\right]}
\endeq
The resulting curve is shown in figure 17, together with experimental results 
\cite{exper}
and 
the results of Monte Carlo simulations \cite{Moon}, for $g={1\over 3}$. 

\begin{figure}[tbh]
\centerline{\psfig{figure=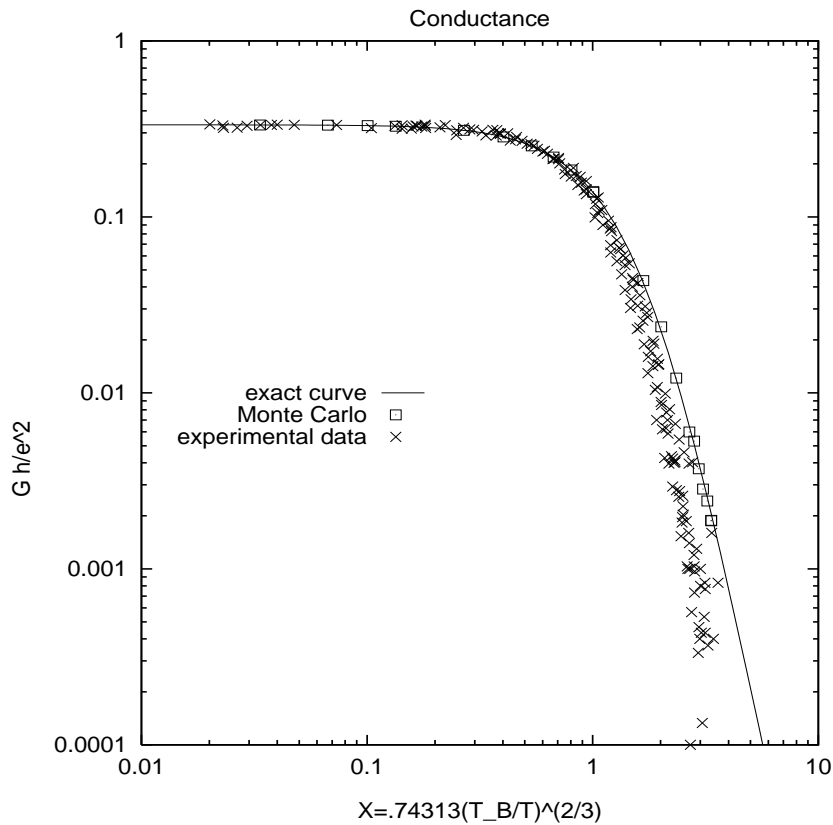,height=4.15in,width=4.5in}}
\caption{Comparison of the field theoretic result with MC simulations and experimental
data for $g={1\over 3}$.}
\end{figure}

The agreement with the simulations is clearly very  good (the is one and only one fitting parameter - the horizontal scale - , accounting for the unknown, non universal
 ratio of  the experimental gate voltage (a ``bare'' quantity)
 to  the parameter $\lambda$ in our renormalized field theory). As far as the experimental
data go, it is also very satisfactory, except 
in the strong backscattering regime. Recall however that the 
field theoretic prediction holds true only in the scaling limit: 
the experimental data are still quite scattered for low values of $G$, indicating that
this limit is not reached yet - actually the ``noise'' is of the same order of magnitude
as the discrepancy from the theoretical curve, as reasonably expected.

\smallskip
\noindent{\bf Exercise:} The problem had been solved previously \cite{KF} in the simplest
case of $g={1\over 2}$, where one can refermionize the hamiltonian for the boson $\phi^e$.
Look at this solution, and compare with what we have just done: what is the meaning
of kink and antikink, what is the bulk scattering, the boundary scattering?

\vfill
\eject

\huge \noindent{\bf Conclusions: further reading and open problems}

\normalsize

\bigskip
\bigskip

I am now leaving you at the 
beginning of a very exciting domain. Let me suggest some further reading and open problems.

\smallskip

These lectures have stopped short of really tackling the problem 
of boundary fixed points classification. Equipped with what you learned here,
you should not have much difficulty reading the paper of J. Cardy on boundary states 
\cite{Johnbdrst}. This paves the way to questions that are still open. For instance,
the problem we have studied at length generalizes,
for  tunneling 
in quantum wires where the spin of the electrons has to be taken into account, to a 
 ``double sine-Gordon problem'', involving two bosons. Surprisingly it has
been shown  \cite{KF} that new non trivial fixed points do exist in that case, besides the 
obvious Neumann and Dirichlet 
possibilities. With a few exceptions
\cite{YK}, \cite{AOS},
nobody knows how these fixed points precisely look like!  As I will mention again below,
what we have discussed is also very close to the Kondo problem. You can learn more
about fixed points and conformal invariance by reading the papers of Affleck and Ludwig 
on the multichannel Kondo problem \cite{ALreview}. There, you will also discover an aspect 
that I have neglected by lack of space: how multipoint correlators can be evaluated at
fixed points by further using conformal invariance  \cite{ALgreens}.

\smallskip

The integrable approach can also be pushed further to 
allow the computation of AC properties, together with space and time dependent Green functions,
in the cross-over regime.
The idea here still relies on massless scattering; but now, one has to evaluate 
matrix elements of physical operators, and these are usually pretty complicated. Moreover,
an infinity of these matrix elements are a priori needed: for instance, the current operator
is able to create {\sl any} neutral configurations of quasiparticles out of the vacuum! 
It turns out however that, first, the matrix elements can be determined by algebraic techniques 
\cite{smirnov},  \cite{giuseppeetal} (the latter reference is recomended as a first reading;
the first is a bit hard to read), and second, in many cases, only a few of these matrix elements
are required to obtain controlled accuracy all the way from the UV to the IR fixed point.  Using that 
technique, for instance the current current correlator itself can be evaluated,
at least at $T=V=0$ \cite{FLeSk} (there does remain a non trivial dependence upon space, time, and the coupling $\lambda$).
Determining correlators with a finite temperature or voltage is still more difficult; some progress in that
direction has been made \cite{LeLeSaSa}, \cite{LSVT}, but a lot remains to be done.

\smallskip

In another direction, for those of you who are more formally oriented,
it should be clear that what I just described is the tip of an iceberg
of beautiful mathematical structures: see \cite{FLeS},\cite{FLeSii} and the series 
\cite{BLZ}. Let me just mention here that the Kondo problem, which 
 would be described by 
\begin{equation}
\label{hamilkondo}
H\equiv H^e=\frac{1}{2}\int_{-\infty}^0 dx \
\left[\Pi^{2}+(\partial_x\Phi)^2\right]+
\lambda \left(S^+e^{-i\beta\Phi(0)/2}+S^-e^{i\beta\Phi(0)/2}\right),
\end{equation}
is just around the corner: it actually does have deep relations with the 
boundary sine-Gordon model, and with the subject of quantum monodromy operators.
Especially exciting results have actually appeared recently, concerning 
an exact duality between the UV and IR regimes of the problem \cite{F},\cite{FS}, and exhibiting 
tantalizing relations with the recent breakthroughs in 4D SUSY gauge theories \cite{SW}.

\smallskip

It is also fair to stress that the methods developed within the context of
quantum impurity problems can be generalized to different systems of physical interest
in $1+1$ dimensions: an example is the amazing recent mapping of the two-ladder
problem onto an $SO(8)$ Gross Neveu model \cite{BFLi}. It is very likely
indeed that more such problems are awaiting us in the near future.

\smallskip

Finally, the traditional question is, can any of this be generalized to
more than $1+1$? Well, the recent excitments in string theory are centered
around somewhat similar ideas in $3+1$, where, roughly, integrability
is replaced by supersymmetry, an incredibly powerful tool. As for $2+1$, I 
don't quite think it's over yet.

\bigskip
\noindent{\bf Acknowledgments}: The material in these notes is partly based on collaborations 
with I. Affleck, P. Fendley, A. Leclair, 
F. Lesage, A. Ludwig, M. Oshikawa, P. Simonetti, S. Skorik and N. Warner; I also 
greatly benefitted from questions by the school students, the organizers, and my fellows
co-lecturers - many thanks to all of them. The work was supported by the USC, the DOE, 
the NSF (through the NYI program), and the 
David and Lucile Packard Foundation.  

\bigskip
\noindent{\bf Dedication}: Many friends from ``Le plateau'' have disappeared
since I left France,  and their memory weighs  very much on my mind. I am especially thinking of Heinz Schulz in this early december:  it is to him  that I would like to dedicate these notes.

\bigskip
\bigskip
\large
\noindent{\bf Appendix: Kernels}
\normalsize

\bigskip

We use the following convention for Fourier transform:
$$\tilde f(y)=
\int^{\infty}_{-\infty} {d\t\over 2\pi} e^{i 2 \gamma\t y/\pi }
f(\t),$$
where $\gamma=t-1$. The bulk kernels $K_{jk}$ are well known; they can be written in the  
form ($\pm$ stand for kink and antikink)
\begeqar
\tilde K_{jk}&=&\delta_{jk}
-2{\cosh y \cosh(\gamma-j)y \sinh ky
\over \cosh \gamma y \sinh y}
\qquad j,k=1\dots \gamma-1;\ j\ge k\nonumber\\
\tilde K_{j,\pm}&=&-{\cosh y
\sinh  jy \over \cosh \gamma y
\sinh y}\nonumber\\
\tilde K_{\pm,\pm}&=&\tilde K_{\pm,\mp}
=-{\sinh  (\gamma-1)y \over
2 \cosh \gamma y \sinh y}\\
\endeqar
with $K_{jk}=K_{kj}$. The boundary kernels are
\begeqar
\tilde \kappa_j&=&
{\sinh j y \over 2 \sinh y \cosh \gamma y}\nonumber\\
\tilde \kappa_-&=& {\sinh(\gamma-1)y \over 2 \sinh 2y\cosh \gamma y }+ 
{1\over 2 \cosh y} \nonumber\\
\tilde \kappa_+&=& {\sinh(\gamma-1)y \over 2 \sinh 2y\cosh \gamma y }.\\
\endeqar
Finally, $\tilde{s}={1\over 2\cosh y}$. 

\vfill
\eject


\begin{thebibliography}{999}

\bibitem{hewson} A.C. Hewson, {\sl The Kondo Problem to Heavy Fermions}, Cambridge University Press,
1997

\bibitem{Glattli} L. Saminadayar, D.C. Glattli, Y. Jin, B. Etienne,
cond-mat/9706307,  Phys. Rev. Lett. {\bf 75} (1997) 2526.

\bibitem{Weizmann} R. de-Picciotto, M. Reznikov, M. Heiblum,
V. Umansky, G. Bunin, D. Mahalu, Nature {\bf 389} (1997) 162.

\bibitem{Sarma} {\sl Perspectives in Quantum Hall Effects}, S. Das Sarma and A. Pinczuk Eds., 
Wiley, 1997

\bibitem{sudiprmp} A. J. Leggett, S. Chakravary, A.T. Dorsey, M. P. A.
Fisher, A. Garg, and W. Zwerger, Rev. Mod. Phys. {\bf  59} (1987) 1.

\bibitem{uli} U. Weiss, {\sl Dissipative Quantum Mechanics}, World Scientifc, Singapore, 1998


\bibitem{andrei} N. Andrei,
K. Furuya, J. Lowenstein, Rev. Mod. Phys. {\bf  55} (1983) 331.

\bibitem{wiegmann} P. B. Wiegmann, A. M. Tsvelick, JETP Lett. {\bf 38} (1983)
591

\bibitem{Shankarrmp} R Shankar, Rev Mod Phys {\bf 66} (1994) 129.

\bibitem{KF} C L Kane and M P A Fisher, Phys. Rev. {\bf B46} (1992)
15233; {\bf B46} (1992) 7268.

\bibitem{Hald}F.D.M. Haldane, J. Phys. {\bf  C14} (1981) 2585.

\bibitem{Wen}X.G. Wen, Phys. Rev. {\bf B41} (1990) 12838;
Phys. Rev.  {\bf B43} (1991) 11025.

\bibitem{exper}F.P. Milliken, C.P. Umbach, R.A. Webb, Solid State
Comm. {\bf 97}, (1996) 309.

\bibitem{Moon}K. Moon, H. Yi,  C.L. Kane,
S.M. Girvin and M.P.A. Fisher, Phys. Rev. Lett. {\bf 71} (1993) 4381.

\bibitem{Nozieres}P Nozi\`eres, J. Low Temp. Phys. {\bf 17} (1974) 31;
P Nozi\`eres and A Blandin, J de Physique (Paris)
{\bf 41} (1980) 193.

\bibitem{ALreview} For reviews, see
I. Affleck, in {\it Correlation Effects
in Low-Dimensional Electron Systems}; Eds.: A. Okiji and
N. Kawakami (Springer-Verlag, Berlin, 1994), cond-mat/9311054;
A. W. W. Ludwig, Int. J. Mod. Phys. {\bf B 8} (1994) 347; in Proceedings of
the ICTP Summer School on {\it Low Dimensional
Quantum Field Theories for Condensed  Matter Physicists},
Trieste (Italy), Sept. 1992
 Editors:  S. Lundqvist, G. Morandi and Yu Lu
(World Scientific, New Jersey, 1995); Physica {\bf B 199\&200}
(1994) 406 (Proceedings of the International
Conference on Strongly Correlated Electron Systems,
San Diego, 1993).


\bibitem{ALnpbii}I. Affleck and A. W. W. Ludwig, Nucl. Phys. {\bf B 352}
(1991) 849;
{\it ibid} {\bf B 360} (1991) 641.

\bibitem{ALgreens} 
  A. W. W. Ludwig and I. Affleck,
 Nucl. Phys. {\bf B 428} (1994)  545.

\bibitem{houches88} Les Houches, session XLIX, 1988, {\sl Fields, Strings and
Critical Phenomena}, Eds. E. Bre\'zin and J. Zinn-Justin, Elsevier, New York, 1989.

\bibitem{Cardyreview} J. L. Cardy,
{\sl Conformal Invariance}, in {\sl Phase Transitions},
Eds. C. Domb and J. L. Lebowitz, vol. 11, Academic Press, New York, 1987.

\bibitem{polchinski} J. Polchinski, in Proceedings of the 1994 Les Houches Summer School,
hep-th/9411028.

\bibitem{philippe} P. Di Francesco, P. Mathieu, D. Se\'ne\'chal, {\sl Conformal
Field Theory}, Springer, New York, 1997.

\bibitem{tsvelikbook} A. M. Tsvelik, {\sl Quantum Field Theory in Condensed Matter
Physics}, Cambridge University Press, Cambridge, 1995. 

\bibitem{KC} P.Kadanoff and H.Ceva,
Phys. Rev. {\bf  B11} (1971) 3918.

\bibitem{Stone} M. Stone, {\sl Bosonization}, World Scientific, Singapore, 1994.

\bibitem{GSW} M. Green, J. Schwarz and E. Witten, {\sl Superstring Theory}, 
Cambridge University Press, 1987.

\bibitem{zeta} E. T. Whittaker, G. N. Watson, {\sl A Course of Modern Analysis},
Cambridge University Press, 1990.

\bibitem{johnbdr} J. L. Cardy, Nucl. Phys. {\bf  B240} (1984) 512. 

\bibitem{Apostol} T. Apostol, {\sl Modular Functions and Dirichlet Series in Number Theory},
Springer (New York) 1990.


\bibitem{Ishibashi}N. Ishibashi, Mod. Phys. Lett. {\bf A4} (1989) 251.

\bibitem{IanMasaki} I. Affleck and M. Oshikawa, Nucl. Phys. {\bf  B495} (1997) 533.

\bibitem{Johnbdrst} J. Cardy, Nucl. Phys. {\bf  B324} (1989) 581.


\bibitem{ALprldegeneracy} I. Affleck and A. W. W. Ludwig, Phys. Rev.  Lett.
 {\bf 67} (1991) 161.

\bibitem{Paulkondo} P. Fendley, Phys. Rev. Lett. {\bf  71}, (1993) 2485.

\bibitem{cthm}A B Zamolodchikov, JETP Lett
{\bf 43} (1986) 730.

\bibitem{FLeS} P. Fendley, F. Lesage and H. Saleur,
J. Stat. Phys. {\bf 79} (1995) 799, hep-th/9409176.

\bibitem{Jack}I.G. Macdonald, {\sl Symmetric Functions and Hall
Polynomials},
Clarendon Press (1979); R.P. Stanley, Adv. in Math. {\bf 77} (1989) 76.

\bibitem{schmid} A. Schmid, Phys. Rev. Lett. {\bf 51} (1983) 1506.

\bibitem{FLSprli} P Fendley, A W W Ludwig, H Saleur,
Phys. Rev. Lett.  {\bf 74} (1995) 3005, cond-mat/9408068.

\bibitem{andre} D. Bernard, A. Leclair, Comm. Math.
Phys. {\bf  142} (1991) 99. 

\bibitem{sashaising} A.B. Zamolodchikov, Adv. Stud. Pure Math {\bf 19} (1989) 1.

\bibitem{SSW} H. Saleur, S. Skorik, N. P. Warner, 
Nucl. Phys. {\bf B441} (1995) 412.

\bibitem{ZZ}A B Zamolodchikov and AL B Zamolodchikov,
Ann Phys (N.Y.) {\bf 120} (1979) 253L.

\bibitem{YBreview} M. Jimbo, {\sl Yang-Baxter Equation in Integrable Systems}, 
Adv. in Mathematical Physics,
vol. 10, World Scientific (Singapore).

\bibitem{giuseppephysrep} G. Mussardo, Phys. Rep. {\bf  218} (1992) 215.

\bibitem{patrickbook} P. Dorey, {\sl Exact S matrices in Two Dimensional Quantum
Field Theory}, Cambridge University Press (1996). 

\bibitem{FT} L.D.Faddeev, L.A.Takhtajan, Phys.
 Lett. {\bf  85A} (1981) 375.

\bibitem{FStrieste} P. Fendley, H. Saleur,
{\sl Massless Integrable
Quantum Field Theories and Massless
Scattering in 1+1 Dimensions}, Proceedings of the Trieste Summer School
on High Energy Physics and Cosmology, July 1993, World Scientific (Singapore).

\bibitem{RS} N.Yu. Reshetikhin and H. Saleur, Nucl. Phys. {\bf B419} (1994)
507.

\bibitem{ZZmassless} A. B. Zamolodchikov, Al. B. Zamolodchikov, Nucl. Phys. {\bf B379} 
(1992) 602.

\bibitem{GZ} S Ghoshal and A B Zamolodchikov, Int. J. Mod.
Phys. {\bf A9} (1994) 3841, hep-th/9306002.

\bibitem{YY} C N Yang and C CP Yang, J Math Phys {\bf 10} (1969)
1115.

\bibitem{Alyoshatba} Al B Zamolodchikov,
Nucl Phys {\bf B342} (1991) 695.

\bibitem{PaulKen}P. Fendley and k. Intriligator, Nucl.Phys. {\bf B372} (1992) 533 

\bibitem{vladimir} V E Korepin, N M Bogoliubov,
A G Izergin, {\sl Quantum Inverse Scattering Method
and Correlation Functions}, Cambridge Univ. Press, Cambridge, 1993.

\bibitem{dilogs} A.N.Kirillov, N.Yu Reshetikhin,
J.Phys. {\bf A20} (1987) 1565, 1587.

\bibitem{FSnoise} P. Fendley, H. Saleur, Phys. Rev. {\bf B54} (1996)  10845. 

\bibitem{LSdual} F. Lesage and H. Saleur, {\sl Duality and IR Perturbation Theory
in Quantum Impurity Problems}, cond-mat/9812045

\bibitem{smirnov} F.A. Smirnov, {\sl Form 
Factors in Completely
Integrable Models of Quantum Field Theory}, World Scientific
(Singapore) and references therein.

\bibitem{giuseppeetal} J. Cardy, G. Mussardo, Nucl. Phys. B410 (1993), 451;
G. Delfino, G. Mussardo, P. Simonetti, Phys. Rev.
{\bf D51}, 6620 (1995).

\bibitem{FLeSk} F. Lesage, H. Saleur, S. Skorik, Nucl. Phys. {\bf B474},
(1996) 602.

\bibitem{LeLeSaSa} A. Leclair, F. Lesage, S. Sachdev and  H. Saleur, Nucl. Phys. {\bf B483} (1996) 579.

\bibitem{LSVT} F. Lesage and  H. Saleur, Nucl. Phys. {\bf  B493} (1997) 613.


\bibitem{YK} H. Yi and C. Kane, 
{\sl Quantum Brownian Motion in a Periodic Potential and the Multi Channel Kondo Problem},
cond-mat/9602099.

\bibitem{AOS} I. Affleck, M. Oshikawa and  H. Saleur, {\sl Boundary
Critical Phenomena in the Three State Potts Model}, cond-mat/9804117.

\bibitem{FLeSii} P. Fendley, F. Lesage and H. Saleur, J. Stat. Phys. {\bf 85} (1996) 211,
cond-mat/9510055.

\bibitem{BLZ} V. Bazhanov, S. Lukyanov and A.B. Zamolodchikov, 
Comm. Math. Phys. {\bf 177} (1996) 381, hep-th/9412229; 
Comm. Math. Phys. {\bf 190} (1997) 247, hep-th/9604044; Nucl.Phys. {\bf B489}
(1997) 487, hep-th/9607099.

\bibitem{F} P. Fendley, {\sl Duality Without Supersymmetry}, hep-th/9804108.

\bibitem{FS} P. Fendley, H. Saleur, {\sl Self-duality in Quantum Impurity 
Problems}, cond-mat/9804173, Phys. Rev. Lett. to appear

\bibitem{SW} N. Seiberg and E. Witten, Nucl. Phys. {\bf  B426} (1994) 19,
hep-th/9407087; Nucl. Phys. {\bf B431} (1994) 484, hep-th/9408099.


\bibitem{BFLi} H. Lin, L. Balents and  M.P. Fisher, {\sl Exact $SO(8)$ Symmetry in the 
Weakly Interacting Two-leg Ladder}, cond-mat/9801285. 





\end{thebibliography}
\end{document}